\documentclass[11pt,a4paper]{article}
\usepackage[a4paper, margin=2.5cm]{geometry}
\usepackage{authblk}
\usepackage[utf8]{inputenc}
\usepackage[T1]{fontenc}
\usepackage{stackengine}
\usepackage{amsmath}
\usepackage{amsfonts}
\usepackage{amssymb}
\usepackage{bbm}
\usepackage{mathtools}
\usepackage{array}
\usepackage{breqn}

\usepackage{floatrow}
\usepackage{graphicx}
\usepackage{tabularx}
\usepackage{multirow}
\usepackage{wrapfig}
\usepackage{titling}
\usepackage{rotating}
\usepackage{float}
\usepackage{color}	
\usepackage[numbers, sort&compress]{natbib}
\usepackage[font=footnotesize,labelfont=bf]{caption}
\usepackage{subcaption}
\usepackage{url}
\usepackage{rotating}

\graphicspath{{./graphics/}}
\usepackage[normalem]{ulem}					

\def\added#1{#1}

\newcommand{\rzero}{\mathcal{R}_0}

\newsavebox\solbox
\makeatletter
\newenvironment{sol}%
{
	\@parboxrestore%
	\begin{lrbox}{\solbox}%
		\begin{minipage}{0.9\linewidth}
		}
		{\end{minipage}\end{lrbox}%
	\begin{center}
		\framebox{\usebox\solbox}
	\end{center}
}
\makeatother

\title{Spatially--Structured Models of Viral Dynamics:\\ A Scoping Review}
\author{Thomas Williams$^1$, James M. McCaw$^{1,2}$ and James M. Osborne$^1$}
\date{{\footnotesize$^1$School of Mathematics and Statistics, University of Melbourne, Australia, $^2$Centre for Epidemiology and Biostatistics, Melbourne School of Population and Global Health, University of Melbourne, Australia\\ Address correspondence to James M. Osborne, jmosborne@unimelb.edu.au}}

\begin{document}
	
	\maketitle
	
	\vspace{-1.0cm}
	
	\tableofcontents
	
	\newpage
	
	\begin{abstract}
		There is growing recognition in both the experimental and modelling literature of the importance of spatial structure to the dynamics of viral infections \added{within the host}. Aided by the evolution of computing power and motivated by recent biological insights, there has been an explosion of new, spatially--explicit models for within--host viral dynamics in recent years. This development has only been accelerated in the wake of the COVID--19 pandemic. Spatially-structured models offer improved biological realism and can account for dynamics which cannot be well-described by conventional, mean-field approaches. However, despite their growing popularity, spatially--structured models of viral dynamics are underused in biological applications. One major obstacle to the wider application of such models is the huge variety in approaches taken, with little consensus as to which features should be included and how they should be implemented for a given biological context. Previous reviews of the field have focused on specific modelling frameworks or on models for particular viral species. Here, we instead apply a scoping review approach to the literature of spatially--structured viral dynamics models as a whole to provide an exhaustive update of the state of the field. Our analysis is structured along two axes, methodology and viral species, in order to examine the breadth of techniques used and the requirements of different biological applications. We then discuss the contributions of mathematical and computational modelling to our understanding of key spatially-structured aspects of viral dynamics, and suggest key themes for future model development to improve robustness and biological utility. 
	\end{abstract}

	\section{Introduction}
	\label{sec:intro}
	
	There is a growing recognition in the virus dynamics literature of the role of spatial structure in influencing viral spread. Following the rapid evolution of computing power, and especially since the outbreak of the COVID--19 pandemic, there has been an explosion of new, spatially--explicit models for within--host viral dynamics. In line with the rapid advancement of sophistication and biological realism in modelling for viral infections, the incorporation of spatial structure enables a description of  dynamics that cannot be well described by mean--field models, such as neighbour--based infection \cite{kumberger_et_al_accounting_for_space_cell_to_cell, imle_et_al_hiv_3d_collagen_matrix, saeki_and_sasaki_spatial_heterogeneity, kreger_et_al_synaptic_cell_to_cell, yakimovich_et_al_infectio}, or immune cell taxis \cite{levin_et_al_T_cell_search_influenza, kadolsky_and_yates_spatial_immune_surveillance, sego_et_al_covid_model, getz_rapid_community_driven_covid_model}. Spatially--structured models have also been shown to exhibit different dynamics to their mean--field counterparts under the same parameter values, suggesting that spatial structure may be necessary to improve estimates of biologically useful parameters \cite{bauer_et_al_agent_based_models_virus, graw_and_perelson_spatial_hiv, strain_et_al_spatiotemporal_hiv}.
	
	However, despite their advantages, spatial models have been underused in an application context. Spatially--structured models of viral dynamics are routinely cited far less frequently and are much less commonly used in biological applications compared to mean--field compartmental models. While this may be in part due to the simplicity of the latter approach (including the difficulty in calibrating spatial models to experimental data), this is also likely due to the variety of approaches taken to modelling spatial structure in viral infections, and a lack of consensus on what features should be included and how they should be implemented. Mean--field models are more or less united in the common language of ODEs (possibly with some slight modifications to account for, for example, age structure or stochasticity), yet spatially--structured models span a huge variety of methods: from PDEs to compartmental approaches, to agent--based approaches, or combinations thereof. To add to the complexity, the actual nature of the spatial structure of infections by different viruses in different host tissues is also widely varied, depending on whether the infection is \emph{in vivo} or \emph{in vitro}, whether the host cells are motile or packed in tissue, how permissive the \added{extracellular environment} is to virion diffusion, and so on. There is therefore a need for a wide--reaching, intersectional analysis of the full breadth of spatially--structured viral dynamics models to make sense of the breadth of techniques used and the requirements of different applications.
	
	Earlier reviews of spatially--structured models of virus dynamics have generally focused on specific viruses or model methodologies. Graw and Perelson reviewed spatial effects in the dynamics of HIV and hepatitis viruses \cite{graw_and_perelson_spatial_hiv, graw_and_perelson_modeling_viral_spread}. Their work identified direct cell--to--cell spread of infection as a distinct motivator for the use of spatial models, and advocated for the use of agent--based models \cite{graw_and_perelson_modeling_viral_spread}. Bocharov \emph{et al.}, by contrast, focused on a PDE modelling approach specifically in HIV, and criticised a lack of models (as of 2012) which took into account tissue heterogeneity in HIV infection \cite{bocharov_review}. Bauer, Beauchemin and Perelson reviewed the use of agent--based models \cite{bauer_et_al_agent_based_models_virus}. Their work highlighted the difference in dynamics seen in spatially--structured compared to mean--field models, and explored the challenges in calibrating and applying agent-based models \cite{bauer_et_al_agent_based_models_virus}. They pointed out, that due to the motility of the target cells in HIV, this virus is somewhat of a special case, requiring a different approach to the treatment of spatial structure \cite{bauer_et_al_agent_based_models_virus}. The most recent survey of the literature on modelling the spatial spread of viral infections was carried out by Gallagher \emph{et al.} in 2018, in a review that mainly focused on influenza A infection \cite{gallagher_spatial_spread}. As with Graw and Perelson, the authors explored mechanisms and implementation of the cell--to--cell infection route, and explained the key role this has in determining not only the spatial structure but also the global dynamics of infection \cite{gallagher_spatial_spread}. In contrast to Graw and Perelson, the authors advocated \emph{against} the use of agent--based modelling at least for the immune response, preferring instead a compartmental approach \cite{gallagher_spatial_spread}.
	
	While these early reviews provided an important overview and insight into approaches to modelling the spatial structure of viral infection in varying domains, there has so far yet to be a review spanning the full scope of modelling methods and viral species. Moreover, there is a need to take stock of the rich body of work that has emerged in recent years, especially following the emergence of SARS--CoV--2. Here, we employ a scoping approach to the literature on spatially--structured models of viral dynamics in order to systematically and exhaustively review the full range of approaches used.

	\subsection{\added{How to read this paper}}
	The remainder of this paper is structured as follows. In Section~\ref{sec:methods} we present the methodology used for our scoping review, including the inclusion and exclusion criteria. We then present an overview of the key publications in Section~\ref{sec:overview}, before moving on to our detailed review of the field. \added{We do so by first examining the literature from the perspective of the methods and techniques used (in Section~\ref{sec:papers_by_methodology}), and then from the perspective of the biological insights these models have generated (in Section~\ref{sec:papers_by_virus}). We discuss distinct topics in these two sections, and in instances where a publication is cited more than once, it is to discuss different aspects of that publication: for example, its mathematical implementation in one section, and its contribution to biological knowledge in another. These sections are structured by model type and viral species modelled, respectively, for easy reference.} \added{Next, in Section~\ref{sec:innately_spatially_structured_dynamics}, we identify some of the main aspects of viral infection dynamics which have been identified in the literature as being inherently spatially structured, and discuss progress made in the modelling literature in describing and understanding these processes.} Finally, in Section~\ref{sec:future}, we present emerging trends for future development of \added{spatially--structured models of viral dynamics}, before closing the paper in Section~\ref{sec:conclusion}.

	\section{Methods} \label{sec:methods}
	
	\subsection{Search string}
	We conducted a search of the literature on September 2nd, 2024, using the search query
	\begin{sol}
		\vspace{1em}
		\hspace{1em}(``virus'' \emph{OR} ``viral'') \emph{AND} (``dynamic$^*$'' \emph{OR} ``kinetic$^*$'') ...\\
		\phantom{x}\hspace{9.2em}\emph{AND} (``spati$^*$''  \emph{OR} ``space'')...\\
		\phantom{x}\hspace{9.2em}\emph{AND} (``math$^*$'' \emph{OR} ``comp$^*$'' \emph{OR} ``sim$^*$'')...\\
		\phantom{x}\hspace{9.2em}\emph{AND} ``model$^*$''
		\vspace{1em}
	\end{sol}
	
	\noindent where \emph{AND} and \emph{OR} are the usual boolean operators and $^*$ is the wildcard token. We queried Scopus and Web of Science using this search string against article titles, abstracts and keywords. We also queried the search string against Scopus's pre--print database (this functionality was not available from Web of Science).

	\subsection{Inclusion and exclusion criteria}
	\label{sec:criteria}
	
	After merging results from both databases and screening for duplicates, we analysed the combined pool of studies and pruned for relevance. Our criterion for inclusion was studies that \emph{introduce or apply a spatially--explicit mathematical or computational model of within--host viral dynamics}. \added{While many models we found considered the viral infection of tissues within a single host, we also found a number of studies of interest which considered only \emph{in vitro} systems, or bacterial hosts. We deemed these studies relevant and useful to our discussion, and therefore adopted a broad interpretation of ``within--host viral dynamics'' as the dynamics of viral infection at a between--cell scale. Given that neither ``within--host'' nor ``between--cell'' featured in our search string, this interpretation will not affect the studies included in the initial search.}
	
	We excluded articles that met one or more of the following criteria:
	
	\begin{itemize}
		\item \emph{Review articles.} We omitted review articles from our search such as those mentioned in the previous section \cite{graw_and_perelson_modeling_viral_spread, graw_and_perelson_spatial_hiv, gallagher_spatial_spread, bocharov_review, bauer_et_al_agent_based_models_virus}, preferring instead to limit our focus to original modelling applications.
		\item \emph{Studies focused purely on within--cell dynamics.} Some studies modelled only the within--cell life cycle of viruses, especially hepatitis B and C \cite{ghaemi_et_al_whole_cell_model_hbv, nakabayashi_et_al_hbv_in_cell, knodel_et_al_surface_PDEs_HCV}. Since our focus is on models of viral infection spread \emph{between} host cells, we excluded these studies from our analysis.
		\item \emph{Studies without an explicit viral or infected population.} This includes some computational models which described only the immune system, sometimes with a general ``antigen'' population \cite{celada_and_seiden_comp_immune_model, grebennikov_et_al_spatial_lymphocytes, pinotti_et_al_three_player_antigen}.
		\item \emph{Studies with a mainly analytical focus.} Our search returned a substantial body of studies focused on the mathematical properties of spatial models of viral dynamics such as existence and smoothness of solutions, stability properties and conditions for Turing patterns (in fact, our database search found more studies that met this description than studies that were retained for analysis). While these works offer a substantial contribution to the collective literature on mathematical modelling of viral infections and are worthy of review in their own right, our work here is focused primarily on the application of such models and as such, we excluded these studies from our analysis. Note that while the determination of a study as ``analytically--focused'' as opposed to ``application--focused'' is difficult to articulate objectively, we found that the distinction between works with a mathematical readership in mind and those with a more interdisciplinary audience was usually very clearly determined. In the few instances where this distinction was unclear, we opted for including the article.
	\end{itemize}
	
	We conducted three iterations of screening of the studies found by our search, applying the inclusion criterion to first the titles, then the abstracts, then the full text of the remaining studies, discarding those which clearly did not meet the inclusion criterion. We only applied the exclusion criteria at the full text level. Of the 3939 unique studies returned by our search, we initially found 93 that were relevant to this review. The largest subset of publications which were removed on the basis of the exclusion criteria above were articles which were deemed to have an analytical focus (we found 111).  
	
\begin{figure}[h!]
    	\centering
    	\includegraphics[width=0.8\textwidth]{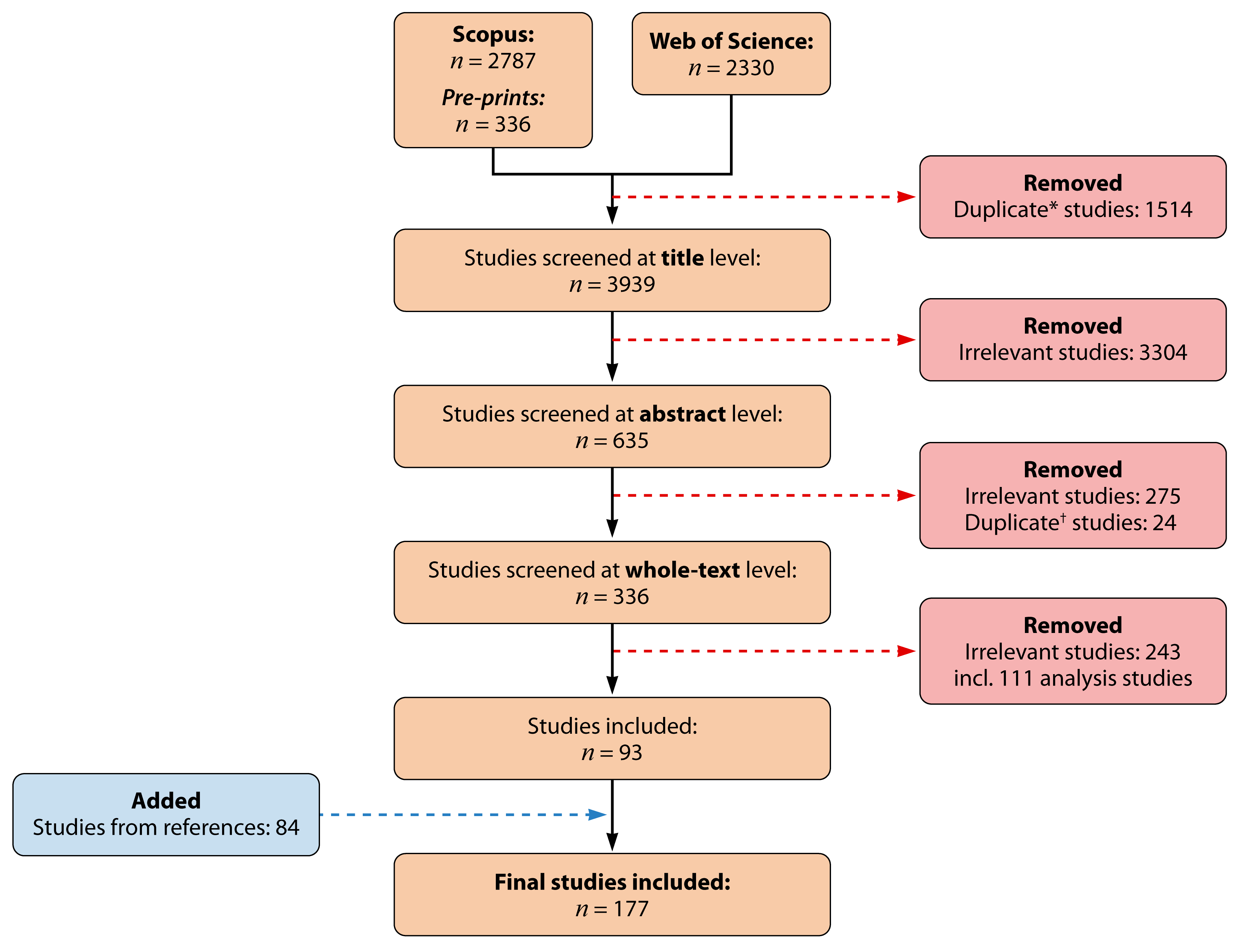}
    	\caption{Workflow for the scoping review. \added{We queried Scopus and Web of Science using our search string to find an initial list of potentially relevant studies, then sequentially refined the list of studies based on our inclusion and exclusion criteria. We then repeated this approach on the reference lists of the included studies to end up with our final list of accepted publications.} $^*$Duplicates automatically detected in JabRef. \mbox{$^{\dagger}$Duplicates} manually detected.}
    	\label{fig:search_schematic}
    \end{figure}

	After having applied our selection criteria, we then screened the references of each of the accepted articles to find any articles missed by our database search. This process found a further 84 relevant studies for a total of 177. We outline the general workflow of the literature search process in Figure~\ref{fig:search_schematic}. We conducted our search in English and only considered articles written in English. We included pre--prints in our search.

	\subsection{Categorisation}
	\label{sec:categorisation}
	
\begin{figure}[h!]
    	\centering
    	\includegraphics[width=0.85\textwidth]{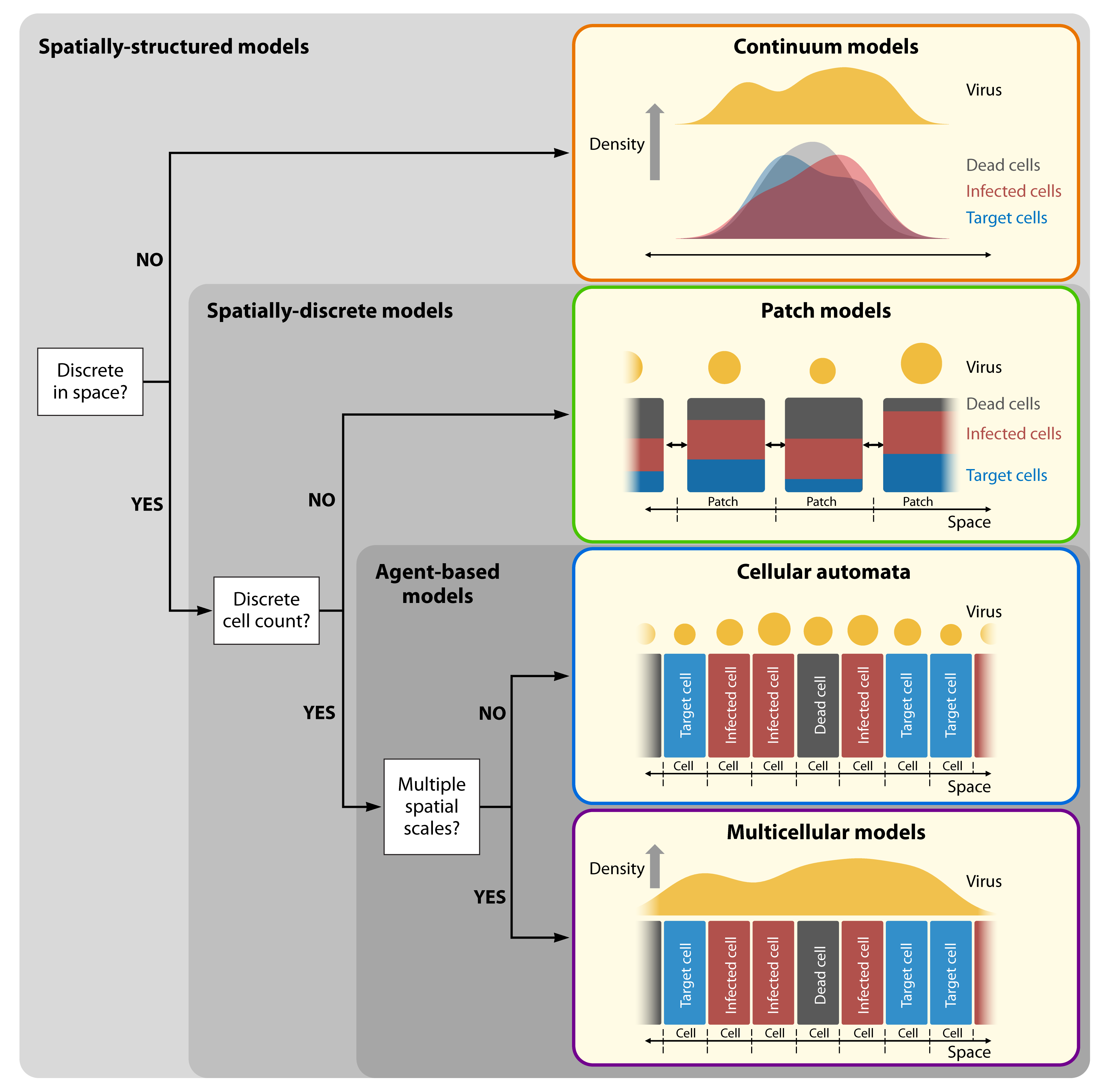}
    	\caption{\added{Classification for modelling frameworks discussed in this review. For each of the model frameworks discussed in this review (continuum models, patch models, cellular automata and multicellular models) we show an indicative diagram of how space is represented in the system. In continuum models, all quantities are modelled as densities in continuous space; in patch models, space is discretised into regions within which cells and other system quantities are represented as densities; in cellular automata, space is discretised into (typically smaller) regions, within which cells are represented by discrete counts. Commonly, as we depict here, discrete regions correspond to the cells themselves. Multicellular models, like cellular automata, include discretisation of space and track discrete counts of cells (again commonly discretising space into the regions occupied by the cells themselves) but also describe some other spatial scale, such as a continuous virus density.}}
    	\label{fig:model_classification}
    \end{figure}

	In order to analyse high--level trends in the literature, we collected for each accepted study the viral species modelled (where one was specified) and the model framework used. We considered \textbf{continuum models}, \textbf{patch models}, \textbf{cellular automata}, and \textbf{multicellular models} in this review, which we define as follows. Models in which the spatial domain and all model components are continuous in space (that is, they are represented by densities, rather than discrete agents) are classified as \textbf{continuum} models. These are overwhelmingly PDE models, especially reaction--diffusion systems. Of the models which include spatially--discrete regions, we define \textbf{patch} models as those which represent cells as a density. Such models typically comprise a lattice of connected spatial regions, each of which contains a (possibly stochastically--simulated) ODE system. Models which included only a few spatial regions which were hard--coded to represent specific organs or tissue regions were not included in our search. The remaining models to be classified comprise spatially--discrete regions and a discrete number of cells and can collectively be considered agent--based models. We subdivide these models into \textbf{cellular automata} and \textbf{multicellular models} based on whether the model considers a single spatial scale or multiple, respectively. For instance, a model in which the cells are taken to be discrete spatial regions and where the extracellular viral load is tracked as the quantity of viral density at each cell would be considered a cellular automaton, since the cell--scale is the only spatial scale considered by the model. By contrast, if extracellular viral load were represented as an overlying field governed by a PDE, for example, the model would be considered multicellular. Note that we consider models with a global virus reservoir which has equal access to all spatial regions as having a secondary spatial scale and therefore multicellular. We summarise this classification in Figure~\ref{fig:model_classification}, \added{and provide illustrations of how spatial structure is typically --- but by no means exclusively --- represented within each model framework}. We assign to each study a single modelling framework: for those which make use of multiple models of different frameworks, we categorise the article by the model with the highest level of spatial resolution used (according to our classification, which corresponds to the bottom of the tree in Figure~\ref{fig:model_classification}). For each study, we also recorded its number of citations according to Google Scholar as of September 2nd, 2024.

	\section{An overview of the use of spatially--structured \mbox{models} in viral dynamics}
	\label{sec:overview}
	
\begin{figure}[h!]
    	\centering
    	\includegraphics[width=\textwidth]{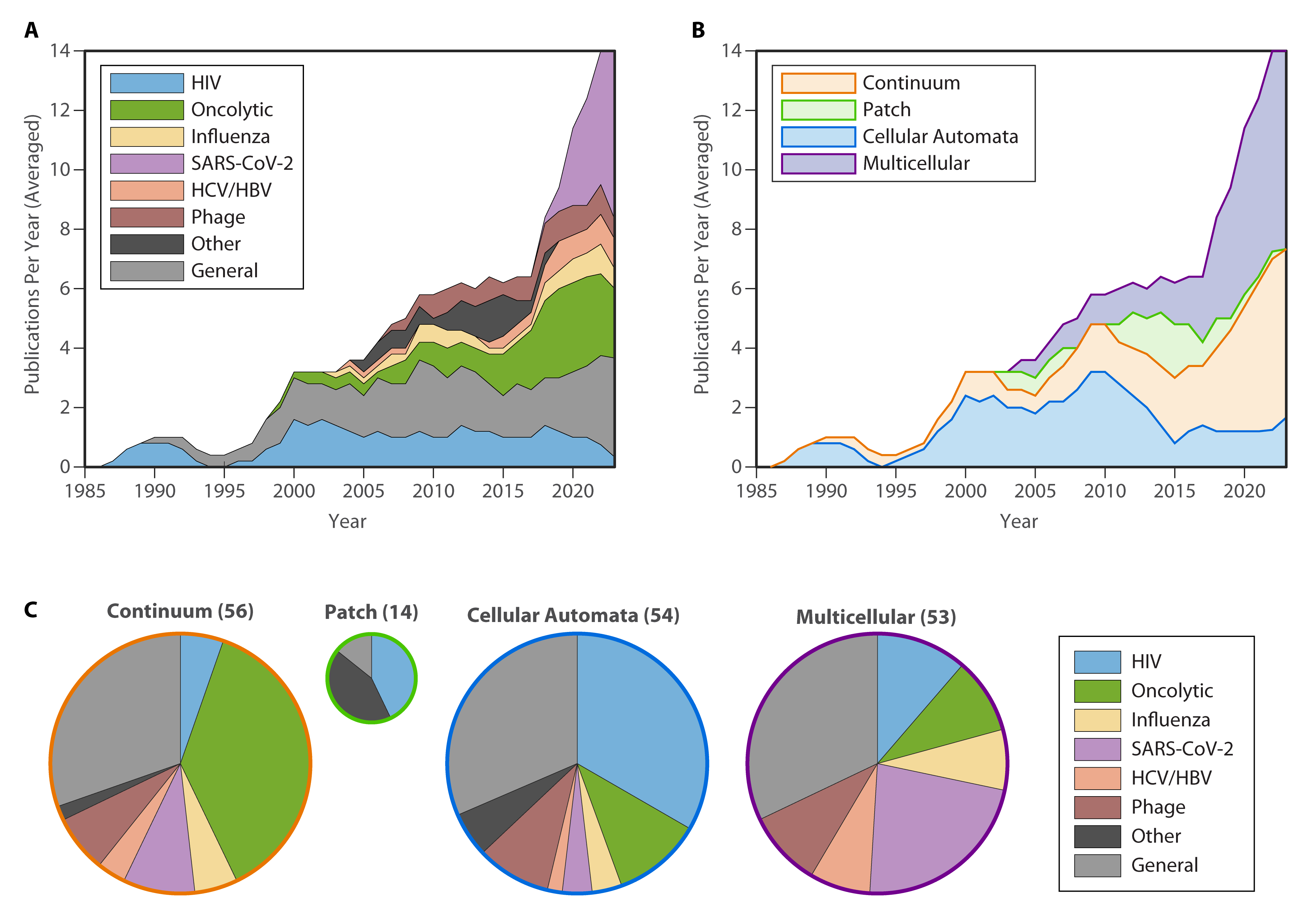}
    	\caption{Spatially--structured viral dynamics models over time (a) by virus species and (b) by methodology. \added{Here we plot the five--year moving average number of publications per year (for 2022 and 2023, we show the average number of publications from two years prior, up to 2023; we do not show data for 2024).} (c) Overall split of models by methodology. \added{The number of publications for each method is given in brackets}.}
    	\label{fig:lit_review_results}
    \end{figure}

	Our search procedure returned 177 publications, spanning 60 years of research across a diverse group of virus species and a wide array of methodologies. The final annotated list of publications retained for analysis can be found on FigShare\footnote{\url{https://doi.org/10.26188/27085984.v2}}. In Figure~\ref{fig:lit_review_results} we present a broad overview of the distribution of the studies found by our search over time, and by virus modelled and method used. Figures~\ref{fig:lit_review_results}(a)--(b) shows the averaged number of publications per year over time, computed by taking the average number of publications per year across the five--year interval centred on a given year. For 2022 and 2023 we compute the average over a shorter interval (from two years prior, up to 2023). We do not show data from 2024. In Figure~\ref{fig:lit_review_results}(a) we stratify these studies by virus species modelled and in Figure~\ref{fig:lit_review_results}(b) we stratify by methodology. In Figure~\ref{fig:lit_review_results}(c), we show the overall split of methodologies used and for which virus species.
	
	To our knowledge, the earliest spatial description of viral spread in tissue with mathematical models was by Koch in 1964, who used the diffusion (heat) equation to analyse the growth of viral plaques \emph{in vitro} \cite{koch_growing_viral_plaques}. However, the first ``model'' in a more recognisable sense, and in some ways the birth of the field of spatially--structured modelling for viral dynamics, was proposed some decades later by Pandey in 1989 \cite{pandey_1989_CA_model_HIV}. Pandey's model used a cellular automaton approach for the dynamics of HIV, and most of the spatial viral dynamics models in the early 1990s followed in this vein, as can be seen in Figures~\ref{fig:lit_review_results}(a)--(b) \cite{pandey_1989_CA_model_HIV}. Figures~\ref{fig:lit_review_results}(a)--(b) also show significant growth and diversification of the literature on spatial spread of viral infections in the first half of the 2000s, with the first spatially--structured models specifically for influenza, hepatitis B and C (HBV and HCV), oncolytic viruses and others, along with the first patch and multicellular models all published around a similar time. Since 2020 and the outbreak of the COVID--19 pandemic, there has been a spike in the development of spatial models for viral infections, with the average yearly number of publications roughly doubling from less than six in the mid--'10s to approximately 13 in 2023. Curiously, this cannot be entirely attributed to models for SARS--CoV--2, with models for the virus accounting for less than a third of recent publications (19 of 63 publications since 2020).

	As shown in Figure~\ref{fig:lit_review_results}(c), continuum models formed the largest proportion of the studies found by our search with 56, although between cellular automaton (54) and multicellular (53) approaches, we observed that agent--based models in general were a more popular approach (107 publications combined). Early agent--based models were almost exclusively cellular automata, although this approach fell out of favour in the 2010s, where multicellular approaches rose to prominence, as can be seen in Figure~\ref{fig:lit_review_results}(b). Patch models were the least popular approach with only 14 publications since 1985. These were mostly models for HIV and herpes simplex virus 2 (HSV--2) with six publications each, including the entire spatial modelling literature specifically focused on HSV--2. The other modelling approaches were also highly specific to particular viral species with the largest proportion of continuum models focused on oncolytic viruses and the largest share of cellular automaton models developed for HIV. Multicellular models, by contrast, were most commonly proposed for general viral infections and not a particular viral species.
	
\begin{figure}[h]
   	\centering
   	\includegraphics[width=0.85\textwidth]{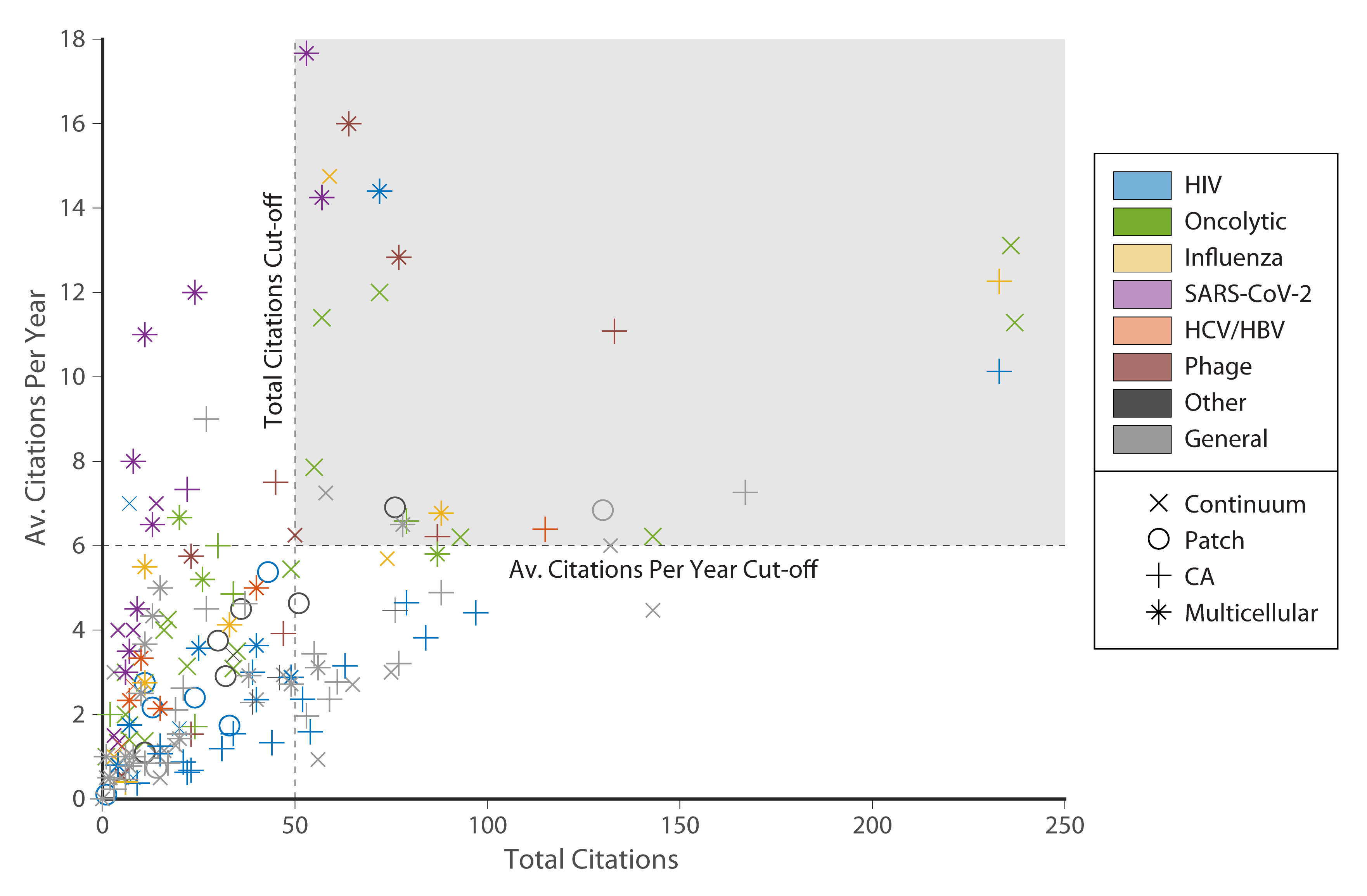}
   	\caption{\added{Identification of papers as highly--cited}. We show each accepted study for our scoping review by total number of citations and average citations per year since publication. Cutoffs for \added{definition as highly--cited} in our analysis are shown. \added{Citation data was collected from Google Scholar on September 2nd, 2024.}}
   	\label{fig:papers_scatter}
   \end{figure}

\begin{figure}[h!]
    	\centering
    	\includegraphics[width=\linewidth]{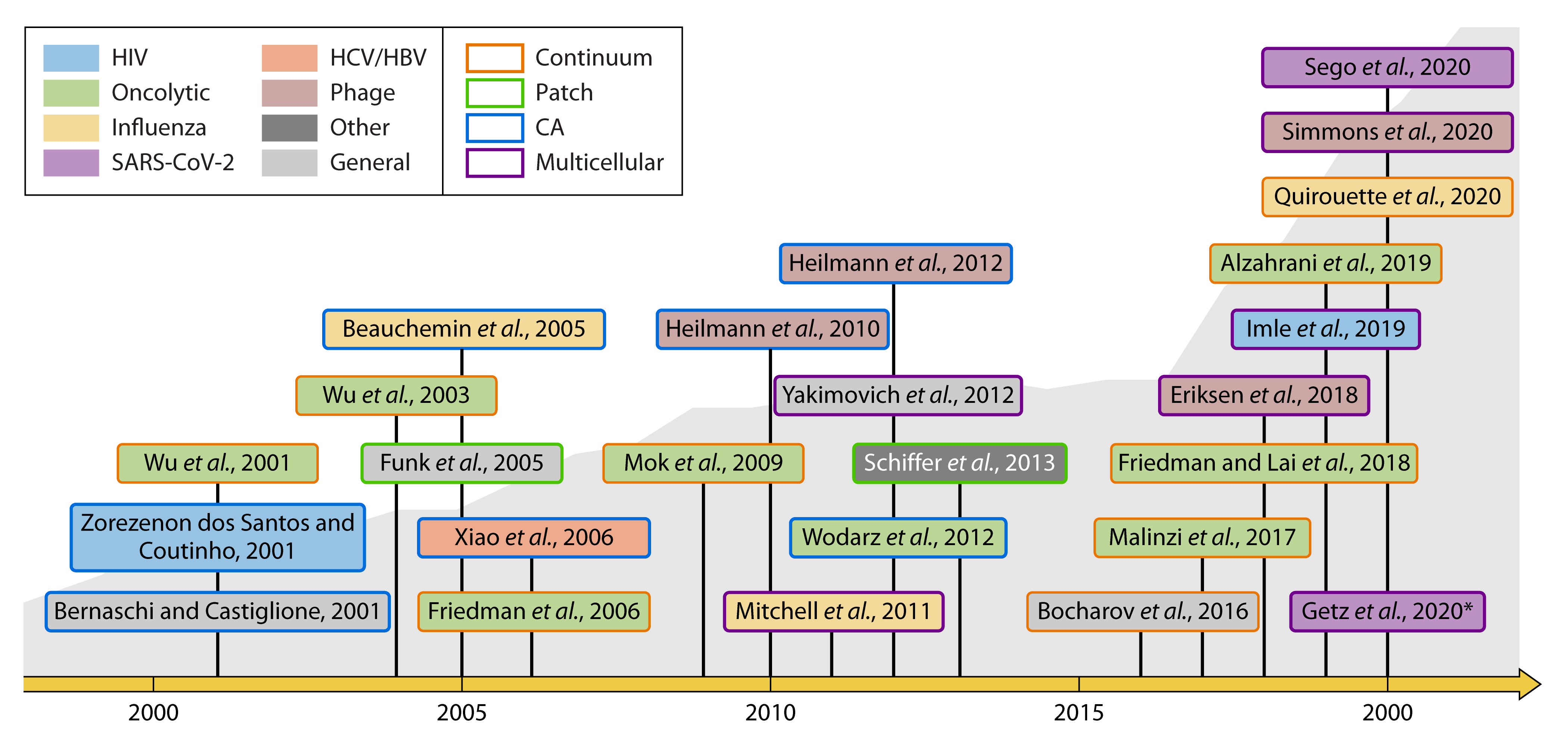}
    	\caption[\added{Highly--cited} spatially--structured viral dynamics models over time.]{\added{Highly--cited} spatially--structured viral dynamics models over time. Criteria for inclusion are more than 50 total citations and more than 6 citations per year on average since publication. $^*$Pre--print. \added{Highly--cited} publications are as follows: \textbf{Bernaschi and Castiglione} \cite{bernaschi_and_castiglione_design_and_implementation}, \textbf{Zorzenon dos Santos and Coutinho} \cite{dos_santos_and_countinho_dynamics_of_HIV_CA}, \textbf{Wu \emph{et al.}} \cite{wu_et_al_modelling_and_analysis_oncolytic}, \textbf{Wein \emph{et al.}} \cite{wein_et_al_validation_and_analysis_oncolytic}, \textbf{Funk \emph{et al.}} \cite{funk_et_al_spatial_models_virus_immune}, \textbf{Beauchemin \emph{et al.}} \cite{beauchemin_et_al_simple_cellular_automaton_model}, \textbf{Friedman \emph{et al.}} \cite{friedman_et_al_glioma_virotherapy}, \textbf{Xiao \emph{et al.}} \cite{xiao_et_al_cellular_automaton_HBV}, \textbf{Mok \emph{et al.}} \cite{mok_et_al_herpes_simplex_in_solid_tumour}, \textbf{Heilmann \emph{et al.} (2010)} \cite{heilmann_et_al_2010_sustainability_of_virulence}, \textbf{Mitchell \emph{et al.}} \cite{mitchell_et_al_high_replication_efficiency_H1N1}, \textbf{Wodarz \emph{et al.}} \cite{wodarz_et_al_complex_spatial_dynamics_oncolytic}, \textbf{Yakimovich \emph{et al.}} \cite{yakimovich_et_al_infectio}, \textbf{Heilmann \emph{et al.} (2012)} \cite{heilmann_et_al_2012_coexistence_phage_and_bacteria}, \textbf{Schiffer \emph{et al.}} \cite{schiffer_et_al_rapid_localised_spread}, \textbf{Bocharov \emph{et al.}} \cite{bocharov_et_al_spatiotemporal_dynamics_virus}, \textbf{Malinzi \emph{et al.}} \cite{malinzi_2017_spatiotemporal_chemovirotherapy}, \textbf{Friedman and Lai} \cite{friedman_and_lai_combination_oncolytic_checkpoint}, \textbf{Eriksen \emph{et al.}} \cite{eriksen_et_al_growing_microcolony}, \textbf{Imle \emph{et al.}} \cite{imle_et_al_hiv_3d_collagen_matrix}, \textbf{Alzahrani \emph{et al.}} \cite{alzahrani_2019_multiscale_cancer_response}, \textbf{Getz \emph{et al.}} \cite{getz_rapid_community_driven_covid_model}, \textbf{Quirouette \emph{et al.}} \cite{quirouette_et_al_influenza_localisation_model}, \textbf{Simmons \emph{et al.}} \cite{simmons_et_al_biofilm_structure}, and \textbf{Sego \emph{et al.}} \cite{sego_et_al_covid_model}.}
    	\label{fig:influential_papers}
    \end{figure}

	We sought to identify which of the publications found had been the most impactful. \added{Though clearly not a complete picture of a study's impact within the field, we chose to quantify this using citation data in order to limit editorial bias in the selection.} To account for the time since publication, we \added{considered} both a study's overall number of citations, and the average number of citations it had received per year since publication, and imposed thresholds on both quantities. These constraints \added{were intended to identify publications that have reached} a specified degree of influence overall, \added{but have also been} influential relative to the number of years since publication. In Figure~\ref{fig:papers_scatter}, we plot each publication found in our search against its overall number of publications and its average number of publications per year. Publications are colour--coded by virus and markers indicate methodology used. We opted for thresholds on the overall and per--year citation counts of 50 and 6, respectively, and found 25 studies that passed both thresholds. We show a timeline of these \added{highly--cited} publications in Figure~\ref{fig:influential_papers}. In the proceeding analysis, we make particular mention of these \added{highlighted} publications. In this review, articles that were included in our search are referenced alongside the authors' names, and \added{highly--cited} publications are indicated in bold.

	\section{Spatially--structured models by methodology} \label{sec:papers_by_methodology}
	
	We begin our analysis by discussing the modelling methodologies used across the literature on the spread of viral infections in space. The choice of methodology used in a given model has significance beyond just the modeller's preference: depending on the length scale of interest, or the specific dynamical features to be explored, certain modelling frameworks may be more, or less suitable. Moreover, there is an inherent trade--off between simpler models --- which are more amenable to analysis and more readily calibrated to experimental data --- and larger, more detailed models, which offer greater biological precision and realism. As such, a model's structure and design is essential to understanding the dynamics that model predicts, and the specificity and robustness of inferences made from such a model. In the following section, we discuss the methodological approaches used within the four modelling categories described in Section~\ref{sec:methods}, that is, continuum models, patch models, cellular automata and multicellular models.

	\subsection{Continuum models}

           \begin{figure}[h!]
       	\centering
       	\includegraphics[width=\textwidth]{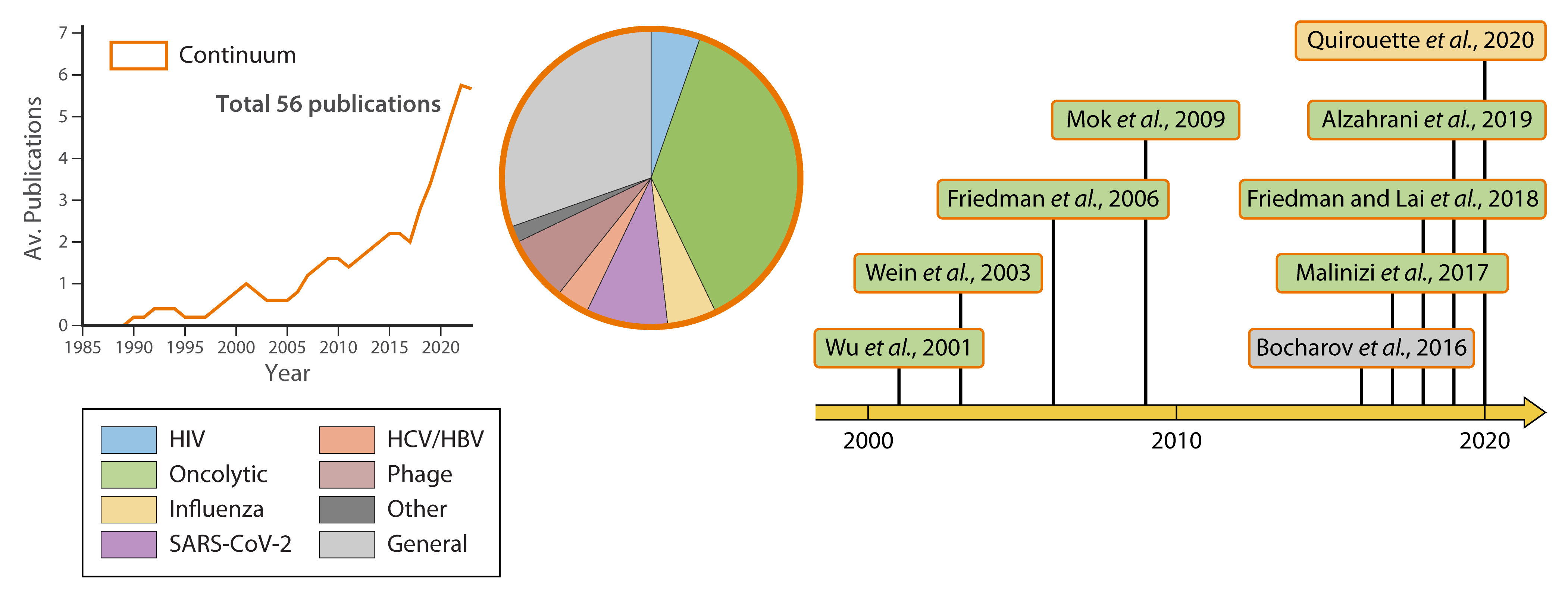}
       	\caption{Continuum models summary. \added{Left: Average number of publications using a continuum model per year. Centre: Continuum models by viral species studied. Right: Highly--cited continuum model publications over time.} \added{Highly--cited} publications are as follows: \textbf{Wu \emph{et al.}} \cite{wu_et_al_modelling_and_analysis_oncolytic}, \textbf{Wein \emph{et al.}} \cite{wein_et_al_validation_and_analysis_oncolytic}, \textbf{Friedman \emph{et al.}} \cite{friedman_et_al_glioma_virotherapy}, \textbf{Mok \emph{et al.}} \cite{mok_et_al_herpes_simplex_in_solid_tumour}, \textbf{Bocharov \emph{et al.}} \cite{bocharov_et_al_spatiotemporal_dynamics_virus}, \textbf{Malinzi \emph{et al.}} \cite{malinzi_2017_spatiotemporal_chemovirotherapy}, \textbf{Friedman and Lai} \cite{friedman_and_lai_combination_oncolytic_checkpoint}, \textbf{Alzahrani \emph{et al.}} \cite{alzahrani_2019_multiscale_cancer_response}, and \textbf{Quirouette \emph{et al.}} \cite{quirouette_et_al_influenza_localisation_model}.}
       	\label{fig:continuum_summary}
       \end{figure}
	
	Continuum models represent the spatial domain and all model components as continuous in space. For the models we found in our search, this meant systems of PDEs, possibly coupled to ODEs, governing the dynamics of density distributions of cells, viruses, cytokines, and so on. The methodological simplicity of this approach also makes continuum models conducive to analytical approaches not possible under other frameworks, such as travelling wave and stability analysis. In representing all model components --- and cells in particular --- as density distributions in space, such models implicitly assume a macroscopic length scale relative to the size of the cell, yet the continuous spatial domain still offers a fine degree of spatial resolution. These properties make continuum models a popular choice for modelling the diffusion and possibly advection of virus and cytokine densities, for example, and the mechanical properties of tissues at the whole--tissue scale. It is perhaps for this reason that continuum models are a particularly popular modelling choice for oncolytic viruses (21 of the 56 continuum models found), where it is of interest to track the the spread of infection through cancer cells at tumour--scale, and also to take into account the mechanical properties of the growing tumour.
	
	The simplest continuum models use a single PDE to track viral spread in space. We mentioned in the previous section Koch's analysis of plaque spread using properties of the diffusion equation \cite{koch_growing_viral_plaques}; similar approaches have since been employed by others. Rodrigo \emph{et al.} also used properties of the diffusion equation, in addition to a stochastic model for multiple infection foci, to study the growth of plaques in viral infections of plants \cite{rodrigo_plant_viruses}. Baabdulla \emph{et al.} used a homogenised Fisher--KPP equation to analyse the spread of influenza A viral load across a tissue culture comprising a structured mix of infection--prone and infection--resistant cells, and found that a good fit was obtained only when the structure of the tissue was well--represented in the homogenisation process \cite{baabdulla_et_al_homogenisation_IAV}. \textbf{Bocharov \emph{et al.}} also employed a Fisher--KPP model for the spread of viral density in permissive tissues, with the addition of a general time--delayed, immune--mediated decay term \cite{bocharov_et_al_spatiotemporal_dynamics_virus}. Even with this simple model, the authors demonstrated complex travelling wave dynamics whose speed and structure depend on the size of the initial viral inoculum \cite{bocharov_et_al_spatiotemporal_dynamics_virus}. Moreover, the authors demonstrated the possibility of spatially--homogeneous or oscillatory dynamics depending on the properties of the viral diffusion and immune response \cite{bocharov_et_al_spatiotemporal_dynamics_virus}.
	
	Many continuum models are very closely related to well--established ODE systems. Such models bear the same target cell limited structure as many ODE systems but in spatially--explicit conditions, mediated by the spatial spread of one or more model components. Often, this is simply encoded via viral diffusion: Frank \cite{frank_within_host_spatial_dynamics} used this approach to study the spread of replication--competent and defective--interfering viruses; others have used such models to track the spreading speed of infections (Ait Mahiout \emph{et al.} \cite{mahiout_2023_respiratory_viral_infections}, Amor and Fort \cite{amor_and_fort_virus_infection_speeds}, de Rioja \emph{et al.} \cite{de_rioja_et_al_front_propagation_T7}), possibly in the context of competing viral strains (Ait Mahiout \emph{et al.} \cite{mahiout_2022_virus_replication_and_competition}, Tokarev \emph{et al.} \cite{tokarev_et_al_fastest_autowave} and Holder \emph{et al.} \cite{holder_et_al_assessing_the_in_vitro_fitness}). In some contexts, cell density is also assumed to diffuse in space. Alzahrani made this assumption in a model of SARS--CoV--2 infection where both healthy and infected cells were assumed to be motile \cite{alzahrani_spatiotemporal_sars_cov_2}, although diffusion of cell density is more commonly found in models of oncolytic viruses, where cellular diffusion accounts for the expansion of the tumour in space (\textbf{Alzahrani \emph{et al.}} \cite{alzahrani_2019_multiscale_cancer_response}, \textbf{Friedman and Lai} \cite{friedman_and_lai_combination_oncolytic_checkpoint}, \textbf{Malinzi \emph{et al.}} \cite{malinzi_2017_spatiotemporal_chemovirotherapy}, for example). 
	
	Model dimensionality is also an important consideration for the dynamical and analytical properties of continuum models. Models with a single spatial dimension have been popular since, despite their simplicity, a single spatial variable is often sufficient to describe a key axis of spatial heterogeneity. \textbf{Quirouette \emph{et al.}} \cite{quirouette_et_al_influenza_localisation_model}, Vimalajeewa \emph{et al.} \cite{vimalajeewa_virus_particle_propagation_respiratory_tract} and Ait Mahiout \emph{et al.} \cite{mahiout_2023_respiratory_viral_infections} have all employed PDE models of respiratory viral infections with a single spatial variable representing depth along the respiratory tract; Holder \emph{et al.} \cite{holder_et_al_assessing_the_in_vitro_fitness} and Haseltine \emph{et al.} \cite{haseltine_et_al_image_guided_growth} have used radial spatial variables to describe the growth of radially--symmetric viral plaques. In models of oncolytic viruses, the three--dimensional tissue environment is frequently collapsed down to a single spatial axis, either representing a travelling front of cancer cells along one spatial dimension (Malinzi \emph{et al.} \cite{malinzi_oncolytic_virus_telegraph_equation, malinzi_et_al_2015_virotherapy_solid_tumour_invasion}), or distance from the centre of a radially--symmetric tumour (\textbf{Wu \emph{et al.}} \cite{wu_et_al_modelling_and_analysis_oncolytic}, \textbf{Wein \emph{et al.}} \cite{wein_et_al_validation_and_analysis_oncolytic}, \textbf{Friedman \emph{et al.}} \cite{friedman_et_al_glioma_virotherapy}, for example). Dunia and Bonnecaze also used this radial symmetry approximation but to represent generalised viral infection in spherical organs \cite{dunia_and_bonnecaze_spherical_organs}. One--dimensional models have the additional benefit of being more amenable to analytical approaches, including the calculation of wave speeds for the infection front (\textbf{Bocharov \emph{et al.}} \cite{bocharov_et_al_spatiotemporal_dynamics_virus}, de Rioja \emph{et al.} \cite{de_rioja_et_al_gliobastomas}, for example). Amor and Fort \cite{amor_and_fort_virus_infection_speeds, amor_and_fort_cohabitation_reac_diff}, Fort and M\'{e}ndez \cite{fort_and_mendez_time_delayed_spread} and Bessonov \emph{et al.} \cite{bessonov_et_al_intracellular_spreading_viral_infection} have also both shown that delays associated with the within--cell viral life cycle strongly impact the rate of speed of the infection wave. Beyond a single spatial dimension, others have also modelled the spread of infections using continuum methods in two dimensions (Graziano \emph{et al.} \cite{graziano_et_al_finite_element_HIV_model}, \textbf{Alzahrani \emph{et al.}} \cite{alzahrani_2019_multiscale_cancer_response}, Reisch and Langemann \cite{reisch_and_langemann_automative_selection}, Liang \emph{et al.} \cite{liang_et_al_patch_formation_virus_DIPs}, for example) and, less commonly, three dimensions (\textbf{Friedman and Lai} \cite{friedman_and_lai_combination_oncolytic_checkpoint}, Bailey \emph{et al.} \cite{bailey_et_al_radial_expansion_intratumoural}). The advantage of the higher--dimensional (and higher--complexity) approach is the ability to capture anisotropic infection distributions and more detailed spatial heterogeneity than is possible in a single spatial dimension.  Liang \emph{et al.} employed a reaction--diffusion system with stochastic effects at low density to study the role of defective interfering particles in viral infection spread, and demonstrated the formation of symmetry--breaking ``patchy'' infection plaques \cite{liang_et_al_patch_formation_virus_DIPs}. Mbopda \emph{et al.} showed varying spatial patterning as a function of the viral diffusion parameter in a PDE model of hepatitis B dynamics \cite{mbopda_et_al_pattern_formation}.
	
	In addition to diffusion, continuum models have also regularly been used to describe anisotropic transport of both viruses and cells. Anekal \emph{et al.} analysed the dispersal of virions under the influence of fluid flow in two spatial dimensions, and showed the formation of a variety of spatial patterns depending on the adsorptive properties of the virions and the characteristics of the fluid \cite{anekal_et_al_virus_spread_fluid_flow}. Directed transport is also a key feature in models of oncolytic viral dynamics due to the moving and growing environment. Timalsina \emph{et al.}, for example, included advection of the components of their model with the spread of tumour density \cite{timalsina_et_al_math_comp_tumour_virotherapy}. Another key aspect of oncolytic viral dynamics is the directed migration of both cells and free virions towards regions of higher extracellular matrix density, known as \emph{haptotaxis}. \textbf{Alzahrani \emph{et al.}} \cite{alzahrani_2019_multiscale_cancer_response, alzahrani_2020_moving_boundary_fusogenic} and Alsisi \emph{et al.} \cite{alsisi_go_or_grow, alsisi_heterogeneous_ECM, alsisi_nonlocal_multscale_approaches} have described in depth the role of extracellular matrix density and haptotaxis in determining the spreading pattering of tumours under oncolytic virotherapy. In brief, these works collectively show that haptotaxis can function to break the symmetry of the spreading tumour, and moreover that this hampers and directs the spread of free virions \cite{alzahrani_2019_multiscale_cancer_response, alzahrani_2020_moving_boundary_fusogenic, alsisi_go_or_grow, alsisi_heterogeneous_ECM, alsisi_nonlocal_multscale_approaches}.
	
	There are some limitations to the continuum modelling approach. As Graw and Perelson pointed out in their review, continuum models --- both PDE and ODE systems --- fall under the umbrella of mean field approximations, in the sense that interaction between model species in such a model framework requires co--location, and cannot take into account interaction between spatially--adjacent quantities, such as direct viral spread between neighbouring cells \cite{graw_and_perelson_spatial_hiv}. It is also difficult to draw sharp boundaries to infected regions using a continuum approach, although these are clearly observed biologically: the usual assumption of linear diffusion for, say, the virus population results in the instant transport of a low level of viral density and therefore infected cell density everywhere in the spatial domain. Explicit moving tumour boundaries have been applied in models of oncolytic viral infections by \textbf{Alzahrani \emph{et al.}} \cite{alzahrani_2019_multiscale_cancer_response, alzahrani_2020_moving_boundary_fusogenic}, Alsisi \emph{et al.} \cite{alsisi_go_or_grow, alsisi_heterogeneous_ECM, alsisi_nonlocal_multscale_approaches}, Timalsina \emph{et al.} \cite{timalsina_et_al_math_comp_tumour_virotherapy} and Kim \emph{et al.} \cite{kim_et_al_choindroitinase_oncolytic} as a means of overcoming this limitation. However, even in these cases, the moving \emph{external} boundary does not account for the boundaries of infected regions which are \emph{internal} to the tumour. The moving boundary approach moreover does not allow for the formation of multiple, disparate infected regions, which are observed experimentally \cite{fukuyama_et_al_color_flu, manicassamy_et_al_gfp_reporter_influenza, bray_et_al_molecular_imaging_influenza}. Another shortfall of the continuum approach is its limitation in describing rare stochastic events, such as extinctions or mutations, which occur at low densities. Baabdulla and Hillen, for example, observed oscillations in tumour load in a PDE model for oncolytic virotherapy that occasionally reached very low levels, but, due to the continuum framework of the model, never resulted in extinction \cite{baabdulla_and_hillen_oscillations_spatial_oncolytic}. Some authors have implemented stochastic elements to models otherwise described by PDEs in an aim to include account for low--density phenomena. We have already mentioned the approach taken by Liang \emph{et al.} in stochastically simulating low--density viral and DIP dynamics \cite{liang_et_al_patch_formation_virus_DIPs}; Li \emph{et al.} also used a semi-stochastic scheme for a PDE model to account for viral mutation in HCV infection \cite{li_et_al_spatially_antiviral_hcv}. Bailey \emph{et al.} took a different approach and used a mainly probabilistic model to quantify the efficacy of randomly--distributed, radially--growing infection foci in oncolytic virotherapy \cite{bailey_et_al_radial_expansion_intratumoural}. While this approach sacrifices much of the mechanistic properties of other continuum models (the size of the infected population per plaque is assumed to grow at a constant rate, for example), the trade--off is a quantification of the inherent variability in efficacy of oncolytic virotherapy \cite{bailey_et_al_radial_expansion_intratumoural}.

	\subsection{Patch models}
	\label{sec:patch_models}

   \begin{figure}[h!]
   	\centering
   	\includegraphics[width=\textwidth]{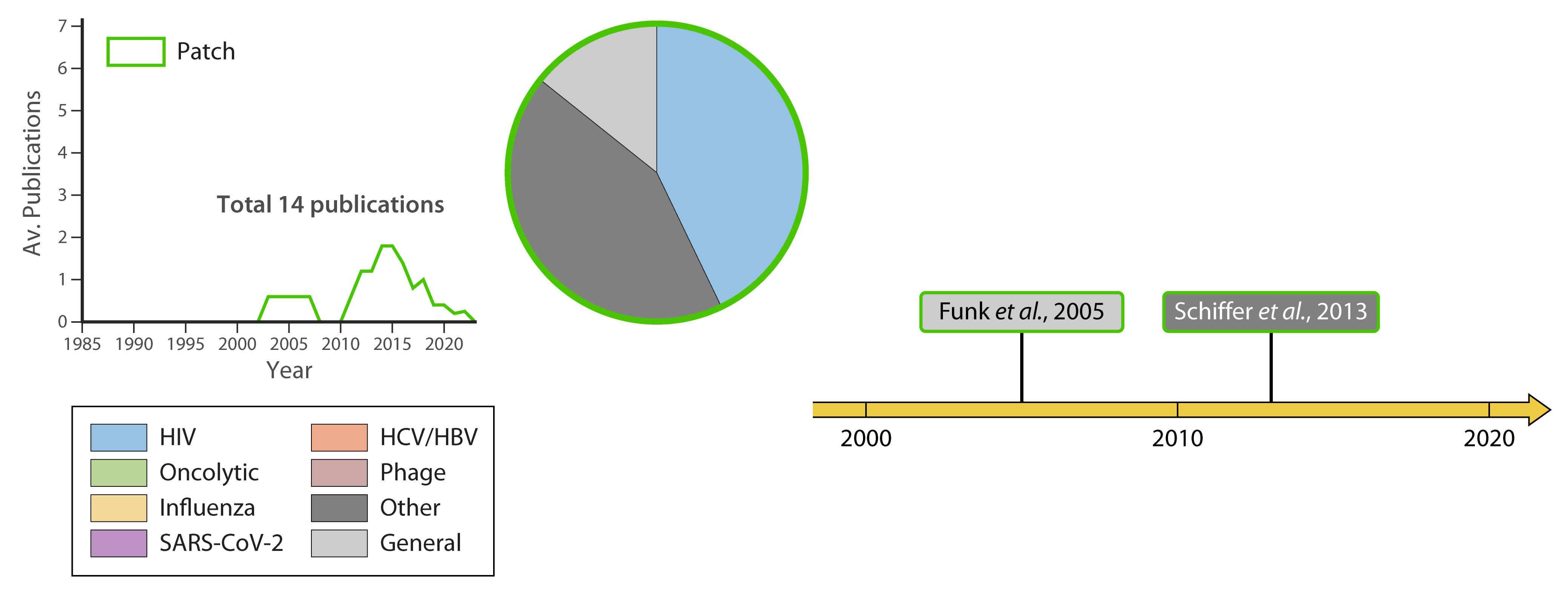}
   	\caption{Patch models summary. \added{Left: Average number of publications using a patch model per year. Centre: Patch models by viral species studied. Right: Highly--cited patch model publications over time.} \added{Highly--cited} publications are as follows: \textbf{Funk \emph{et al.}} \cite{funk_et_al_spatial_models_virus_immune}, and \textbf{Schiffer \emph{et al.}} \cite{schiffer_et_al_rapid_localised_spread}.}
   	\label{fig:patch_summary}
   \end{figure}

	Patch models divide the spatial domain into discrete, interconnected regions (or ``patches''), within each of which conditions are assumed to be well--mixed. Since cell populations under the patch framework are represented as densities (as opposed to discrete counts, which we defined as the distinction between patch and cellular automaton models), this construction relies on each patch containing sufficiently many cells that the density assumption remains valid. As such, patch models, like continuum models, are best suited for macro--scale modelling. In their simplest and most common form, patch models are defined as a network of regions, each governed by a set of ODEs, where connectivity between the regions is implemented through the flow of one or more model components, usually virus density. Assuming the model governing each patch is identical aside from the patch connectivity, such models can be thought of as closely related to both PDE models --- in a sense, this setup is a reaction--diffusion PDE system with spatial averaging over the patch regions --- and also mean--field ODE models: the overall dynamics of a patch system are governed by the within--patch model with the addition of a spatial delay as the infection spreads between patches. The earliest patch models were proposed by Orive \emph{et al.} \cite{orive_et_al_viral_infection_in_internally_structred_hosts} and \textbf{Funk \emph{et al.}} \cite{funk_et_al_spatial_models_virus_immune}, both of whom derived key results about the role of infection spread in patch networks. Orive and colleagues showed that increased migration between patches reduced the chance of the infection taking hold \cite{orive_et_al_viral_infection_in_internally_structred_hosts}. Funk and coworkers found that extent of coupling between compartments affected the global rate of infection, as well as the long--time dynamics of the system \cite{funk_et_al_spatial_models_virus_immune}. The authors demonstrated conditions under which the ability of the system to form an equilibrium depended on the extent of viral flow between compartments \cite{funk_et_al_spatial_models_virus_immune}.
	
	The patch approach can be considered an extension of compartmental ODE models. For example, some authors have considered distinct ODE systems for the upper and lower regions of the respiratory tract, where the two compartments are coupled by the flow of virions \cite{ke_perelson_et_al_covid_upper_and_lower_rt, goyal_remdesivir_macaques}; others considered more complex networks of interconnected organ compartments \cite{voutouri_in_silico_dynamics_covid} (note that we do not consider such models explicitly spatial and thus did not include these in our analysis). Patch models extend this approach to an arbitrary network or lattice of compartments. Among the patch models found by our search, a popular approach was to employ a regular structure to the patches, with either rectangular (\textbf{Funk \emph{et al.}} \cite{funk_et_al_spatial_models_virus_immune}, Byrne \emph{et al.} \cite{byrne_et_al_spatiotemporal_hsv2_and_hiv}, Zhang \emph{et al.} \cite{zhang_et_al_massively_multi_agent_system}) or hexagonal packing of the patches (\textbf{Schiffer \emph{et al.}} \cite{schiffer_et_al_hsv2_suppression_pritelivir, schiffer_et_al_increased_hsv2_shedding, schiffer_et_al_rapid_localised_spread, schiffer_et_al_rapid_viral_expansion, schiffer_hsv2_specific_cd8, dhankani_et_al_hsv2_genital_tract_shedding}). The patch approach has the property that its spatial structure is only defined implicitly via the strength and network of the connectivity between the patches. As such, depending on the structure of the patch connectivity, such models can reflect complex biological topologies which are not easily described by continuum models, for example. Nakaoka \emph{et al.} proposed a patch model for the spread of HIV infection between lymph nodes, explicitly representing the complex interconnectivity of the lymphoid tissue network, while treating the dynamics within each lymph node as well--mixed \cite{nakaoka_et_al_HIV_in_lymphoid_tissue_network}. Cardozo \emph{et al.} \cite{cardozo_et_al_compartmental_analysis, cardozo_et_al_cryptic_viremia} and Jagarapu \emph{et al.} \cite{jagarapu_et_al_integrated_spatial_dynamics_pharmacokinetic}, by contrast, applied a patch model at the within--lymph node scale. Their model considered distinct lymphoid follicles, interconnected via a free blood and lymph compartment, where the spherical lymphoid follicles were in turn subdivided into radial shell patches \cite{jagarapu_et_al_integrated_spatial_dynamics_pharmacokinetic, cardozo_et_al_compartmental_analysis, cardozo_et_al_cryptic_viremia}. \added{Recent experimental and modelling results have lent credibility to this ``patch'' representation of secondary lymphoid tissues by identifying clear distinctions in HIV viral load and turnover in follicular and extrafollicular compartments, which may also influence viral evolutionary dynamics \cite{ollerton_et_al_follicular, chung_et_al_secondary_lymphoid}.}
	
	The within--patch dynamics of patch models need not be governed by deterministic systems. A model put forward by \textbf{Schiffer \emph{et al.}} and used in a number of applications for HSV--2 infection considered a hexagonally--packed lattice of patches governed by stochastically--simulated ODE systems \cite{schiffer_et_al_hsv2_suppression_pritelivir, schiffer_et_al_increased_hsv2_shedding, schiffer_et_al_rapid_localised_spread, schiffer_et_al_rapid_viral_expansion, schiffer_hsv2_specific_cd8, dhankani_et_al_hsv2_genital_tract_shedding}. The authors employ stochasticity due to the random nature of the initiation of viral shedding and clearance \cite{schiffer_hsv2_specific_cd8}; as discussed above, these low--density phenomena are not well--captured by continuum approaches. Another stochastic patch system was proposed by Zhang \emph{et al.} for the dynamics of HIV infection \cite{zhang_et_al_massively_multi_agent_system}. In their model, which borders on a cellular automaton approach, infection and death of cells are modelled as probabilistic events which occur between virions and cells in the same lattice site, while diffusion is rendered as a deterministic, continuous flow between adjacent compartments \cite{zhang_et_al_massively_multi_agent_system}. 
	
	The patch modelling approach is also subject to shortcomings, and is by far the least used methodology for spatial viral dynamics modelling among the frameworks we considered. ODE patch models suffer from the same drawback as continuum models in that the continuous flow of virions between compartments prevents the determination of distinct boundaries to infected regions. In fact, the boundaries of infected regions are even less clearly defined for patch models since these models only possess spatial resolution at the patch scale. On the other hand, patch models do not have the benefit of being especially conducive to mathematical analysis. Some theory for within--host patch modelling was developed by Orive \emph{et al.} \cite{orive_et_al_viral_infection_in_internally_structred_hosts} and \textbf{Funk \emph{et al.}} \cite{funk_et_al_spatial_models_virus_immune}, although only for two--patch populations or approximate models. Moreover, with the exception of network--structured models, such as that of Nakaoka \emph{et al.} \cite{nakaoka_et_al_HIV_in_lymphoid_tissue_network}, the regions of patch models do not have any specific meaning. These patches instead represent somewhat arbitrary mesoscopic regions of cells which are large compared to the size of the cell (such that a density description is valid), yet small compared to the overall size of the tissue (in order to impose spatial structure).

	\subsection{Agent--based modelling: general approaches}
	\label{sec:agent_based}
	
	The class of models characterised by the presence of discrete spatial regions and a discrete cell population (these spatial regions may or may not be the cells themselves) can broadly be described as agent--based. We consider two sub--types of agent--based models in this analysis: cellular automata, where there is only one spatial scale present, and multicellular models, which include more than one spatial scale. These two sub--types have unique nuances which we articulate in the following sections, but there are substantial overlaps between the two approaches which we discuss here.
	
	Agent--based models which treat individual cells as spatially--discrete regions permit the possibility of integrating the within--cell viral dynamics of each cell. This is generally incorporated using a patch--like approach, where each cell is assigned an ODE system governing its spatially--homogeneous internal dynamics (Akpinar \emph{et al.} \cite{akpinar_et_al_spatiotemporal_viral_amplification}, \textbf{Sego \emph{et al.}} \cite{sego_et_al_covid_model}, Goyal and Murray \cite{goyal_and_murray_CCT_in_HBV}, Bouchnita \emph{et al.} \cite{bouchnita_et_al_towards_multiscale_HIV}, for example). Note that incorporating within--cell dynamics in this way does not explicitly add an additional spatial scale, and thus can be implemented in a cellular automaton approach as we have defined it (Akpinar \emph{et al.} \cite{akpinar_et_al_spatiotemporal_viral_amplification} does so), however, within--cell dynamics are more commonly found in more complex, explicitly multiscale multicellular models. There are a range of benefits from including within--cell viral dynamics. For one, explicitly representing the within--cell viral life cycle offers a more mechanistic account of the well--established lag observed between the time at which a cell is infected and the time at which a cell begins releasing virions (the eclipse phase), usually accounted for by a blanket latently--infected phase \cite{sego_et_al_covid_model, smith_validated_models_of_immune_response, yin_and_redovich_kinetic_modelling_virus_growth, holder_et_al_design_considerations_influenza}. This added detail moreover allows for a more detailed description of virus uptake and release by target cells (\textbf{Sego \emph{et al.}} \cite{sego_et_al_covid_model}, \textbf{Getz \emph{et al.}} \cite{getz_rapid_community_driven_covid_model}, Rowlatt \emph{et al.} \cite{rowlatt_et_al_modelling_within_host_covid}). Virus uptake is otherwise usually accounted for only under a general viral ``decay'' term \cite{marzban_et_al_hybrid_PDE_ABM_model, bartha_et_al_in_silico_evalutation_paxlovid, blahut_et_al_hepatitis_c_two_modes_of_spread, smith_validated_models_of_immune_response, yin_and_redovich_kinetic_modelling_virus_growth}. Furthermore, in accounting for components of the viral life cycle in more detail, models can be made to more closely represent experimental systems. In hepatitis virus dynamics, for example, multiple sources of experimental data track quantities of intracellular components such as intracellular viral RNA or covalently closed circular DNA (cccDNA) and models that account for such quantities are therefore better equipped to model the experimental findings \cite{graw_and_perelson_modeling_viral_spread, goyal_and_murray_CCT_in_HBV}.
	
	A popular approach in building agent--based models for viral infections has been in the development of generalised, often highly complex computational platforms which are modular in structure, such that a core modelling framework can readily be extended to incorporate the specifics of a particular application. Such platforms are sometimes self--described as ``simulators'' instead of models. One of the earlier instances of such an approach was adapted from an immune simulator proposed by Caleda and Seiden \cite{celada_and_seiden_comp_immune_model} (in its original statement, the model did not include an explicit viral population and was therefore not included in our analysis). This was model was later developed by the authors and others --- including \textbf{Bernaschi and Castiglione} --- into the ``ImmSim'' platform \cite{castiglione_et_al_EBV_immsim, bezzi_et_al_transition_between_immune_and_disease_states, kohler_et_al_systematic_vaccine_complexity, bernaschi_and_castiglione_escape_mutants, castiglione_et_al_mutation_fitness_viral_diversity, baldazzi_et_al_enhanced_agent_based_model}, and an offshoot, ``ParImm'' \cite{bernaschi_and_castiglione_design_and_implementation}. The central model of these systems describes an exceptionally complex virtual immune system powered by a so--called ``bit--string'' methodology, where each all modelled molecules, virions, and cell binding sites are represented as binary strings, and where interaction between these agents is governed by similarity between strings \cite{bernaschi_and_castiglione_escape_mutants, bernaschi_and_castiglione_design_and_implementation} (this bit--string approach is rarely used among the spatial viral dynamics literature, although Zhang \emph{et al.} \cite{zhang_et_al_massively_multi_agent_system} and Fachada \emph{et al.} \cite{fachada_et_al_simulating_antigenic_drift_influenza_A} used similar ideas). The underlying immune system model of ImmSim and its derivatives has been adapted to be applied to generalised viral infections \cite{bezzi_et_al_transition_between_immune_and_disease_states, kohler_et_al_systematic_vaccine_complexity} or specific viral species such as HIV \cite{bernaschi_and_castiglione_escape_mutants, castiglione_et_al_mutation_fitness_viral_diversity, baldazzi_et_al_enhanced_agent_based_model} or Epstein--Barr virus (EBV) \cite{castiglione_et_al_EBV_immsim}. In addition to ImmSim, Meier--Schellersheim and Mack proposed an alternative cellular automaton immune simulator, ``SIMMUNE,'' although, to our knowledge, this has not been applied to modelling viral infections since the original statement of the model \cite{meierschellersheim_and_mack_SIMMUNE}. \textbf{Yakimovich \emph{et al.}} developed another modelling platform called ``Infectio'', which coupled a discrete cell grid to a particle strength exchange model for viral flow \cite{yakimovich_et_al_cell_free_human_adenovirus, yakimovich_et_al_infectio}. Other authors have built agent--based simulation models catered to more specific scenarios. Shapiro \emph{et al.} developed a computational model for EBV infection called ``PathSim'' which explicitly incorporated the structure of Waldeyer's ring \cite{shapiro_et_al_virtual_EBV_simulation_mechanism, duca_et_al_virtual_EBV_biological_interpretations}, and more recently, Moses \emph{et al.} proposed ``SIMCov'', a model for SARS--CoV--2 infection in a virtual human lung \cite{moses_et_al_lung_covid_model}.
	
	In recent years, large simulator models have again become popular, although recent approaches are more commonly multicellular in structure and account for multiple spatial scales. Perhaps the most substantial among these was proposed by \textbf{Getz \emph{et al.}}, which introduced an enormously detailed ``tissue simulator'' for SARS--CoV--2 infection \cite{getz_rapid_community_driven_covid_model, islam_et_al_agent_based_lung_fibrosis}. This model incorporated a detailed description of virus--host interactions, as well as a complex immune system, within--cell dynamics, and even tissue mechanics over time \cite{getz_rapid_community_driven_covid_model, islam_et_al_agent_based_lung_fibrosis}. Moreover, the model was developed with the explicit intention of modularity, such that further sub--models might be later incorporated into its architecture \cite{getz_rapid_community_driven_covid_model}. An alternative multicellular simulator of acute respiratory infection, coined the ``Cellular Immunity Agent--Based Model (CIABM)'' was suggested by An \emph{et al.}, which the authors used to demonstrate heterogeneous viral dynamics between different hosts \cite{becker_et_al_CIABM, cockrell_and_an_comparative_computational_modelling}. Rowlatt \emph{et al.} proposed a similarly detailed model of respiratory infection, which they used to isolate the role of specific immune components in SARS--CoV--2 infection, especially macrophages \cite{rowlatt_et_al_modelling_within_host_covid}. Recent modular platforms have increasingly been integrated with existing software packages for tissue dynamics. The Getz \emph{et al.} model \cite{getz_rapid_community_driven_covid_model, islam_et_al_agent_based_lung_fibrosis} was developed in PhysiCell \cite{ghaffarizadeh_et_al_physicell}, for instance, while a recent series of models by Glazier and colleagues have been developed mainly with CompuCell3D \cite{sego_et_al_cellularisation, sego_et_al_covid_model, sego_et_al_multicellular_influenza, aponte_serrano_RNA_virus_replication, gianlupi_et_al_multiscale_model_of_antiviral_timing}. Among these, one model by \textbf{Sego \emph{et al.}} \cite{sego_et_al_covid_model} explicitly demonstrated cross--platform compatibility by being implemented in CompuCell3D \cite{swat_et_al_compucell3D}, and also being implemented in Chaste \cite{mirams_et_al_chaste, pitt_francis_et_al_Chaste}, and Morpheus \cite{staruss_et_al_morpheus} for later comparison work.
	
	In all, these highly complex computational models serve as simulation ``engines'', such that modellers or other end--users can simulate and visualise infection dynamics with a high degree of biological detail, with the possibility of building in further layers of model detail specific to a given application. The drawback to this approach is that the high degree of complexity significantly increases the difficulty of analysis, both of the dynamics of the model, but also of parameter identifiability. This is exacerbated by the enormous computational costs associated with these models, which can include millions of individual agents tracked over complex spatial domains. As such, for some of these models it is difficult to study the interaction of different components to any degree of detail, and virtually impossible to conduct statistically robust parameter estimations for such models (if sufficient data were even to exist). Parameter values for such models can therefore only be extracted from other models which may have varying relevance to the model to which they are applied. Clearly, there is a distinct application in mind for such models compared to, say, lightweight mean--field models for viral dynamics. While immune and tissue simulators are useful for hypothesis generation and simulation, for these to be useful for precision exercises such as parameter estimation, further advances in computational and statistical inference techniques are needed.

	\subsection{Agent--based modelling: cellular automata}

            \begin{figure}[h!]
        	\centering
        	\includegraphics[width=\textwidth]{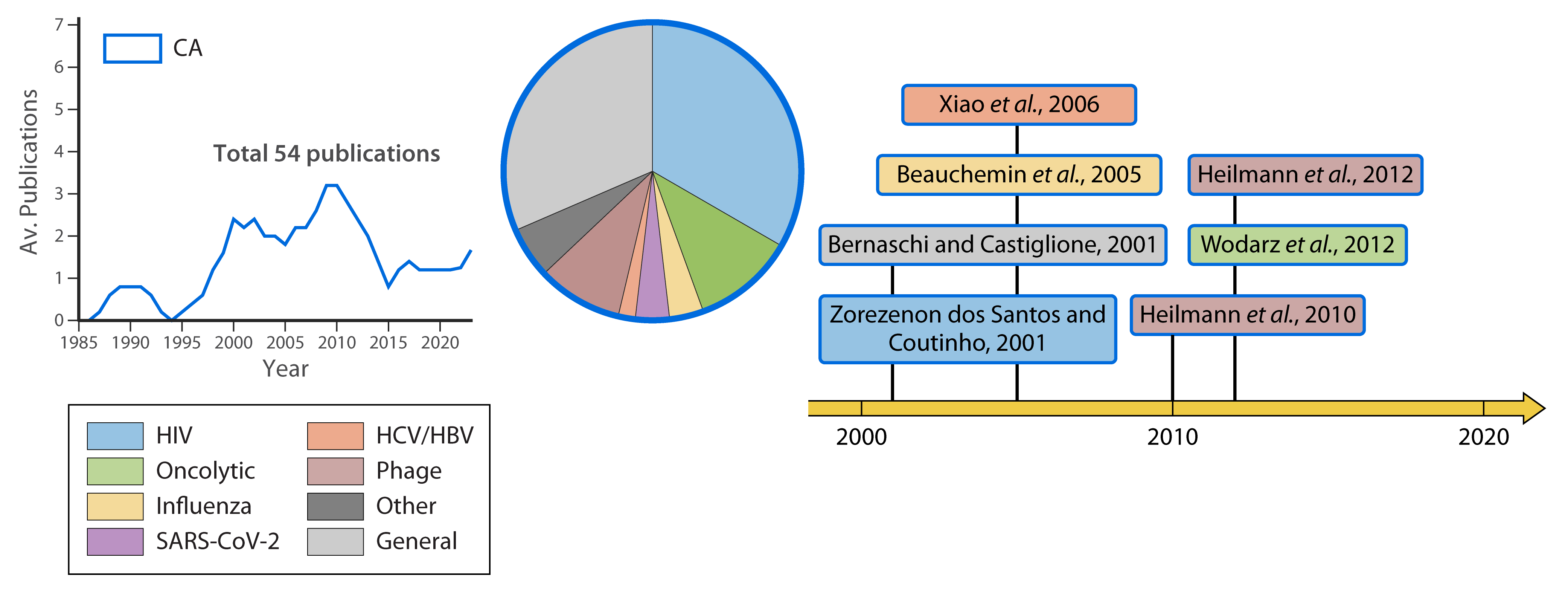}
        	\caption{Cellular automata summary. \added{Left: Average number of publications using cellular automata per year. Centre: Cellular automaton publications by viral species studied. Right: Highly--cited cellular automaton publications over time.} \added{Highly--cited} publications are as follows: \textbf{Bernaschi and Castiglione} \cite{bernaschi_and_castiglione_design_and_implementation}, \textbf{Zorzenon dos Santos and Coutinho} \cite{dos_santos_and_countinho_dynamics_of_HIV_CA}, \textbf{Beauchemin \emph{et al.}} \cite{beauchemin_et_al_simple_cellular_automaton_model}, \textbf{Xiao \emph{et al.}} \cite{xiao_et_al_cellular_automaton_HBV}, \textbf{Heilmann \emph{et al.} (2010)} \cite{heilmann_et_al_2010_sustainability_of_virulence}, \textbf{Wodarz \emph{et al.}} \cite{wodarz_et_al_complex_spatial_dynamics_oncolytic}, and \textbf{Heilmann \emph{et al.} (2012)} \cite{heilmann_et_al_2012_coexistence_phage_and_bacteria}.}
        	\label{fig:cellular_automata_summary}
        \end{figure}
    
	Cellular automata are a class of discrete computational models. Like patch models, cellular automata involve discretising the spatial domain into a generally regular grid of explicit regions, however, the point of departure is that in cellular automata, the cell population is also discrete. Since this modelling approach involves tracking individual cells, cellular automata are commonly used for meso-- or microscopic spatial scales, although cellular automata with vast cell populations are known (Moses \emph{et al.} \cite{moses_et_al_lung_covid_model} and Shapiro \emph{et al.} \cite{shapiro_et_al_virtual_EBV_simulation_mechanism, duca_et_al_virtual_EBV_biological_interpretations}, for example). Cellular automata are governed by distinct, often stochastic and adjacency--based rules, and as such generate dynamics which are notably distinct from those of continuum models. Lavigne \emph{et al.} and Clark \emph{et al.} have compared infection dynamics using analogous PDE and cellular automaton models, and observed qualitatively different infection structures and outcomes between the two approaches \cite{lavigne_et_al_interferon_signalling_ring_vaccination}.
	
	There is a great diversity in the implementation of cellular automata for viral dynamics within the host. Some authors follow a patch--like approach to discretisation of space, such that each spatial region may simultaneously contain many cells (\textbf{Bernaschi and Castiglione} \cite{bernaschi_and_castiglione_design_and_implementation}, Pandey \cite{pandey_1989_CA_model_HIV}, Rodriguez--Brenes \emph{et al.} \cite{rodriguez_brenes_et_al_virus_spread_from_low_moi}, Shapiro \emph{et al.} \cite{shapiro_et_al_virtual_EBV_simulation_mechanism}, for example). While the standard approach under this construction is to keep track of the actual number of cells at lattice sites, in some early approaches, authors have instead considered a binary ``high'' or ``low'' concentration of each cell species in each spatial region ( Pandey \emph{et al.} \cite{pandey_1989_CA_model_HIV, pandey_1991_CA_interacting_cellular_network, pandey_and_stauffer_metastability, mielke_and_pandey_fuzzy_interaction, mannion_et_al_effect_of_mutation, mannion_et_al_monte_carlo_population_dynamics}, Kougias and Schulte \cite{kougias_and_schulte_simulating_immune_response_to_HIV}). As an alternative to patch--like discretisation, spatial regions may represent the cells themselves (\textbf{Xiao \emph{et al.}} \cite{xiao_et_al_cellular_automaton_HBV}, \textbf{Beauchemin \emph{et al.}} \cite{beauchemin_et_al_simple_cellular_automaton_model}, \textbf{Wodarz \emph{et al.}} \cite{wodarz_et_al_complex_spatial_dynamics_oncolytic}, \textbf{Zorzenon dos Santos and Coutinho} \cite{dos_santos_and_countinho_dynamics_of_HIV_CA}, for example). In these cases, rather than each lattice site tracking a count of cells of different types, each lattice site describes a cell in a certain state which may change depending on whether that cell is susceptible to infection, productively infected, and so on. These cell--regions are often assumed to pack densely and tessellate the entire spatial domain, however some authors have also allowed for empty sites (\textbf{Wodarz \emph{et al.}} \cite{wodarz_et_al_complex_spatial_dynamics_oncolytic, wodarz_and_levy_hiv_two_modes_of_spread}, Strain \emph{et al.} \cite{strain_et_al_spatiotemporal_hiv}, Reis \emph{et al.} \cite{reis_et_al_in_silico_evolutionary_tumour_virotherapy}). Bankhead III \emph{et al.} randomly populated a mostly--empty cell lattice with cell islands whose degree of clustering and confluency was determined from \emph{in vitro} imaging \cite{bankhead_et_al_simulation_framework_SARS}.
	
	There are also highly varied approaches to the representation of viruses in cellular automaton systems. Some authors treat virions, like cells, as discrete agents, and track the movement of individual agents as they spread across the spatial grid (Fachada \emph{et al.} \cite{fachada_et_al_simulating_antigenic_drift_influenza_A}, Shapiro \emph{et al.} \cite{shapiro_et_al_virtual_EBV_simulation_mechanism, duca_et_al_virtual_EBV_biological_interpretations}, \textbf{Bernaschi and Castiglione} \cite{bernaschi_and_castiglione_design_and_implementation}, Jafelice \emph{et al.} \cite{jafelice_et_al_CA_with_fuzzy_parameters}, for example). Other approaches involve tracking the spread of viral density between lattice sites (Kumberger \emph{et al.} \cite{kumberger_et_al_accounting_for_space_cell_to_cell}, Immonem \emph{et al.} \cite{immonem_et_al_hybrid_stochastic_deterministic}, Moses \emph{et al.} \cite{moses_et_al_lung_covid_model}, for example). Note that we allow such an approach under the cellular automaton classification provided that \emph{cells} are treated as discrete objects and viral density is described only as a property of the lattice sites. For cellular automata with an explicit viral population, infection is usually implemented (as in the patch framework) via co--location, that is, between virions and susceptible cells which occupy the same lattice site. As an alternative, there is a body of cellular automaton models --- particularly in cases where the spatial lattice consists of the cells themselves --- which do not consider a free viral population at all (Morselli \emph{et al.} \cite{morselli_et_al_agent_based_and_continuum_models}, Jarrah \emph{et al.} \cite{jarrah_et_al_optimal_control_in_vitro_virus}, \textbf{Wodarz \emph{et al.}} \cite{wodarz_et_al_complex_spatial_dynamics_oncolytic}, Precharattana \emph{et al.} \cite{precharattana_et_al_investigation_of_spatial_pattern_formation}, for example). In such cases, infection is usually implemented not via \emph{co--location} of infecting and susceptible agents, but by \emph{adjacency}. For example, this may be encoded as the probability of a cell--region changing state from uninfected to infected, where the probability is dependent on the number of neighbouring cells in an infected state, as in \textbf{Xiao \emph{et al.}} \cite{xiao_et_al_cellular_automaton_HBV} and \textbf{Beauchemin \emph{et al.}} \cite{beauchemin_et_al_simple_cellular_automaton_model}. For some, the assumption of neighbour--based infection is considered a proxy for highly diffusion--limited viral spread (\textbf{Xiao \emph{et al.}} \cite{xiao_et_al_cellular_automaton_HBV}, \textbf{Wodarz \emph{et al.}} \cite{wodarz_et_al_complex_spatial_dynamics_oncolytic}, for example). Adjacency--based interactions are generally limited to neighbouring cells, although Pandey and collaborators have considered complex mixing operations between lattice sites across the entire grid, in order to represent the migration of cells involved in HIV--host interactions and complex infection spreading dynamics as a result \cite{pandey_1989_CA_model_HIV, pandey_1991_CA_interacting_cellular_network, pandey_and_stauffer_metastability, mannion_et_al_effect_of_mutation, mannion_et_al_monte_carlo_population_dynamics, mielke_and_pandey_fuzzy_interaction}.
	
	The discrete nature of cellular automata makes them a useful language for describing stochastic dynamics (although it is worth mentioning that these models are not necessarily stochastic in nature; \textbf{Zorzenon dos Santos and Coutinho} considered a deterministic neighbour--based infection rule \cite{dos_santos_and_countinho_dynamics_of_HIV_CA}). Stochastic dynamics are especially relevant to models at the microscopic scale, such as where the lattice sites are the cells themselves. \textbf{Wodarz \emph{et al.}} studied infection spreading patterns for oncolytic viruses \cite{wodarz_et_al_complex_spatial_dynamics_oncolytic}. Therein, the authors used deterministic approximations of their stochastic cellular automaton model to define regions of parameter space corresponding to different spread patterns, but observed empirically that for parameter values near the boundaries of these regions the cellular automaton the cellular automaton could probabilistically generate different patterns \cite{wodarz_et_al_complex_spatial_dynamics_oncolytic}. The authors also found that when the size of the model grid was increased but the initial infection inoculum size is held fixed, there was an increase in the variety of stochastic outcomes of the cellular automaton for the same region of parameter space \cite{wodarz_et_al_complex_spatial_dynamics_oncolytic}. \textbf{Beauchemin \emph{et al.}}, by contrast, showed that when the initial infection inoculum is scaled with the increase in size of the spatial grid, it has the effect of \emph{reducing} stochastic variability of the system \cite{beauchemin_et_al_simple_cellular_automaton_model}.
	
	This sensitivity to implementation represents a substantial challenge in applying cellular automata. In addition to the finding by Wodarz and Beauchemin \emph{et al.} that the dynamics of their models were dependent on the size of the spatial lattice \cite{beauchemin_et_al_simple_cellular_automaton_model, wodarz_et_al_complex_spatial_dynamics_oncolytic}, Moses \emph{et al.} also showed that their SIMCov model generated significantly different dynamics on a two or three dimensional spatial lattice compared to their lung lattice model \cite{moses_et_al_lung_covid_model}. The fact of this sensitivity to implementation is not necessarily a limitation of the cellular automaton approach --- since it enables the role of structural features in dynamics to be explored --- however, this implementation--dependence means that a high degree of care must be taken in approaching the design of cellular automaton models.

	\subsection{Agent--based modelling: multicellular models} \label{sec:agent_based_multicellular}

        \begin{figure}[h!]
    	\centering
    	\includegraphics[width=\textwidth]{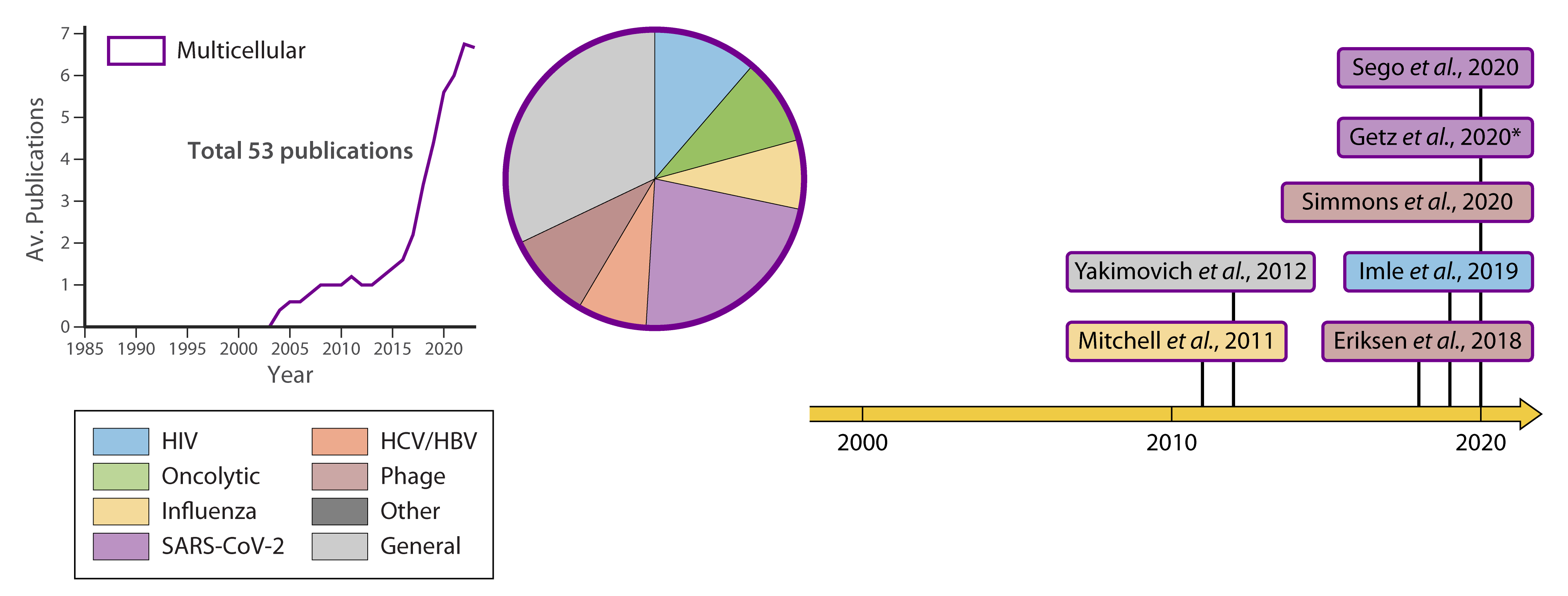}
    	\caption{Multicellular models summary. \added{Left: Average number of publications using multicellular models per year. Centre: Multicellular models by viral species studied. Right: Highly--cited multicellular model publications over time.} \added{Highly--cited} publications are as follows: \textbf{Mitchell \emph{et al.}} \cite{mitchell_et_al_high_replication_efficiency_H1N1}, \textbf{Yakimovich \emph{et al.}} \cite{yakimovich_et_al_infectio}, \textbf{Eriksen \emph{et al.}} \cite{eriksen_et_al_growing_microcolony}, \textbf{Imle \emph{et al.}} \cite{imle_et_al_hiv_3d_collagen_matrix}, \textbf{Getz \emph{et al.}} \cite{getz_rapid_community_driven_covid_model}, \textbf{Simmons \emph{et al.}} \cite{simmons_et_al_biofilm_structure}, and \textbf{Sego \emph{et al.}} \cite{sego_et_al_covid_model}.}
    	\label{fig:multicellular_summary}
    \end{figure}
	
	Multicellular models --- to follow the nomenclature of Glazier and colleagues \cite{sego_et_al_cellularisation, sego_et_al_covid_model, sego_et_al_multicellular_influenza, gianlupi_et_al_multiscale_model_of_antiviral_timing, aponte_serrano_RNA_virus_replication}, for example --- are agent--based systems closely related to cellular automata. As with cellular automata, multicellular models involve discrete spatial regions (for example, cells), track discrete numbers of cells of different types, and typically describe meso-- or microscopic dynamics. However, the defining feature of the multicellular framework is the presence of multiple spatial scales. These distinct scales typically correspond to the cell scale and a finer--resolution viral scale, for example a discrete cell grid coupled to a continuous virus density surface. There is an important subtlety here: while in practice, such a viral density may be \emph{simulated} at the cell--scale (that is, in the same manner as a cellular automaton), we consider the model a multicellular system provided that the governing viral dynamics are \emph{defined} on an alternate spatial scale to the cell lattice (such as by a PDE). This hybrid approach in simultaneously incorporating model components on different length scales has rapidly grown in popularity in recent years among the viral dynamics literature, and is in line with similar approaches in oncology and tissue dynamics, for example \cite{macklin_et_al_multicellular_review, fletcher_and_osborne_seven_challenges, okuda_et_al_3D_vertex_model}. Note that in principle, as we have defined it, this framework could allow for discrete spatial regions to contain groups of cells, however we only found two instances where this was the case. Paiva \emph{et al.} \cite{paiva_et_al_multiscale_oncolytic_virotherapy} allowed cancer cells to crowd into a single lattice site; the other model which permitted multiple cells within a single site, by Garrido Zornoza \emph{et al.} \cite{garrido_zornoza_et_al_stochastic_microbial_dispersal}, is a phage model defined at ecological scale which is a special case among the models we discuss, as we explore in Section~\ref{sec:other_viruses}. All the other models retained in our analysis treated cells as individual spatial regions.
	
	The simplest approach under the multicellular framework is in coupling a cellular automaton to a global virus population. This construction has been used by authors in a variety of applications (Cangelosi \emph{et al.} \cite{cangelosi_et_al_multiscale_HBV} and Goyal and Murray \cite{goyal_and_murray_CCT_in_HBV} in HBV infection, Blahut \emph{et al.} \cite{blahut_et_al_hepatitis_c_two_modes_of_spread} in HCV, Kreger \emph{et al.} \cite{kreger_et_al_synaptic_cell_to_cell} and Shi \emph{et al.} \cite{shi_et_al_viral_load_based_HIV} in HIV, for example) and generally follows from an assumption that free viruses spread quickly relative to the size of the model tissue \cite{cangelosi_et_al_multiscale_HBV, goyal_and_murray_CCT_in_HBV, blahut_et_al_hepatitis_c_two_modes_of_spread, kreger_et_al_synaptic_cell_to_cell}. A similar idea was applied by Howat \emph{et al.}, who considered the spread of free virus at the cell scale, but type I interferon (comprised of molecules far smaller than virions which therefore diffuse much more quickly) as a global quantity \cite{howat_et_al_modelling_dynamics_of_type_I_IFN}. \added{Similarly, Durso-Cain \emph{et al.} considered a diffusing viral population but a global antibody population \cite{durso_cain_hcv_dual_spread}.} Models which consider only infection from a global viral reservoir do not give rise to spatially--structured infections, and as such, the global viral population is incorporated into the model dynamics in different ways. Shi \emph{et al.} considered neighbour--based infection at a rate related to the global viral load \cite{shi_et_al_viral_load_based_HIV}. More typically, however, models with a global virus population generally consider two modes of viral infection: unstructured infection due to the global population of free virions, and structured local infection arising from direct cell--to--cell infection between neighbouring cells (Saeki and Sasaki \cite{saeki_and_sasaki_spatial_heterogeneity}, Kreger \emph{et al.} \cite{kreger_et_al_synaptic_cell_to_cell}, Goyal and Murray \cite{goyal_and_murray_CCT_in_HBV}, Blahut \emph{et al.} \cite{blahut_et_al_hepatitis_c_two_modes_of_spread}). Such models therefore describe two spatial scales of infection, which are generally not well--described by cellular automata. We describe modelling of the cell--to--cell infection route in further detail in Section~\ref{sec:innately_spatially_structured_dynamics}.
	
	A typical approach in multicellular modelling is to embed discrete cells within an underlying continuous spatial domain. Over this domain, authors commonly define the virus population as a density distribution in space governed by PDEs (\textbf{Mitchell \emph{et al.}} \cite{mitchell_et_al_high_replication_efficiency_H1N1}, \textbf{Yakimovich \emph{et al.}} \cite{yakimovich_et_al_cell_free_human_adenovirus}, \textbf{Imle \emph{et al.}} \cite{imle_et_al_hiv_3d_collagen_matrix}, \textbf{Sego \emph{et al.}} \cite{sego_et_al_covid_model}, \textbf{Getz \emph{et al.}} \cite{getz_rapid_community_driven_covid_model}, for example). Others include additional small, diffusive species like cytokines or immune components, described in a similar manner (\textbf{Sego \emph{et al.} \cite{sego_et_al_covid_model}}, \textbf{Getz \emph{et al.}} \cite{getz_rapid_community_driven_covid_model}, Cai \emph{et al.} \cite{cai_et_al_spatial_dynamics_of_immune_repsonse}, Aristotelous \emph{et al.} \cite{aristotelous_et_al_covid_model} for example). While cellular automata and patch systems cannot easily account for anisotropic spread of virions and other small species, by encoding the dynamics of these components in the language of PDEs, it is natural to account for more complex flows. \textbf{Yakimovich \emph{et al.}} \cite{yakimovich_et_al_cell_free_human_adenovirus, yakimovich_et_al_infectio} and Chen \emph{et al.} \cite{chen_sars_cov_2_in_lung, chen_et_al_antibody_protection, zhang_et_al_computational_modelling_extereme_heterogeneity, pearson_et_al_modelling_variability_in_sars_cov_2, zhang_et_al_global_sensitivity_analysis} added advection to their description of viral density spread and showed the formation of ``comet''--shaped plaques (as were observed by, for example, Anekal \emph{et al.} \cite{anekal_et_al_virus_spread_fluid_flow} in their continuum model). Chen \emph{et al.}, in particular, showed that the varied rates of viral advection by the mucociliary escalator in different generations of the lung led to dramatic changes in the spatiotemporal dynamics of infection \cite{chen_sars_cov_2_in_lung}.
	
	This discrete--continuum coupling also enables tracking of the motion of discrete agents. This is of particular relevance in descriptions of the adaptive immune response, which is largely mediated by cytotoxic CD8$^+$ T cells. These cells are relatively few in number and therefore are not well described by a density distribution \cite{levin_et_al_T_cell_search_influenza, kadolsky_and_yates_spatial_immune_surveillance}. As such, \textbf{Sego \emph{et al.}} \cite{sego_et_al_covid_model}, \textbf{Getz \emph{et al.}} \cite{getz_rapid_community_driven_covid_model}, Levin \emph{et al.} \cite{levin_et_al_T_cell_search_influenza} and Cai \emph{et al.} \cite{cai_et_al_spatial_dynamics_of_immune_repsonse} all explicitly modelled discrete CD8$^+$ T cells which migrate towards sites of infection (we discuss immune cell migration in greater depth in Section~\ref{sec:innately_spatially_structured_dynamics}). Some authors have also modelled the individual migration of host cells in the infection of motile cell populations. \textbf{Imle \emph{et al.}} \cite{imle_et_al_hiv_3d_collagen_matrix}, Bouchnita \emph{et al.} \cite{bouchnita_et_al_towards_multiscale_HIV} and Seich Al Basatena \cite{al_basatena_et_al_non_lytic_CD8} all modelled HIV infection among populations of migrating host cells; Jenner \emph{et al.} \cite{jenner_et_al_oncolytic_voronoi} also accounted for a spatially--evolving tissue in a model of oncolytic viral infection. In addition to individual cell migration, models which account for the motion of individual virions have also been proposed (Itakura \emph{et al.} \cite{itakura_et_al_reproducibility_chronic_virus_infection_model}, Chen \emph{et al.} \cite{chen_sars_cov_2_in_lung, chen_et_al_antibody_protection, zhang_et_al_computational_modelling_extereme_heterogeneity, pearson_et_al_modelling_variability_in_sars_cov_2, zhang_et_al_global_sensitivity_analysis}).
	
	One method for implementing this coupled discrete--continuum framework is the Cellular Potts Model \cite{graner_and_glazier_CPM}. Under this framework, the spatial domain and continuum quantities are discretised over a grid of subcellular voxels. Cells are represented as discrete agents without a fixed shape or lattice: instead, they are characterised by a given target size (in terms of number of voxels), as well as a range of other energy constraints, such as minimisation of boundary length. Over time, the model is simulated by applying stochastic noise at the voxel scale and minimising the energy state of the cell grid. This model therefore accounts for realistic tissue mechanics and packing over time, while providing a natural discretisation for the continuum layer of the model. This framework has been applied to the spatial dynamics of viral infections by \textbf{Imle \emph{et al.}} \cite{imle_et_al_hiv_3d_collagen_matrix} and Glazier and colleagues \cite{sego_et_al_cellularisation, sego_et_al_covid_model, sego_et_al_multicellular_influenza, aponte_serrano_RNA_virus_replication, gianlupi_et_al_multiscale_model_of_antiviral_timing}. \textbf{Getz \emph{et al.}} \cite{getz_rapid_community_driven_covid_model} take a similar approach for viral transmission but use an off lattice point force model for cell mechanics. 
	
	The multicellular framework permits the simultaneous representation of multiple biological length scales, which in some ways combines the benefits of continuum and cellular automaton modelling: small particles, especially virions, can be described in terms of density dynamics at the tissue--scale, while the modelling of discrete cells imposes sharp infection boundaries and naturally accounts for cell--scale stochasticity. However, this framework also comes with significant challenges. The substantial leap in model complexity compared to either continuum or lightweight cellular automaton models means that multicellular systems are far more computationally expensive to run. Moreover, this increase in modelling detail is not always accompanied by similarly detailed available biological data. Levin \emph{et al.}, for example, remarked on a lack of data in describing CD8$^+$ T cell motion \cite{levin_et_al_T_cell_search_influenza}. This lack of data is especially important since many large multicellular models (\textbf{Sego \emph{et al.}} \cite{sego_et_al_covid_model} and \textbf{Getz \emph{et al.} \cite{getz_rapid_community_driven_covid_model}}, for example) do not (and cannot realistically) calibrate parameter values to biological data using rigorous statistical methods, instead adapting estimated parameter values from elsewhere in the literature. The relevance of such parameter values may vary depending on their context within the model from which they are derived and the biological system to which they are applied. 
	
	An additional challenge associated with multicellular systems lies in their implementation. Multicellular models offer a fine resolution of tissues at the cell--scale, however, there is little or no consensus on how such cells should be arranged within the model tissue, for example, or to what extent this influences the dynamics. Some authors use square packing for the cells (Marzban \emph{et al.} \cite{marzban_et_al_hybrid_PDE_ABM_model, bartha_et_al_in_silico_evalutation_paxlovid} and Chen \emph{et al.} \cite{chen_et_al_antibody_protection, chen_sars_cov_2_in_lung, aristotelous_et_al_covid_model, zhang_et_al_computational_modelling_extereme_heterogeneity, pearson_et_al_modelling_variability_in_sars_cov_2, zhang_et_al_global_sensitivity_analysis}, for example), some use hexagonal packing (\textbf{Yakimovich \emph{et al.}} \cite{yakimovich_et_al_cell_free_human_adenovirus, yakimovich_et_al_infectio}, Levin \emph{et al.} \cite{levin_et_al_T_cell_search_influenza}, Beauchemin \emph{et al.} \cite{beauchemin_et_al_modeling_influenza_in_tissue}, Whitman \emph{et al.} \cite{whitman_et_al_spatiotemporal_host_virus_influeza}, for example), while other consider irregular, time--varying cell packing (\textbf{Sego \emph{et al.}} \cite{sego_et_al_covid_model}, \textbf{Getz \emph{et al.}} \cite{getz_rapid_community_driven_covid_model}, \textbf{Imle \emph{et al.}} \cite{imle_et_al_hiv_3d_collagen_matrix}). In addition to implementation of the cell sheet, there is also a wide variety in approaches to the representation of continuum quantities relative to the cell grid. We have already discussed the voxel length scale of the Cellular Potts Model; many other authors simply discretise viral and other densities at the cell--scale (\textbf{Mitchell \emph{et al.}} \cite{mitchell_et_al_high_replication_efficiency_H1N1}, Beauchemin \emph{et al.} \cite{beauchemin_et_al_modeling_influenza_in_tissue}, Holder \emph{et al.} \cite{holder_et_al_design_considerations_influenza}, Howat \emph{et al.} \cite{howat_et_al_modelling_dynamics_of_type_I_IFN}, for example). Others use alternative discretisations independent of the cell grid (\textbf{Getz \emph{et al.}} \cite{getz_rapid_community_driven_covid_model, islam_et_al_agent_based_lung_fibrosis}, Itakura \emph{et al.} \cite{itakura_et_al_reproducibility_chronic_virus_infection_model}, Jenner \emph{et al.} \cite{jenner_et_al_oncolytic_voronoi}). This point is important, since, in some models, biological mechanisms occur over small spatial scales, such as cell--to--cell infection between migrating cells (\textbf{Imle \emph{et al.}} \cite{imle_et_al_hiv_3d_collagen_matrix}, Seich Al Basatena \emph{et al.} \cite{al_basatena_et_al_non_lytic_CD8}), or viral diffusion in crowded environments (\textbf{Sego \emph{et al.}} \cite{sego_et_al_covid_model}, Beauchemin \emph{et al.} \cite{beauchemin_et_al_modeling_influenza_in_tissue}, Jenner \emph{et al.} \cite{jenner_et_al_agent_based_glioblastoma}). Analysis has suggested that in multicellular models of viral dynamics in which virus diffuse sufficiently slowly, the difference between cell--scale and fine--resolution discretisation of the viral density results in qualitative differences in model outcomes \cite{williams_et_al_spatial_discretisation}. This suggests that the dynamics of such systems are dependent on biological activity at spatial scales far smaller than the cell, which may defy modelling assumptions \cite{williams_et_al_spatial_discretisation}. Findings such as these point to a need for further discussion and analysis of implementation dependence for multicellular viral dynamics systems.

	\section{Spatially--structured models by virus species} \label{sec:papers_by_virus}
	
	Much of the specifics of viral infection dynamics can vary substantially between viral species. Different viral pathogens possess unique life cycles, target different cell populations with different biological and structural properties within the host, and have varied interactions with the host immune response. There are therefore a range of modelling particularities which are specific to the viral strain being modelled. This is especially true of spatially--structured models since different host tissues possess specific spatial structures and cells with particular motility properties. In this section, we discuss how spatially--structured models have been applied to the dynamics of specific viral species.

	\subsection{HIV}

        \begin{figure}[h!]
    	\centering
    	\includegraphics[width=\textwidth]{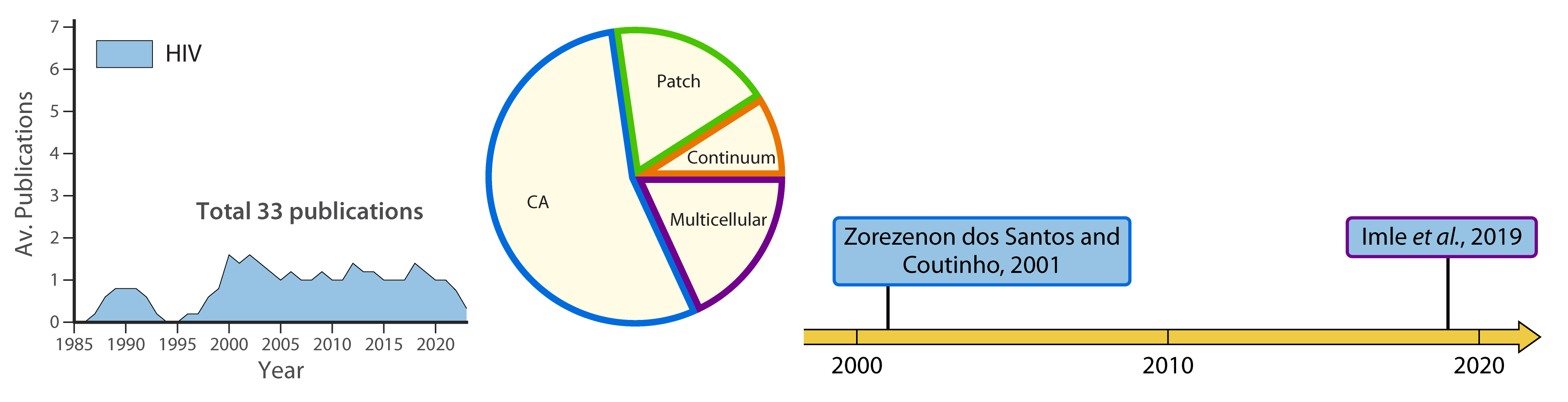}
    	\caption{Spatially--structured models for HIV. \added{Left: Average number of models for HIV per year. Centre: HIV models by model type. Right: Highly--cited publications for HIV models over time.} \added{Highly--cited} publications are as follows: \textbf{Zorzenon dos Santos and Coutinho} \cite{dos_santos_and_countinho_dynamics_of_HIV_CA}, and \textbf{Imle \emph{et al.}} \cite{imle_et_al_hiv_3d_collagen_matrix}.}
    	\label{fig:HIV_summary}
    \end{figure}
	
	HIV is a chronic infection of lymphoid tissue, particularly CD4$^+$ T cells \cite{perelson_hiv_review, nakaoka_et_al_HIV_in_lymphoid_tissue_network}. Although now well--managed with the daily administration of antiretrovirals, if left untreated HIV infection is characterised by three distinct phases: acute (time scale of days), chronic (time scale of years), and AIDS (time scale of weeks), the last of which is fatal \cite{zhang_et_al_massively_multi_agent_system, perelson_hiv_review, graw_and_perelson_spatial_hiv}. HIV modelling represents the oldest and most developed branch of the viral dynamics literature in mean--field as well as spatial modelling (reviewed by Perelson \cite{perelson_hiv_review, graw_and_perelson_spatial_hiv} and Nowak and May \cite{nowak_and_may_virus_dynamics_book}, for example), and is the subject of the bulk of early viral dynamics models. In addition to its unique three--phase temporal structure of dynamics, HIV is unusual in terms of the spatial structure of its infection, since its target cells, CD4$^+$ T cells, are motile \cite{graw_and_perelson_spatial_hiv, imle_et_al_hiv_3d_collagen_matrix}. Moreover, in addition to releasing virions into the extracellular environment, HIV also spreads largely among host cells through virological synapses formed between host and susceptible cells \cite{komarova_et_al_relative_contribution_CC_in_hiv, bocharov_review}.
	
	The target tissue of HIV infection comprises spatially disparate yet interconnected lymph nodes and lymphatic tissue. Therefore, the definition of model domains and their spatial scale are highly important in modelling HIV. Despite this, many authors define non--specific spatial domains (Zhang \emph{et al.} \cite{zhang_et_al_massively_multi_agent_system}, Pandey \emph{et al.} \cite{pandey_1989_CA_model_HIV, pandey_1991_CA_interacting_cellular_network, pandey_and_stauffer_metastability, mielke_and_pandey_fuzzy_interaction, mannion_et_al_effect_of_mutation, mannion_et_al_monte_carlo_population_dynamics}, Raza \emph{et al.} \cite{raza_et_al_numerical_efficient_splitting_HIV}, for example). In these cases, the spatial domain might be interpreted as a generalisation of the entire volume of susceptible tissue within the host or alternatively as a representative region of tissue. However, given the structure of the target tissue for HIV, there is a clear difference between these macro-- and mesoscopic spatial scales of infection, and such models can therefore only be interpreted as abstractions of the spatial reality. A popular target for many modellers is the within--lymph node dynamics of HIV infection (Bernaschi and Castiglione \cite{bernaschi_and_castiglione_escape_mutants}, \textbf{Zorzenon dos Santos and Coutinho} \cite{dos_santos_and_countinho_dynamics_of_HIV_CA}, Marinho \emph{et al.} \cite{marinho_et_al_PDE_HIV}, Strain \emph{et al.} \cite{strain_et_al_spatiotemporal_hiv}, for example). Though only one section of the overall target tissue is involved in HIV infection, this single--lymph node approach has proved a useful model system for reproducing important global dynamics of HIV such as the three stages of infection (\textbf{Zorzenon dos Santos and Coutinho} \cite{dos_santos_and_countinho_dynamics_of_HIV_CA}, Bernaschi and Castiglione \cite{bernaschi_and_castiglione_escape_mutants}, Marinho \emph{et al.} \cite{marinho_et_al_PDE_HIV}). Jagarapu \emph{et al.} \cite{jagarapu_et_al_integrated_spatial_dynamics_pharmacokinetic} and Cardozo \emph{et al.} \cite{cardozo_et_al_compartmental_analysis, cardozo_et_al_cryptic_viremia} used lymph node models to describe the dynamics of chronic HIV infection under antiretroviral treatment. They showed that antiretrovirals were heterogeneously distributed in lymph nodes, and in particular, that the drugs were unable to deeply penetrate into lymphoid follicles, facilitating the ongoing replication of virions \cite{jagarapu_et_al_integrated_spatial_dynamics_pharmacokinetic, cardozo_et_al_compartmental_analysis, cardozo_et_al_cryptic_viremia}. We only found one study, by Nakaoka \emph{et al.}, which explicitly modelled the between--lymph node dynamics of HIV infection \cite{nakaoka_et_al_HIV_in_lymphoid_tissue_network}. In their work, the authors showed that, even for high antiretroviral dosage, there was a limit to the global drug efficacy \cite{nakaoka_et_al_HIV_in_lymphoid_tissue_network}. This was due to the heterogeneity in drug delivery across the lymphoid network, with some lymphoid compartments not receiving sufficient drug load to inhibit viral replication even when high doses were administered \cite{nakaoka_et_al_HIV_in_lymphoid_tissue_network}. Other biological domains have also been modelled for HIV. Jafelice \emph{et al.} \cite{jafelice_et_al_CA_with_fuzzy_parameters} and Benyoussef \emph{et al.} \cite{benyoussef_et_al_dynamics_of_HIV_infection_2D} considered HIV dynamics in the blood, while Immonem \emph{et al.} \cite{immonem_et_al_hybrid_stochastic_deterministic} and \textbf{Imle \emph{et al.}} \cite{imle_et_al_hiv_3d_collagen_matrix} modelled HIV dynamics \emph{in vitro}. These \emph{in vitro} models serve as useful tools for interpreting biological data, although several groups have noted that it is difficult to build laboratory systems for HIV infection which are faithful to the dynamics within the host, due to the migratory properties of the host CD4$^+$ T cells and the three--dimensional structure of host tissues \cite{imle_et_al_hiv_3d_collagen_matrix, iwami_et_al_HIV_CC_estimate, komarova_et_al_relative_contribution_CC_in_hiv}. \textbf{Imle \emph{et al.}} applied a detailed interdisciplinary approach to overcome this disconnect by studying the dynamics of a three--dimensional \emph{in vitro} system which included a complex collagen lattice, designed to simulate the structure of infected tissue within the host \cite{imle_et_al_hiv_3d_collagen_matrix}. The authors found that these structural considerations hampered the diffusion of free virions compared to other \emph{in vitro} systems and promoted direct cell--to--cell infection \cite{imle_et_al_hiv_3d_collagen_matrix}.
	
	Authors have approached the motility of the target cells in HIV infection in different ways. There are several cellular automaton approaches to HIV which ignore CD4$^+$ T cell migration altogether and assume static cells and neighbour--based infection (\textbf{Zorzenon dos Santos and Coutinho} \cite{dos_santos_and_countinho_dynamics_of_HIV_CA}, Sloot \emph{et al.} \cite{sloot_et_al_CA_drug_therapy_HIV}, Precharattana \emph{et al.} \cite{precharattana_et_al_investigation_of_spatial_pattern_formation}, for example), the same as has been implemented for viral species which target non--motile cells (\textbf{Wodarz \emph{et al.}} \cite{wodarz_et_al_complex_spatial_dynamics_oncolytic} and Bhatt \emph{et al.} \cite{bhatt_et_al_modelling_the_spatial_dynamics_oncolytic_virotherapy} in oncolytic viruses, \textbf{Xiao \emph{et al.}} \cite{xiao_et_al_cellular_automaton_HBV} for HBV, for example). Strain \emph{et al.} applied a similar approach, where instead of being explicitly neighbour--based, infection probability was determined by local virion bursts from static infected cells \cite{strain_et_al_spatiotemporal_hiv}. While such models have the benefit of simplicity and have been shown to exhibit temporal dynamics which qualitatively agree with clinical and experimental observations of HIV infections, others, including Graw and Perelson, have criticised this approach for its un--biological construction \cite{graw_and_perelson_spatial_hiv}. Despite the high--level agreement with observed dynamics, this disconnect with the biological reality of HIV--infected tissues means that these models should not be interpreted as (spatially) mechanistic. Pandey \emph{et al.} modelled HIV infection using a static cell grid, but accounted for cell motility implicitly through the rules for infection. The authors used cellular automaton models with a complex (phenomenological) mixing operation between lattice sites to account for experimentally--observed diffuse infection patterns \cite{pandey_1989_CA_model_HIV, pandey_1991_CA_interacting_cellular_network, pandey_and_stauffer_metastability, mielke_and_pandey_fuzzy_interaction, mannion_et_al_effect_of_mutation, mannion_et_al_monte_carlo_population_dynamics}. Other authors have represented the movement of HIV target cells as flux between compartments. In the patch--based approach of Jagarapu \emph{et al.} \cite{jagarapu_et_al_integrated_spatial_dynamics_pharmacokinetic} and Cardozo \emph{et al.} \cite{cardozo_et_al_compartmental_analysis, cardozo_et_al_cryptic_viremia}, CD4$^+$ T cell density had essentially two rates of migration for different regions of the tissue: diffusion--like spread between radial layers of lymph follicles, and instantaneous spread among the (mean--field) free \mbox{blood/lymph} compartment. Bernaschi and Castiglione, by applying the ImmSim model to HIV, tracked the locations of a discrete population of CD4$^+$ T cells as they migrated across a coarse spatial lattice \cite{bernaschi_and_castiglione_escape_mutants, baldazzi_et_al_enhanced_agent_based_model, castiglione_et_al_mutation_fitness_viral_diversity}. Seich Al Basatena \emph{et al.} also tracked the movement of discrete CD4$^+$ T cells across a lattice, but at far higher resolution, such that the cells moved across a lattice of cell--sized, mostly empty sites \cite{al_basatena_et_al_non_lytic_CD8}. More recently, some authors have also modelled the migration of individual T cells in continuous space. Bouchnita \emph{et al.} \cite{bouchnita_et_al_towards_multiscale_HIV} and Grebennikov and Bocharov \cite{grebennikov_and_bocharov_spatially_resolved_modelling} considered spherical cells whose displacement was governed by Newtonian motion. At an even finer level of spatial detail, \textbf{Imle \emph{et al.}} \cite{imle_et_al_hiv_3d_collagen_matrix} also used physics--based rules for CD4$^+$ T cell migration, but also allowed for cellular deformation (through modelling in the Cellular Potts Model framework, discussed in Section~\ref{sec:agent_based_multicellular}).

	\subsection{Oncolytic viruses}
	\label{sec:oncolytic}

        \begin{figure}[h!]
    	\centering
    	\includegraphics[width=\textwidth]{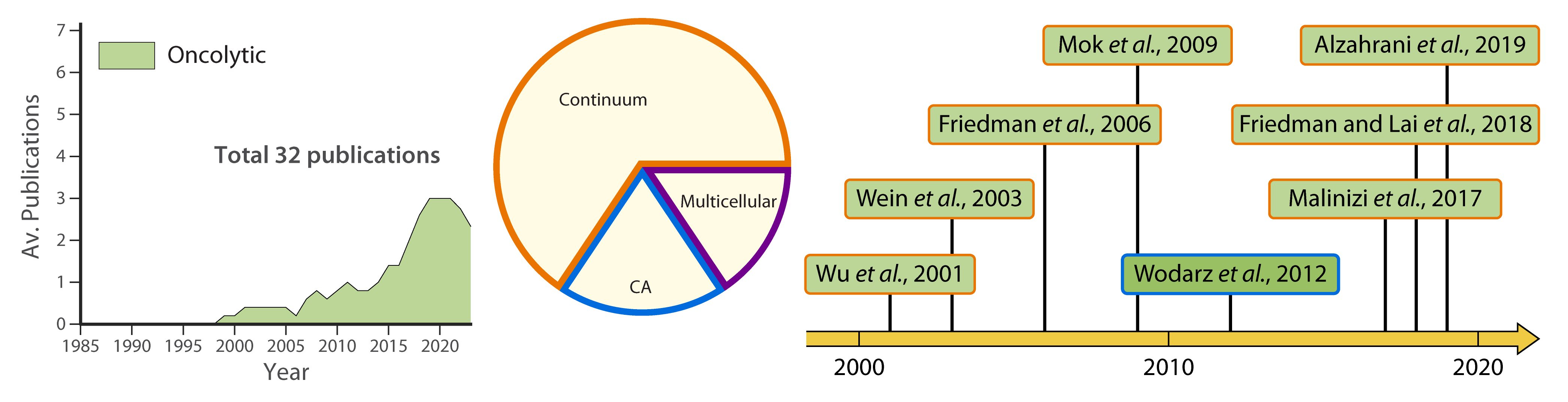}
    	\caption{Spatially--structured models for oncolytic viruses. \added{Left: Average number of models for oncolytic viruses per year. Centre: Oncolytic virus models by model type. Highly--cited publications for oncolytic virus models over time.} \added{Highly--cited} publications are as follows: \textbf{Wu \emph{et al.}} \cite{wu_et_al_modelling_and_analysis_oncolytic}, \textbf{Wein \emph{et al.}} \cite{wein_et_al_validation_and_analysis_oncolytic}, \textbf{Friedman \emph{et al.}} \cite{friedman_et_al_glioma_virotherapy}, \textbf{Mok \emph{et al.}} \cite{mok_et_al_herpes_simplex_in_solid_tumour}, \textbf{Wodarz \emph{et al.}} \cite{wodarz_et_al_complex_spatial_dynamics_oncolytic}, \textbf{Malinzi \emph{et al.}} \cite{malinzi_2017_spatiotemporal_chemovirotherapy}, \textbf{Friedman and Lai} \cite{friedman_and_lai_combination_oncolytic_checkpoint}, and \textbf{Alzahrani \emph{et al.}} \cite{alzahrani_2019_multiscale_cancer_response}.}
    	\label{fig:oncolytic_summary}
    \end{figure}
    
	Oncolytic viruses are a broad group of genetically--engineered viruses that preferentially infect, replicate within, and kill tumour cells, while leaving healthy cells untouched. This, in principle, makes the prospect of oncolytic virotherapy an attractive treatment option for cancers, although experimental and clinical results have so far been mixed, and development is ongoing \cite{lawler_et_al_oncolytic_viruses_review, jebar_et_al_clinical_oncolytics_hepatocellular}. The complex, multiscale dynamics of oncolytic viral infection, as well as the marked spatial constraints within tumours, makes oncolytic viruses well suited to spatially--explicit mathematical modelling, although such viruses are a special case among the viral dynamics literature. For one, since oncolytic viral infections are deliberately introduced into host tissue, the placement and dosage of the viral inoculum are especially important \cite{wein_et_al_validation_and_analysis_oncolytic, wu_et_al_modelling_and_analysis_oncolytic, bailey_et_al_radial_expansion_intratumoural}. Moreover, since infection takes place in pathologically crowded tissue under high pressure, host tissue is subject to constant movement and rearrangement, and free viral diffusion is highly constrained \cite{jenner_et_al_oncolytic_voronoi, morselli_et_al_agent_based_and_continuum_models}.
	
	Spatially--structured models for oncolytic viruses are overwhelmingly implemented as PDE systems. There are several reasons for this. For one, it is clearly of interest to model virus--host interaction at the whole--tumour scale, which is generally well approximated by a density description for the model components. PDE systems also allow description of the mechanics of the intra--tumour density (Morselli \emph{et al.} \cite{morselli_et_al_agent_based_and_continuum_models}, \textbf{Friedman and Lai} \cite{friedman_and_lai_combination_oncolytic_checkpoint}, \textbf{Friedman \emph{et al.}} \cite{friedman_et_al_glioma_virotherapy}) and the tumour--healthy tissue interface (\textbf{Alzahrani \emph{et al.}} \cite{alzahrani_2019_multiscale_cancer_response, alzahrani_2020_moving_boundary_fusogenic}, Alsisi \emph{et al.} \cite{alsisi_go_or_grow, alsisi_heterogeneous_ECM, alsisi_nonlocal_multscale_approaches}). In simpler models, the growth of the tumour is approximated as radially symmetric, reflecting the approximately spherical structure of solid tumours in the early stages of growth (\textbf{Malinzi \emph{et al.}} \cite{malinzi_2017_spatiotemporal_chemovirotherapy}, \textbf{Friedman \emph{et al.}} \cite{friedman_et_al_glioma_virotherapy}, \textbf{Wu \emph{et al.}} \cite{wu_et_al_modelling_and_analysis_oncolytic}, \textbf{Wein \emph{et al.}} \cite{wein_et_al_validation_and_analysis_oncolytic}, \textbf{Mok \emph{et al.}} \cite{mok_et_al_herpes_simplex_in_solid_tumour}). This assumption reduces the model to one--dimensional in space, which simplifies mathematical analysis. The aims of analysis of models of oncolytic viral dynamics are often different to those in analysis of the dynamics of other viruses, and more often focus on the tumour, rather than the infection. While for, say, SARS--CoV--2, authors have derived approximations of the speed of the infection front (Tokarev \emph{et al.} \cite{tokarev_et_al_fastest_autowave}, Ait Mahiout \emph{et al.} \cite{mahiout_2022_virus_replication_and_competition, mahiout_2023_respiratory_viral_infections}), in models of oncolytic viral infections, estimates have instead been made for the speed of the tumour front (\textbf{Malinzi \emph{et al.}} \cite{malinzi_2017_spatiotemporal_chemovirotherapy}, de Rioja \emph{et al.} \cite{de_rioja_et_al_gliobastomas}, \textbf{Wein \emph{et al.}} \cite{wein_et_al_validation_and_analysis_oncolytic}). \textbf{Wu \emph{et al.}} \cite{wu_et_al_modelling_and_analysis_oncolytic} and \textbf{Wein \emph{et al.}} \cite{wein_et_al_validation_and_analysis_oncolytic} also used analytical techniques to derive expressions for tumour extinction thresholds as functions of the model parameters.
	
	A popular application of models for oncolytic viral infections is in testing the impact of different approaches to viral administration to the tumour. \textbf{Wu \emph{et al.}} \cite{wu_et_al_modelling_and_analysis_oncolytic} and \textbf{Wein \emph{et al.}} \cite{wein_et_al_validation_and_analysis_oncolytic} both studied models for radially symmetric tumours with virus initially delivered either to the rim or to the core of the tumour. \textbf{Wu \emph{et al.}} suggested that, taking into account the necrosis of the tumour core due to nutrient diffusion from the rim, core injection was likely to be somewhat redundant, and that rim injection would be more effective in spreading through the tumour \cite{wu_et_al_modelling_and_analysis_oncolytic}. \textbf{Wein \emph{et al.}} found that for oncolytic virotherapy to be effective, the virus must be highly aggressive --- that is, extremely infectious and fast--spreading --- in order to outpace the growth of the tumour \cite{wein_et_al_validation_and_analysis_oncolytic}; this result was recapitulated by Al-Johani \emph{et al.} \cite{al_johani_et_al_spatiotemporal_dynamics_virotherapy} and Simbawa \emph{et al.} \cite{simbabwa_et_al_spatiotemporal_dynamics_oncolytic_radiovirotherapy}. Pooladvand \emph{et al.}, however, has warned that high infectivity is not necessarily sufficient for complete eradication of the tumour \cite{pooladvand_et_al_viral_infectivity_in_oncolytic_virotherapy_outcomes}. Other authors have also found that the success of oncolytic virotherapy depends on viral infection reaching widespread regions of the tumour. Storey and Jackson compared a viral dosage strategy which repeatedly targeted the same location in the tumour to an adaptive strategy which instead delivered oncolytic viral load to the location of highest tumour cell density at that point in time, and found that the adaptive strategy dramatically reduced the final tumour mass after the same length of treatment time \cite{storey_and_jackson_ABM_combination_oncolytic}. Bailey \emph{et al.} proposed a model with randomly seeded foci arising from viral extravasation but with a hard limit on the radius of infection foci, following experimental observations of limited focal spread \cite{bailey_et_al_radial_expansion_intratumoural}. This approach suggested that, in order for virotherapy to be effective, the number of viral foci would have to be extremely high, such that the overall infected volume would cover a sufficiently wide region \cite{bailey_et_al_radial_expansion_intratumoural}. \added{However, since the model is phenomenological, the authors cannot deduce a mechanistic explanation for the failure of the treatment at a reasonable dosage.}
	
	This spatially constrained spread of infection foci within tumours is a recurring theme in explaining the limited \emph{in vivo} efficacy of oncolytic viruses. One reason for this limitation in the spatial spread of infection is due to the fact that the extremely dense tumour environment severely inhibits the diffusion of free virions \cite{jenner_et_al_agent_based_glioblastoma, morselli_et_al_agent_based_and_continuum_models}. As a result, many authors have argued that infection is likely only possible directly between neighbouring cells (Morselli \emph{et al.} \cite{morselli_et_al_agent_based_and_continuum_models}, Bhatt \emph{et al.} \cite{bhatt_et_al_modelling_the_spatial_dynamics_oncolytic_virotherapy}, Reis \emph{et al.} \cite{reis_et_al_in_silico_evolutionary_tumour_virotherapy}). \textbf{Wodarz \emph{et al.}} showed that experimentally--observed infection patterns for oncolytic viruses could be generated from a cellular automaton with adjacency--based infection only \cite{wodarz_et_al_complex_spatial_dynamics_oncolytic}. As a result, it has been widely argued that increased diffusion of free virions would dramatically improve oncolytic virotherapy (Paiva \emph{et al.} \cite{paiva_et_al_multiscale_oncolytic_virotherapy}, Jenner \emph{et al.} \cite{jenner_et_al_agent_based_glioblastoma, jenner_et_al_oncolytic_voronoi}, \textbf{Mok \emph{et al.}} \cite{mok_et_al_herpes_simplex_in_solid_tumour}, for example), however, there is disagreement in the modelling community about the conditions under which this might come about. \textbf{Mok \emph{et al.}} \cite{mok_et_al_herpes_simplex_in_solid_tumour} and Paiva \emph{et al.} \cite{paiva_et_al_multiscale_oncolytic_virotherapy} have argued that a reduction in density of the tumour extracellular matrix would increase the rate of viral diffusion, however, Alsisi \emph{et al.} \cite{alsisi_go_or_grow, alsisi_heterogeneous_ECM, alsisi_nonlocal_multscale_approaches} have suggested that virions preferentially migrate \emph{towards} regions of higher matrix density via haptotaxis. A related problem to the rate of viral diffusion is the relative density of the tumour cells and normal stromal cells. Since the latter cannot be infected by the oncolytic virus, \textbf{Wein \emph{et al.}} \cite{wein_et_al_validation_and_analysis_oncolytic} and Jenner \emph{et al.} \cite{jenner_et_al_agent_based_glioblastoma} have shown that oncolytic viral infections are far more effective in tumours where cancer cells are densely packed, with few interspersed stromal cells. However, in contrast to these findings, Berg \emph{et al.} showed that oncolytic infections spread more slowly in a three--dimensional tumour model compared to an equivalent model in two dimensions, despite cells in the former having a higher number of adjacent tumour cells \cite{berg_et_al_in_vitro_in_silico_multidimensional_oncolytic}. \added{The lack of consensus in these findings points to complex, multifactorial dynamics at play, and suggests a need for an improved understanding of the various mechanisms of oncolytic virus dispersal within tumours, including through cell--to--cell transmission, diffusion, and haptotaxis. Moreover, these results suggest that a realistic representation of the tumour microenvironment, such as the extracellular matrix density, the distribution of cancer cells and non--cancer cells, and perhaps the dimensionality of the tumour, may be necessary to ensure dynamics faithful to oncolytic viral infection \emph{in vivo}.}

	Since the spread of viral infections in tumours is \added{believed to} mainly between neighbouring cells, the arrangement of cells in space is \added{likely} highly important in agent--based modelling of oncolytic viruses. Simpler approaches to the arrangement of tumour cells in space consider a regular lattice of cell--sites which remain fixed over time (\textbf{Wodarz \emph{et al.}} \cite{wodarz_et_al_complex_spatial_dynamics_oncolytic}, Reis \emph{et al.} \cite{reis_et_al_in_silico_evolutionary_tumour_virotherapy}, Paiva \emph{et al.} \cite{paiva_et_al_multiscale_oncolytic_virotherapy} for example). Others have constructed spatially--static networks of cells based on Voronoi tessellations derived from experimental imaging of tumours, which provides a realistic distribution of the number of neighbours between cells across the tumour (Berg \emph{et al.} \cite{berg_et_al_in_vitro_in_silico_multidimensional_oncolytic} and Bhatt \emph{et al.} \cite{bhatt_et_al_modelling_the_spatial_dynamics_oncolytic_virotherapy}). Berg \emph{et al.} compared oncolytic viral dynamics on a regular and Voronoi cell lattice in three dimensions and found that infections on the Voronoi lattice were less likely to result in eradication of the tumour \cite{berg_et_al_in_vitro_in_silico_multidimensional_oncolytic}. \added{Spatiotemporal evolution of the cell grid is also likely important in the dynamics of oncolytic viruses (as compared to other viral species) due to the rapid growth of cancer tissue. Indeed, as we have seen, several authors describing oncolytic viral dynamics at the whole-tumour scale have shown that the growth of the tumour is a significant obstacle to effective treatment (\textbf{Wein \emph{et al.}} \cite{wein_et_al_validation_and_analysis_oncolytic}, Al-Johani \emph{et al.} \cite{al_johani_et_al_spatiotemporal_dynamics_virotherapy}, Simbabwa \emph{et al.} \cite{simbabwa_et_al_spatiotemporal_dynamics_oncolytic_radiovirotherapy}, for example). However, only one model at the cell scale has accounted for growth of the tissue:} Jenner \emph{et al.} considered a time--varying Voronoi lattice model for tumour cells, which also enabled the explicit representation of tumour cell proliferation and remodelling, and moreover facilitated the simulation of time--varying adjacency between cells \cite{jenner_et_al_oncolytic_voronoi}. \added{Further work is needed, however, to more deeply explore the dynamics of infection on not just a realistic static cell packing, but also an evolving one.}

	\subsection{Respiratory viruses}
	\label{sec:respiratory}

   \begin{figure}[h!]
   	\centering
   	\includegraphics[width=\textwidth]{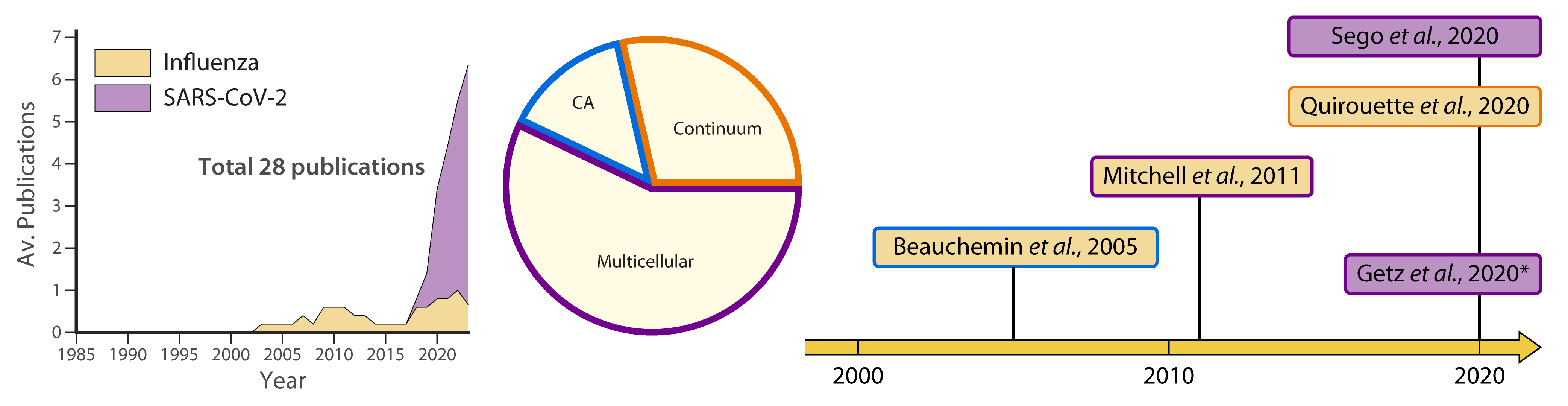}
   	\caption{Spatially--structured models for influenza and SARS--CoV--2. \added{Left: Average number of models for respiratory viruses per year. Centre: Respiratory virus models by model type. Right: Highly--cited publications for respiratory virus models over time.} \added{Highly--cited} publications are as follows: \textbf{Beauchemin \emph{et al.}} \cite{beauchemin_et_al_simple_cellular_automaton_model}, \textbf{Mitchell \emph{et al.}} \cite{mitchell_et_al_high_replication_efficiency_H1N1}, \textbf{Getz \emph{et al.}} \cite{getz_rapid_community_driven_covid_model}, \textbf{Quirouette \emph{et al.}} \cite{quirouette_et_al_influenza_localisation_model}, and \textbf{Sego \emph{et al.}} \cite{sego_et_al_covid_model}.}
   	\label{fig:respiratory_summary}
   \end{figure}

	Viral infections that target the respiratory tract --- including influenza and SARS--CoV--2 --- have had a spike in popularity among the modelling literature following the outbreak of the COVID--19 pandemic. Respiratory viruses, unlike HIV or oncolytic viruses, broadly follow the dynamics of the ``general'' viral infection as modelled in the literature: that is, respiratory viral infections spread among non--motile cells within a single, contiguous body of tissue, across which cell--free virions can diffuse relatively freely \cite{gallagher_spatial_spread, beauchemin_et_al_modeling_influenza_in_tissue, holder_et_al_design_considerations_influenza}. Nonetheless, there are still elements of spatial infection dynamics which are specific to infections of the respiratory tract. We note that, to the level of detail that mathematical models have described them, there is little difference between the dynamics of different respiratory viral species; as such in the following we discuss models for respiratory viral infections in general.
	
	Viruses which infect the respiratory tract usually target the monolayer of epithelial cells which line the airways \cite{chen_sars_cov_2_in_lung, moses_et_al_lung_covid_model, gallagher_spatial_spread}. Thus, to an extent, the natural geometry of respiratory infection is two--dimensional, especially at the microscopic scale, and as a consequence virtually all of the respiratory infection models retained for analysis from our search also consider two spatial dimensions. However, at a larger length scale, there is substantial structural heterogeneity along the length of the respiratory tract. The respiratory tract has a complex, tree--like structure, comprising 24 generations of branching airways, such that upper passages like the trachea are relatively wide (on the order of centimetres), while deeper airways are very narrow (on the order of fractions of millimetres) \cite{makevnina_lung_atlas, chen_sars_cov_2_in_lung}. Ayadi \emph{et al.} \cite{ayadi_et_al_modelling_and_simulation_SARS_CoV_2} and Moses \emph{et al.} \cite{moses_et_al_lung_covid_model} have explicitly considered the dynamics of viral infection across the entire, heterogeneous lung. In the latter case, Moses \emph{et al.} constructed a full--size, three--dimensional digital human lung, complete with a detailed branching structure for the cells \cite{moses_et_al_lung_covid_model}. The authors showed that viral dynamics on this highly complex structure were significantly different to the dynamics of the same infection model on a purely two--dimensional cell grid \cite{moses_et_al_lung_covid_model}. While the use of this hyperrealistic geometry demonstrates the notable differences in the dynamics of infections on flat sheets compared to the lung, there is a substantial conceptual leap between the two geometries. Other authors have instead employed an intermediate approach, studying the spread of viral infections on sections of the \added{respiratory tract} in order to identify specific ways in which the structure and geometry of the \added{respiratory tract} influence the spread of infection. Chen \emph{et al.} \cite{chen_sars_cov_2_in_lung} and Williams \emph{et al.} \cite{williams_et_al_lung} studied the spread of viral infections on individual \added{airway} branches, or on small sections of branching tissue. The latter group showed that tubular and branching tissue geometry imposes rich, spatially--directional infection dynamics \cite{williams_et_al_lung}. These dynamics moreover exhibited an increased sensitivity to the rate of viral diffusion compared to the dynamics of infection on a flat cell sheet \cite{williams_et_al_lung}. \added{While these theoretical works have offered hints into the ways in which the dynamics of respiratory viral infections \emph{in vivo} may differ from infections \emph{in vitro}, a deeper integration with experimental data is now necessary to gain quantitative insights into the spread of infection in the respiratory tract. Obtaining relevant \emph{in vivo} data remains a challenge, however, advances in imaging techniques and genetic engineering (discussed in Section~\ref{sec:future_data}) may soon provide data necessary for such analysis.}
	
	In addition to structural heterogeneity, there is also variability in the cellular composition of upper and lower regions of the respiratory tract. As a consequence, the rate of spread for different viral strains varies for different regions of the lung depending on their cellular tropism \cite{fukuyama_et_al_color_flu, gallagher_spatial_spread}. Tokarev \emph{et al.} \cite{tokarev_et_al_fastest_autowave} and Ait Mahiout \emph{et al.} \cite{mahiout_2022_virus_replication_and_competition} calibrated PDE models to experimental data and used travelling wave methods to compare the dynamics of the omicron and delta variants of SARS--CoV--2. The authors found that the omicron variant had a more rapid invasion speed and therefore a fitness advantage in the early stages of infection \cite{tokarev_et_al_fastest_autowave} and in the upper regions of the respiratory tract \cite{mahiout_2022_virus_replication_and_competition}. However, Tokarev \emph{et al.} also found that in the delta variant spread faster in the \emph{lower} respiratory tract \cite{tokarev_et_al_fastest_autowave}. These findings are consistent with the usual experimental and clinical observation that viruses which spread preferentially in the lower respiratory tract (like avian--adapted influenza strains or the delta variant of SARS--CoV--2) are typically associated with high pathogenicity, while those that spread preferentially in the upper respiratory tract (like seasonal influenza strains or the omicron variant of SARS--CoV--2) are typically more infectious but result in milder disease \cite{smith_validated_models_of_immune_response, gallagher_spatial_spread, ke_perelson_et_al_covid_upper_and_lower_rt, fukuyama_et_al_color_flu}.
	
	Another characteristic feature of the lung is the mucociliary escalator, the directed flow of the mucus lining the respiratory epithelium from lower to upper regions of the respiratory tract. \textbf{Quirouette \emph{et al.}} \cite{quirouette_et_al_influenza_localisation_model}  included the mucociliary escalator in a model of influenza infection, and suggested that infection can effectively only spread up the respiratory tract from the initial point of virion deposition \cite{quirouette_et_al_influenza_localisation_model}. The rate of mucociliary flow is variable at different depths of the respiratory tract, with slower flow in \added{lower airways} \cite{makevnina_lung_atlas, chen_sars_cov_2_in_lung}. Chen \emph{et al.} compared the dynamics of SARS--CoV--2 infection in airway tissue from different generations of the respiratory tract and suggested that infection spread was severely limited in the lower airways \cite{chen_sars_cov_2_in_lung}. The authors postulated that infection propagation in the deep lung must therefore be driven by a large quantity of viral depositions from upper regions of the lung, such as virus--laden droplets or aerosols \cite{chen_sars_cov_2_in_lung}. Chakravarty \emph{et al.} --- who extended  the Quirouette model --- reached a similar conclusion, arguing that aerosolisation of infected nasopharyngeal mucosa (ANM) was necessary for infection to reach the deep lung \cite{chakravarty_et_al_virus_loaded_droplets}. Vimalajeewa \emph{et al.}, by contrast, modelled advection \emph{down} the respiratory tract in the course of SARS--CoV--2 infection due to air flow in the respiratory lumen (the authors accounted for the mucociliary escalator as a drag force and slip correction factor in the flow model) \cite{vimalajeewa_virus_particle_propagation_respiratory_tract}. In addition to its role in disease progression, Zhang \emph{et al.} suggested that the rate of mucus advection also influences the nasal viral load \cite{zhang_et_al_global_sensitivity_analysis}. The authors argue that variation in this rate between individuals could offer one potential explanation for the extreme variation observed in clinical nasal swab titres among COVID--19 patients \cite{zhang_et_al_global_sensitivity_analysis}.

	\subsection{Other viruses} 
	\label{sec:other_viruses}

        \begin{figure}[h!]
    	\centering
    	\includegraphics[width=\textwidth]{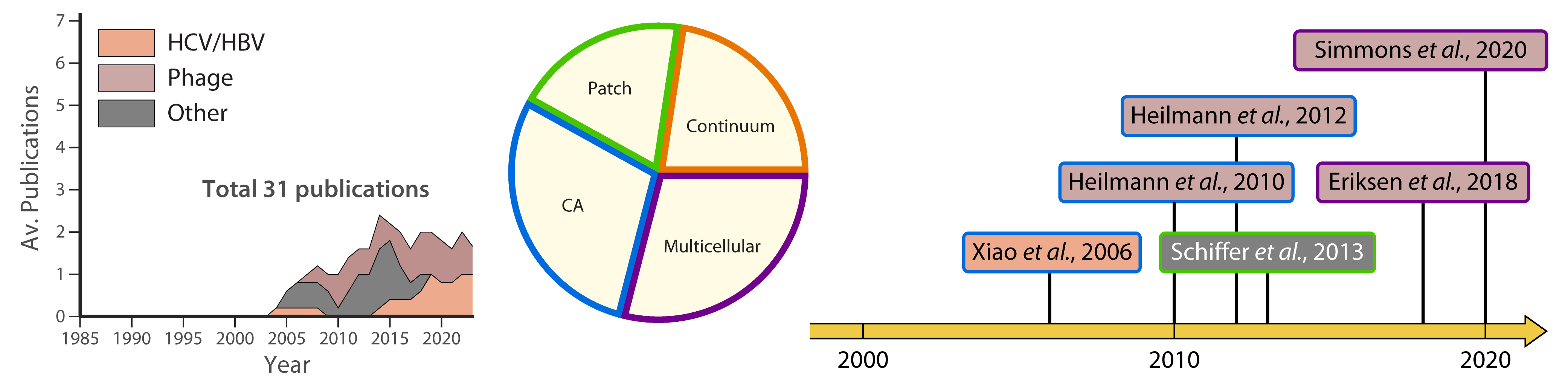}
    	\caption{Spatially--structured models for other specific viruses. \added{Left: Average number of models for other specific viruses per year. Centre: Other specific virus models by model type. Right: Highly--cited publications for other virus models over time.} \added{Highly--cited} publications are as follows: \textbf{Xiao \emph{et al.}} \cite{xiao_et_al_cellular_automaton_HBV}, \textbf{Heilmann \emph{et al.} (2010)} \cite{heilmann_et_al_2010_sustainability_of_virulence}, \textbf{Heilmann \emph{et al.} (2012)} \cite{heilmann_et_al_2012_coexistence_phage_and_bacteria}, \textbf{Schiffer \emph{et al.}} \cite{schiffer_et_al_rapid_localised_spread}, \textbf{Eriksen \emph{et al.}} \cite{eriksen_et_al_growing_microcolony}, and \textbf{Simmons \emph{et al.}} \cite{simmons_et_al_biofilm_structure}.}
    	\label{fig:other_summary}
    \end{figure}
	
	In addition to the viruses which we have discussed above, several other species have been considered. Among these, a substantive body of work has been developed for the modelling of bacteriophage (or simply ``phage'') dynamics. Phage and the bacteria which they infect are both some of the largest biological populations in existence, and their interaction is one of the oldest predator--prey relationships on Earth \cite{simmons_et_al_biofilm_structure, mitarai_et_al_population_dynamics_phage}. Bacteria and phage inhabit and interact within a broad array of environments, including in soil, water, and within larger organisms \cite{heilmann_et_al_2010_sustainability_of_virulence, garrido_zornoza_et_al_stochastic_microbial_dispersal}. There is a major difference between phage dynamics and the dynamics of the other virus species which we have considered in this review, in that \added{while we have so far mainly considered viruses which infect tissues within a single host, the bacteria which phage infect are single--cell organisms. However, there are also many similarities. Phage--bacteria systems are frequently studied --- and modelled --- \emph{in vitro} (Valdez \emph{et al.} \cite{valdez_et_al_heterogeneous_bacteria-phage}, de Rioja \emph{et al.} \cite{de_rioja_et_al_front_propagation_T7}, for example), and the dynamics of these systems can be directly compared to the \emph{in vitro} dynamics of other, human--adapted viruses. Bacteria also frequently form dense, surface--bound microbial communities called biofilms, which can be likened to the crowded environment within tumours \cite{valdez_et_al_heterogeneous_bacteria-phage, simmons_et_al_biofilm_structure, eriksen_et_al_growing_microcolony}. Despite first appearances, modelling of phage--bacteria dynamics is therefore closely interlinked with models for other viral species, and forms an important component of the development of spatially--structured viral dynamics models in general.}
	
	Due to the unique properties of phage--bacteria dynamics, as compared to the dynamics of other viral species we have discussed, modellers have considered the spatially--structured interactions between phage and microbe communities on very large spatiotemporal scales. Garrido Zornoza \emph{et al.} modelled the spread of phage among a microbial community at an ecological spatial scale (the only such model we found in our search), which described spatially explicit microbial habitats, from which both viruses and host microbes were subject to stochastic aerosolisation and advection with wind \cite{garrido_zornoza_et_al_stochastic_microbial_dispersal}. The authors showed that this influx--outflux process among habitats lead to the spatial spread of microbial communities but also resulted in destabilisation and occasional local extinctions in individual habitats \cite{garrido_zornoza_et_al_stochastic_microbial_dispersal}. While phage--bacteria systems exhibit spatially--structured dynamics on this ecological spatial scale, most authors instead model the dynamics of phage--bacteria interactions on far smaller length scales, particularly in biofilms. The crowded nature of these environments requires particular consideration by modellers. Martínez-Calvo \emph{et al.}, for instance, considered density-dependent motility of bacteria and showed that the spread of phage infection among the colony can disrupt bacterial aggregation \cite{martinez-calvo_et_al_pattern_formation_bacteria_phage}. Moreover, as with oncolytic viral infections in tumours, the densely--packed environment of biofilms almost entirely suppress viral diffusion, resulting in highly localised viral spread. Taylor \emph{et al.}, for example, showed that the high cell density found in biofilms promotes the formation of spatial clustering of infection \cite{taylor_et_al_emergence_multiple_infections}. Valdez \emph{et al.} showed that, as a consequence of this highly localised spread, differently--shaped initial distributions of phage among bacterial communities gave rise to spatially distinct colony structures \cite{valdez_et_al_heterogeneous_bacteria-phage}. Modellers have accounted for the crowded nature of biofilms in various ways. Some, including \textbf{Heilmann \emph{et al.} (2012)} \cite{heilmann_et_al_2012_coexistence_phage_and_bacteria} and Mitarai \emph{et al.} \cite{mitarai_et_al_population_dynamics_phage}, have implemented linear diffusion to describe the dispersal of cell--free phage with a very small diffusion coefficient compared to say, the Stokes--Einstein estimate (Akpinar \emph{et al.} \cite{akpinar_et_al_spatiotemporal_viral_amplification}, Beauchemin \emph{et al.} \cite{beauchemin_et_al_modeling_influenza_in_tissue}, \textbf{Sego \emph{et al.}} \cite{sego_et_al_covid_model}, for example, took similar approaches for other viral species). Other authors have instead opted for non--linear forms of the diffusion term: Hunter \emph{et al.}, for example, considered density--dependent diffusion \cite{hunter_et_al_virus_host_travelling_waves}; \textbf{Simmons \emph{et al.}} explicitly considered physical blocking of phage motion by bacteria \cite{simmons_et_al_biofilm_structure}. \added{A direct comparison of these different potential mechanisms is lacking, but could be compared to experimental evidence to provide insight into the characteristics of phage dispersal in the crowded biofilm environment, which may also have consequences for other viral species, especially oncolytic viruses.}
	
	One of the remarkable features of phage--bacteria interactions is the coexistence of both populations over vast time scales, even within a single biofilm. This relationship is a property of the rich and finely--calibrated predator--prey dynamic between phage and host microbes. Several authors have shown that spatial structure in phage distribution within microbial communities is a key contributor to long--term stability of both populations. \textbf{Heilmann \emph{et al.} (2010)} compared coexistence of phage and bacterial populations in a cellular automaton model with either spatially--localised or well--mixed conditions, and found that the former permitted coexistence over a far larger region of parameter space \cite{heilmann_et_al_2010_sustainability_of_virulence}. Authors have proposed several mechanisms by which spatial structure permits coexistence. In their study, \textbf{Heilmann \emph{et al.} (2010)} allowed infection of dead or already--infected bacteria by phage, and demonstrated a ``shielding'' effect, whereby infected or dead regions of the biofilm would act as phage sinks, protecting regions of susceptible bacteria \cite{heilmann_et_al_2010_sustainability_of_virulence}. In a different study, the same group (\textbf{Heilmann \emph{et al.} (2012)}) \cite{heilmann_et_al_2012_coexistence_phage_and_bacteria}, demonstrated the spontaneous formation of these spatial bacterial ``refuges'' in their models, a phenomenon also shown by Bull \emph{et al.} \cite{bull_et_al_phage_bacterial_dynamics}. Others have shown that the spatial structure of infection influences the complex co-evolutionary dynamics between phage and bacteria. Haerter and Sneppen modelled the evolution of phage resistance mediated by the Clustered Regularly--Interspaced Short Palindromic Repeats (CRISPR) mechanism and showed that well-mixed conditions resulted in inefficient and Darwinian evolution of resistance, while spatially--structured conditions resulted in Lamarckian evolution and provided better agreement with experimental evidence \cite{haerter_and_sneppen_spatial_structure_lamarckian}. Others have sought to explain the counter-intuitive experimental observation that phage appear to be able to develop virulence faster than bacteria can develop defences, yet do not drive host populations to extinction \cite{heilmann_et_al_2010_sustainability_of_virulence, heilmann_et_al_2012_coexistence_phage_and_bacteria}. Coberly \emph{et al.} suggested that in competition between moderate and highly--virulent phages, the high selection pressure for resistance to the virulent strain would favour the persistence of the weaker virus, despite the moderate virus being otherwise outcompeted \cite{coberly_et_al_space_time_host_evolution}. Notably, the authors suggested this was a direct consequence of the spatial structure of infection, since they observed ``percolating clusters'' of resistant bacteria which provided spatial refuges for the moderate phage \cite{coberly_et_al_space_time_host_evolution}. In another study, \textbf{Eriksen \emph{et al.}} suggested that the survival of phage populations among spatially--isolated bacterial microcolonies relies on the phage sustainably ``farming'' colonies instead of destroying them \cite{eriksen_et_al_growing_microcolony}.
	
	Hepatitis B and C viruses, which cause chronic infections of the liver, have also been modelled by several authors. Despite the recent development of highly effective direct--acting antivirals for the latter, hepatitis viruses are still globally widespread with relatively high rates of mortality \cite{guedj_et_al_hcv_daclatasvir, dahari_et_al_HCV_in_the_era_of_DAAs}. \textbf{Xiao \emph{et al.}} \cite{xiao_et_al_cellular_automaton_HBV} and Cangelosi \emph{et al.} \cite{cangelosi_et_al_multiscale_HBV} have argued that spatial heterogeneity of infection may play a role in immune evasion, supporting viral persistence in the liver. Goyal and Murray \cite{goyal_and_murray_CCT_in_HBV} and Blahut \emph{et al.} \cite{blahut_et_al_hepatitis_c_two_modes_of_spread} have also argued that direct cell--to--cell transmission of hepatitis viruses also contributes to the chronicity of infection (we discuss modelling for cell--to--cell infection in greater depth in Section~\ref{sec:innately_spatially_structured_dynamics}). Moreover, within--cell dynamics appears to be key in hepatitis B dynamics. Cangelosi \emph{et al.} showed that viral cccDNA represents one of the most robust obstacles to hepatitis B treatment, and suggested that even a single cccDNA molecule would be sufficient to reinfect the entire liver \cite{cangelosi_et_al_multiscale_HBV}. \added{Interestingly, although the non--spatial modelling literature for hepatitis B and C is fairly extensive (reviewed by Dahari \emph{et al.} \cite{dahari_et_al_HCV_in_the_era_of_DAAs} and Canini and Perelson \cite{canini_and_perelson_viral_kinetic}, for example), there are relatively few (seven) articles which account for spatial structure. This is despite clear biological evidence of the role of spatial dynamics in the spread of hepatitis B and C infection, such as the presence and suspected high rate of cell--to--cell transmission \cite{sattenau_avoiding_the_void, goyal_and_murray_CCT_in_HBV, blahut_et_al_hepatitis_c_two_modes_of_spread}, and experimentally observed clusters of infected cells in hepatitis C--infected livers \cite{canini_and_perelson_viral_kinetic, graw_et_al_inferring_HCV}. There is therefore substantial scope for future model development for hepatitis viruses which takes into account spatial effects in the dynamics.}
	
	Some authors have also modelled the spatial dynamics of herpes simplex virus (HSV--2). HSV--2 infections target the genital tract and are also chronic \cite{schiffer_et_al_HSV2_review}. \textbf{Schiffer \emph{et al.}} showed that, during viral shedding episodes, HSV--2 replicates extraordinarily quickly, and moreover rapidly spreads in space, effectively ``outrunning'' the initial immune response and prolonging the episode \cite{schiffer_et_al_rapid_localised_spread}. Schiffer \cite{schiffer_hsv2_specific_cd8} and Dhankani \emph{et al.} \cite{dhankani_et_al_hsv2_genital_tract_shedding} have also highlighted the highly complex and spatially--structured interaction between HSV--2 infections and host immune reponses. Dhankani \emph{et al.} suggested that while the dynamics of HSV--2 episodes could be partially predicted in the short term based on the spatial distribution of immune cell density, on the time scale of weeks or months the dynamics of the episode became impossible to predict \cite{dhankani_et_al_hsv2_genital_tract_shedding}. HSV--2 infections have also been modelled in co--infection with HIV. Modelling has shown a synergistic relationship between the two species: Byrne \emph{et al.} suggested that HSV--2--infected women were up to 8.6 times more likely to acquire HIV infection than healthy hosts \cite{byrne_et_al_spatiotemporal_hsv2_and_hiv}. On the other hand, Schiffer \emph{et al.} showed that, as a result of CD4$^+$ T cell depletion, HIV--infected individuals were far more susceptible to HSV--2 infection \cite{schiffer_et_al_increased_hsv2_shedding}.
	
	We also found a single publication which modelled plant viruses (Rodrigo \emph{et al.} \cite{rodrigo_plant_viruses}), along with a handful of models for Epstein--Barr virus (EBV). EBV is an extremely common viral infection in humans which causes generally mild illness but remains dormant in hosts and may be reactivated, occasionally causing long term complications \cite{shapiro_et_al_virtual_EBV_simulation_mechanism, duca_et_al_virtual_EBV_biological_interpretations}; it has also been associated with the development of cancers \cite{castiglione_et_al_EBV_immsim}. EBV targets the tonsils of Waldeyer's ring: Shapiro \emph{et al.} used a cellular automaton model which explicitly accounted for the entire anatomy of Waldeyer's ring and its peripheral circulation \cite{shapiro_et_al_virtual_EBV_simulation_mechanism, duca_et_al_virtual_EBV_biological_interpretations}. In a separate model, Castiglione \emph{et al.} showed that the persistence of EBV in the host was due to latently--infected B cells spreading to this peripheral pool, where they were not subject to immune surveillance \cite{castiglione_et_al_EBV_immsim}.

	\section{Spatially--structured dynamics} \label{sec:innately_spatially_structured_dynamics}
	
\begin{figure}[h!]
    	\centering
    	\includegraphics[width=\textwidth]{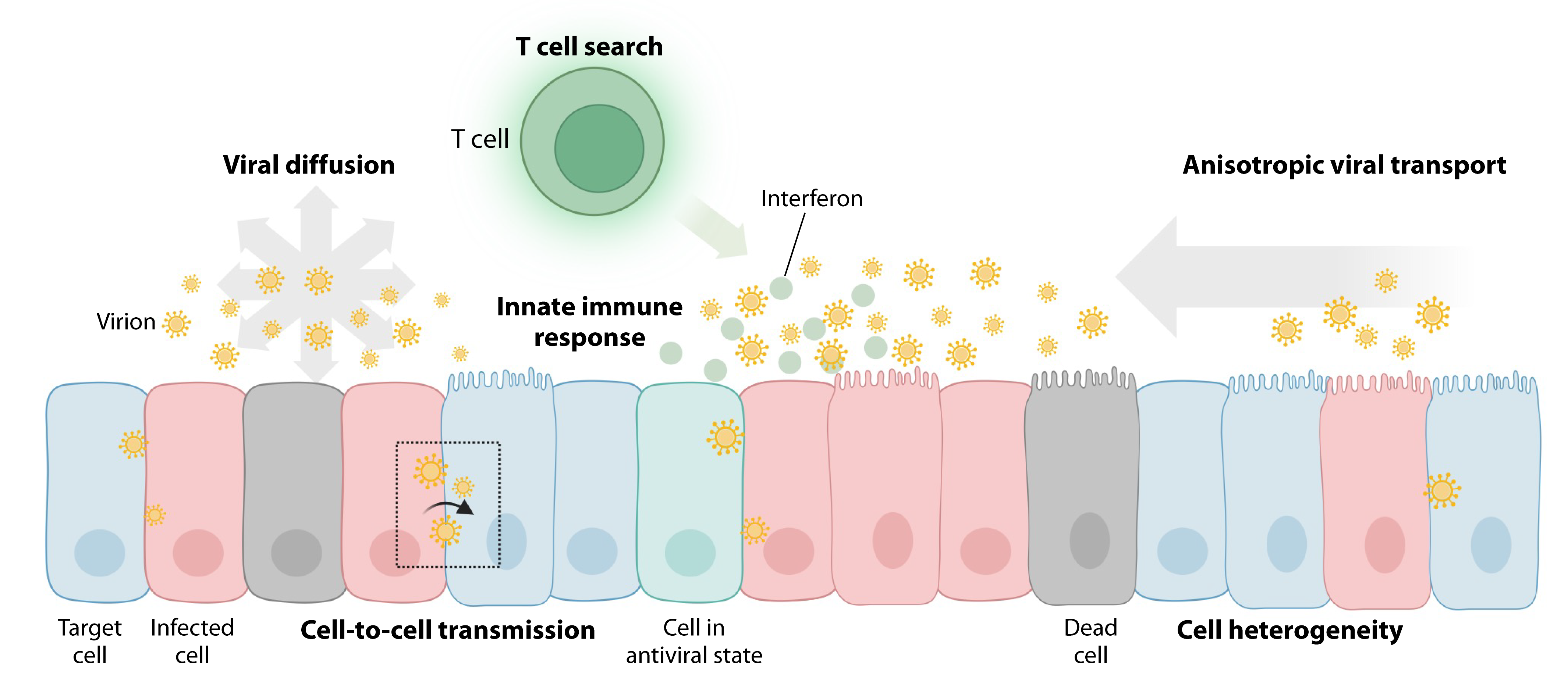}
    	\caption{Sources of spatially-structured viral dynamics discussed in Section~\ref{sec:innately_spatially_structured_dynamics}.}
    	\label{fig:sources_of_spatial_structure}
    \end{figure}

	Mean--field models, especially ODE models, have a long history of application to the dynamics of viral infections and have offered substantial insight (reviewed in \cite{perelson_hiv_review, smith_and_perelson_influenza_review, nowak_and_may_virus_dynamics_book, smith_validated_models_of_immune_response}, for example). However, there are a range of aspects of viral infection dynamics which inherently involve a spatial mode of action, such as the action of the immune system, or the spread of infection in crowded tissue. The mismatch between the spatially--structured biological reality and these non--spatial models means that there is a limit to the precision possible with such models in representing accurate infection dynamics, or in accurately determining parameter values \cite{bauer_et_al_agent_based_models_virus, gallagher_spatial_spread}. In this section, we review how spatially--structured models have accounted for spatially--acting aspects of viral infection, and discuss how these dynamics differ from descriptions by non--spatial models. \added{Figure~\ref{fig:sources_of_spatial_structure} illustrates the sources of spatially--structured viral dynamics discussed in this section in a biological context.}

	\subsection{Spatial heterogeneity and parameter estimation}
	
	One of the simplest improvements offered by spatially--structured modelling is in allowing for spatially--heterogeneous distribution of infection. \added{Heterogeneity of the host tissue is illustrated in Figure~\ref{fig:sources_of_spatial_structure}.} We have already discussed several examples of heterogeneous tissues, including heterogeneity in pressure and density within tumours for oncolytic viruses (Morselli \emph{et al.} \cite{morselli_et_al_agent_based_and_continuum_models}, Alsisi \emph{et al.} \cite{alsisi_go_or_grow, alsisi_heterogeneous_ECM, alsisi_nonlocal_multscale_approaches}), variable viral advection rate at different depths in the lung due to the mucociliary escalator (\emph{et al.} \cite{chen_sars_cov_2_in_lung, zhang_et_al_computational_modelling_extereme_heterogeneity, chen_et_al_antibody_protection, pearson_et_al_modelling_variability_in_sars_cov_2}, Vimalajeewa \emph{et al.} \cite{vimalajeewa_virus_particle_propagation_respiratory_tract}), and structural heterogeneity in the arrangement of cells (Bhatt \emph{et al.} \cite{bhatt_et_al_modelling_the_spatial_dynamics_oncolytic_virotherapy}, Jenner \emph{et al.} \cite{jenner_et_al_oncolytic_voronoi}, Berg \emph{et al.} \cite{berg_et_al_in_vitro_in_silico_multidimensional_oncolytic}). Heterogeneity of tissue has been shown to qualitatively impact the outcome of models compared to homogeneous systems. Chen \emph{et al.} showed that the probability of the same viral inoculum establishing infection varied dramatically at different depths of the respiratory tract, based solely on variations in the rate of mucociliary flow \cite{chen_sars_cov_2_in_lung}. In a patch model for viral dynamics, \textbf{Funk \emph{et al.}} implemented random heterogeneity of model parameters between patches and found that this heterogeneity impacted the dynamics of the equilibrium state of infection \cite{funk_et_al_spatial_models_virus_immune}.
	
	Beyond these qualitative differences in infection patterns, spatially--varying infections have been shown to have a profound quantitative impact on parameter estimation. Funk \emph{et al.} also showed that in one region of parameter space, the patch model predicted viral persistence while an analogous mean--field model predicted extinction \cite{funk_et_al_spatial_models_virus_immune}. More explicitly, Strain \emph{et al.} compared simulations of HIV dynamics using both a cellular automaton and an ODE system, and found that the mean--field model overestimated viral infectiousness more than 10--fold \cite{strain_et_al_spatiotemporal_hiv}. \textbf{Mitchell \emph{et al.}} obtained a similar result in influenza \cite{mitchell_et_al_high_replication_efficiency_H1N1}. In their study, the authors calibrated both a cellular automaton and an ODE system to experimental data, and found that the predicted basic reproduction number, $\rzero$, for pandemic--strain influenza was estimated about 5--10 times higher using the ODE model \cite{mitchell_et_al_high_replication_efficiency_H1N1}. Clearly, this has dramatic consequences for the characterisation of the virus.

	\subsection{Cell--to--cell viral transmission}
	\label{sec:cell-to-cell}
	
	ODE models for viral dynamics rely on an assumption that cells and virions are well--mixed. Under this construction, virions have equal access to all cells, and all uninfected cells are equally likely to become infected, regardless of their position in space. Under the standard assumption that infection is propagated through the dispersal of cell--free virions, this corresponds to an assumption that virions spread quickly across the model tissue (Holder \emph{et al.} \cite{holder_et_al_design_considerations_influenza}, Beauchemin \cite{beauchemin_well_mixed_assumption}); we have already discussed some instances where this assumption might be appropriate (Cangelosi \emph{et al.} \cite{cangelosi_et_al_multiscale_HBV}, Blahut \emph{et al.} \cite{blahut_et_al_hepatitis_c_two_modes_of_spread}, Kreger \emph{et al.} \cite{kreger_et_al_synaptic_cell_to_cell}, for example). However, there has been growing recognition, both in the biological and modelling literature, of the importance of the cell--to--cell route of infection \cite{jansens_et_al_virus_long_distance_TNTs, kumar_influenza_TNTs, kongsomros_et_al_trogocytosis_influenza, tiwari_et_al_TNTs_review, gallagher_spatial_spread}. \added{Cell--to--cell transmission is illustrated in the bottom left of Figure~\ref{fig:sources_of_spatial_structure}.} This mode of transmission has been observed in infections by virtually all medically--important viruses, including retroviruses, herpesviruses, coronaviruses and influenza viruses \cite{tiwari_et_al_TNTs_review, sattenau_avoiding_the_void}. Since cell--to--cell transmission relies on adjacency between cells, spatial structure is a central aspect of its mode of action. Crucially, although this phenomenon has been described using ODE methods, this approach has had limited success. Kumberger \emph{et al.} generated synthetic time series data of viral infection using a cellular automaton which accounted for both modes of infection, then attempted to refit the data using ODEs \cite{kumberger_et_al_accounting_for_space_cell_to_cell}. The authors found that ODE models could not accurately describe the dynamics unless additional terms were included to artificially account for the spatial structure \cite{kumberger_et_al_accounting_for_space_cell_to_cell}. A similar conclusion was reached by Graw and colleagues in an ODE modelling study (not included in our search since this was a non--spatial model), who attempted to fit experimental data on cell--to--cell infection and also needed to artificially account for the spatial structure of the infection to obtain a reasonable fit \cite{graw_et_al_hcv_cell_to_cell_stochastic}. Williams \emph{et al.} conducted simulation--estimation studies on both ODE and multicellular models for viral dynamics with both modes of infection and found that the balance of the two modes of spread in the ODE model was not practically identifiable in the presence of observational noise \cite{williams_et_al_inference}. This finding suggests that the two mechanisms are relatively interchangeable in these non--spatial models \cite{williams_et_al_inference}.
	
	Spatially--structured modelling has provided insight in characterising the role of the two modes of infection in overall viral infection dynamics. Blahut \emph{et al.} fit a multicellular model to experimental data for hepatitis C and concluded that cell--to--cell infection could account for as much as $99$\% of infection events, however cell--free infection was crucial to the overall progression of the virus by providing access to uninfected regions of the tissue away from the infection focus \cite{blahut_et_al_hepatitis_c_two_modes_of_spread}. \added{Durso-Cain \emph{et al.}, who also modelled hepatitis C dynamics, reported a similar synergistic effect between the two modes of transmission, and also noted that the cell--to--cell mechanism was dominant, especially early in the infection \cite{durso_cain_hcv_dual_spread}.} \added{By contrast}, Goyal and Murray, argued that \added{in hepatitis B}, cell--to--cell infection was \emph{not} especially important to the overall dynamics in the early stages of infection, but was crucial in the chronic stage of infection through T cell exhaustion \cite{goyal_and_murray_CCT_in_HBV}. On this longer time scale, Saeki and Sasaki \cite{saeki_and_sasaki_spatial_heterogeneity} and Kreger \emph{et al.} \cite{kreger_et_al_synaptic_cell_to_cell} have also implicated cell--to--cell infection in the evolutionary dynamics of viral infections (we discuss evolutionary dynamics in greater depth in Section~\ref{sec:immune_activity_in_space}). 
	
	However, while modelling has offered qualitative insights into the role of cell--to--cell transmission, there remains substantial quantitative uncertainty as to the actual \emph{amount} of cell--to--cell transmission (relative to cell--free infection) in the spread of different viral infections. This represents a substantial modelling challenge for several reasons: for one, this quantity is likely to be substantially different for different viral species \cite{zeng_et_al_sars_cov_2_cell_to_cell, blahut_et_al_hepatitis_c_two_modes_of_spread, kongsomros_et_al_trogocytosis_influenza, imle_et_al_hiv_3d_collagen_matrix}, and moreover, the balance of the two roles of transmission is expected to vary between \emph{in vitro} and \emph{in vivo} contexts \cite{dixit_and_perelon_HIV_patients, durso_cain_hcv_dual_spread}. Blahut \emph{et al.} \cite{blahut_et_al_hepatitis_c_two_modes_of_spread} and Durso--Cain \emph{et al.} \cite{durso_cain_hcv_dual_spread} estimated the relative contribution of the two modes of viral spread by comparing experimental data with and without the presence of extracellular virus--neutralising antibody. This approach, however, is limited to \emph{in vitro} settings, and incurs additional complexity by introducing the possibility of inhibition or enhancement of the cell--to--cell route in the presence of cell--free infection blockage. Kumberger \emph{et al.} \cite{kumberger_et_al_accounting_for_space_cell_to_cell} and Williams \emph{et al.} \cite{williams_et_al_inference}, by contrast, each attempted to estimate the relative proportion of the two modes of infection using only a single simulated dataset, describing uninhibited viral dynamics. Both groups determined that global infection metrics alone were insufficient to practically identify the balance of the two modes of infection, however, Williams \emph{et al.} \cite{williams_et_al_inference} showed that, in principle, data describing the spatial structure of the infection would be sufficient to enable this quantity to be inferred.
	
	Due to the migration of target cells in space, the role of cell--to--cell transmission in HIV is somewhat unique. A detailed discussion was provided by \textbf{Imle \emph{et al.}} \cite{imle_et_al_hiv_3d_collagen_matrix}. The authors simulated HIV dynamics using a Cellular Potts Model calibrated to detailed imaging data to track the movement and contact between CD4$^+$ T cells. Upon comparison with experimental time series data, the authors concluded that cell--to--cell transmission of HIV requires at least $25$ minutes of contact between donor and recipient cells \cite{imle_et_al_hiv_3d_collagen_matrix}. The authors also modelled \emph{in vitro} data collected from cells either in suspension or in a 3D collagen matrix --- which was argued to be representative of conditions \emph{in vivo} --- and found that cell--to--cell infection was significantly promoted in the collagen matrix, compared to the suspension \cite{imle_et_al_hiv_3d_collagen_matrix}. However, this calculation was derived using fits to non--spatial ODE models, and, perhaps as a consequence, estimates for the prevalence of the cell--to--cell mechanism were subject to extremely wide uncertainty \cite{imle_et_al_hiv_3d_collagen_matrix}.

	\subsection{Spatial immune activity}\label{sec:immune_activity_in_space}
	
	In addition to the spatial spread of infection, there are also aspects of the immune response which are inherently spatially--structured. \added{Immune system dynamics in the context of viral infection are illustrated in the upper portion of Figure~\ref{fig:sources_of_spatial_structure}.} Modelling has shown that the presence of spatial structure significantly increases the potency of the interferon--mediated innate immune response (Lavigne \emph{et al.} \cite{lavigne_et_al_interferon_signalling_ring_vaccination}, Segredo--Otero and Sanju\'{a}n \cite{segredo_otero_et_al_spatial_structure_innnate_immunity}). Lavigne \emph{et al.} modelled innate immune response to viral infection using ODE, PDE and cellular automaton techniques \cite{lavigne_et_al_interferon_signalling_ring_vaccination}. The authors demonstrated the formation of buffer regions comprised of interferon--mediated antiviral cells surrounding the infection focus, which they likened to a ``ring vaccination'' strategy as in epidemiology \cite{lavigne_et_al_interferon_signalling_ring_vaccination}. The cellular automaton model, owing to its stochasticity, also facilitated the possibility of temporary viral resurgence, due to rare long--range infection events beyond the protective region \cite{lavigne_et_al_interferon_signalling_ring_vaccination}.
	
	In the adaptive immune response, T cells are recruited from the lymph nodes, then traverse infected tissue along chemotactic gradients to the site of infected cells. This search process has been described by a number of authors (\textbf{Beauchemin \emph{et al.}} \cite{beauchemin_et_al_simple_cellular_automaton_model}, \textbf{Sego \emph{et al.}} \cite{sego_et_al_covid_model}, \textbf{Getz \emph{et al.}} \cite{getz_rapid_community_driven_covid_model}, for example). Levin and colleagues used a multicellular model to characterise the dynamics of the T cell search in the adaptive immune response to influenza A infection, and showed that the window of time which T cells have to locate and clear infection foci is dependent on the viral strain \cite{levin_et_al_T_cell_search_influenza}. A similar conclusion was reached by Morselli \emph{et al.} in the context of oncolytic viral dynamics, who suggested that if the adaptive immune response is activated too quickly following the initiation of virotherapy, the infection has limited effect on the growth of the tumour \cite{morselli_et_al_hybrid_discrete_immune}. Levin \emph{et al.} moreover suggested that the ability of the immune response to effectively clear infection relies on viral diffusion being sufficiently slow: beyond a certain threshold, the infection spreads too quickly for the T cell response to keep up \cite{levin_et_al_T_cell_search_influenza}. This serves to highlight the importance of space in the action of the immune response to viral infections. Kadolsky and Yates, too, showed that under the assumption of chemotactic T cell migration, the infection structure in space is crucial in determining the outcome of the immune response, with highly clustered infections significantly more likely to be cleared \cite{kadolsky_and_yates_spatial_immune_surveillance}. As a consequence, since infection typically spreads patchily within the host, the authors predicted that mean--field models would overestimate the number of T cells needed to clear infection \cite{kadolsky_and_yates_spatial_immune_surveillance}.
	
	Just as spatial structure plays a key role in immune action, it is also important in the evolution of immune--resistant traits among the viral population. Capit\'{a}n \emph{et al.} \cite{capitan_et_al_severe_hindrance_viral_infection_propagation} and Lavigne \emph{et al.} \cite{lavigne_et_al_interferon_signalling_ring_vaccination} have demonstrated that spatial confinement of infection dramatically increases evolutionary pressure on viruses, and Phan and Wodarz \cite{phan_and_wodarz_modelling_multiple_infection} showed that spatial constraints promoted competitive exclusion between two competing viral strains and increased the likelihood of one strain becoming extinct. Yin \cite{yin_spatially_resolved_evolution} showed that spreading of viral plaques in space promotes the development and spatial separation of viral mutants. Segredo--Otero and Sanju\'{a}n discussed the role of spatial structure in the evolution of evasion of the innate immune response specifically \cite{segredo_otero_et_al_spatial_structure_innnate_immunity}. The authors showed that in infection with a mix of interferon--triggering and interferon--evading viral variants, the immune--evading population are selected for and the immune--triggering variants are eliminated only when spatial structure is present \cite{segredo_otero_et_al_spatial_structure_innnate_immunity}. Sardany\'{e}s \emph{et al.} also suggested that spatial structure selects for viral genotypes which are robust against deleterious mutation (so--called ``survival of the flattest'') \cite{sardanyes_et_al_simple_quasispecies_survival_flattest}. Spatially--structured immune dynamics have also been used to explain evolutionary strategies of both viruses and host immune responses. Komarova used a model of viral--antibody interaction to show an evolutionary advantage of lytic replication of viruses, that is, virions which replicate inside host cells but are only released by killing (``bursting'') the cell \cite{komarova_viral_reproductive_strategies}. The model demonstrated a ``flooding'' effect, where the bulk release of virus in bursting events allowed virions to spread far enough, without being neutralised by antibodies, to infect new cells and continue the infection \cite{komarova_viral_reproductive_strategies}. By contrast, the model suggested that a steady, budding release of virions would be more effectively controlled by the antibody response \cite{komarova_viral_reproductive_strategies}. Cuesta \emph{et al.} probed this phenomenon further in the context of evolutionary competition between virus and host, for instance in phage--bacteria dynamics \cite{cuesta_et_al_struggle_for_space}. The authors determined that a large burst size is advantageous for the virus only when host cells are readily available: in spatially--restricted settings this strategy offers no such advantage \cite{cuesta_et_al_struggle_for_space}. Seich Al Basatena \emph{et al.} studied the role of lytic and non--lytic CD8$^+$ T cells in viral escape during HIV infection, and found that while a lytic CD8$^+$ T cell response rapidly selected for an escape variant virus, even a non--lytic response (in the sense of secreting antiviral chemokines that on contact with susceptible cells block any infection) was found to promote viral escape \cite{al_basatena_et_al_non_lytic_CD8}. This effect was due to the spatial structure of infection, since both wild--type and variant viruses were observed to form infections in clusters, only the former of which could be recognised by the CD8$^+$ T cells \cite{al_basatena_et_al_non_lytic_CD8}. As a consequence, the resulting distribution of antiviral chemokines were observed to spatially coincide with wild--type infected regions, while leaving variant--infected regions less likely to be affected \cite{al_basatena_et_al_non_lytic_CD8}.
	
	Saeki and Sasaki \cite{saeki_and_sasaki_spatial_heterogeneity} and Kreger \emph{et al.} \cite{kreger_et_al_synaptic_cell_to_cell} have also discussed the role of the cell--to--cell and cell--free mechanisms of infection in evolutionary dynamics. Saeki and Sasaki used Monte Carlo simulation to predict the evolution of a preference for either mechanism where infection took place on a cell grid, a given proportion of which were selected to be susceptible \cite{saeki_and_sasaki_spatial_heterogeneity}. The authors showed that cell--to--cell spread tended to be selected for the more that cells were clustered together \cite{saeki_and_sasaki_spatial_heterogeneity}. Conversely, Kreger \emph{et al.} considered the evolution of HIV under varying proportions of infections arising from the cell--to--cell route \cite{kreger_et_al_synaptic_cell_to_cell}. The authors found that a cell--to--cell spread was conducive to the prevalence of disadvantageous mutants, which may facilitate the pre--existence of drug--resistant variants within the host \cite{kreger_et_al_synaptic_cell_to_cell}.

	\section[Future development for modelling]{Future development for modelling of spatial dynamics of viral infections}
	\sectionmark{Future development for modelling}
	\label{sec:future}
	
	Spatial structure has come to be recognised as an important factor in the dynamics of viral infections. Here, we have conducted a scoping review to show the huge diversity in approaches to modelling spatial aspects of viral infections. The wide range of methods and approaches used mirrors the significant diversity in modelling requirements between applications, including particularities of viral species or \added{host cell populations}, for example. We have shown that spatial models offer a description of a range of biological phenomena with an intrinsically spatial mode of action, such as cell--to--cell transmission, where an accurate description is not possible using traditional ODE modelling. Spatial models have moreover been shown to demonstrate substantial differences in parameter estimates compared to non--spatial approaches, suggesting that this approach to modelling may be necessary to improve quantitative biological insights as well. However, there remain important challenges to be addressed if spatially--structured models are to become more widely used in biological applications.

	\subsection{Robust implementation}
	\label{sec:future_implementation}
	
	One of the key observations from our scoping review is the presence of an incredibly wide array of approaches to representing spatial structure in viral infections. While it is natural --- indeed, essential --- that there is debate among the modelling community over which components and interactions should be captured in a model, this lack of consensus is particularly prevalent among spatial models and at a much more fundamental level. While traditional, non--spatial models for viral dynamics more or less share the common language of ODEs, there are fundamental differences between, say, PDE models of viral dynamics and cellular automata. Our analysis has shown that this diversity in approaches is not a shortcoming of the field, but a necessity in order to describe the highly varied spatial dynamics of different viruses at different scales. However, future models should be developed with a clearer sense of the scope and limitation of their approach. PDE models, for instance, provide a good macroscopic description of tissue, yet due to a lack of discreteness or stochasticity, should not be relied upon to account for rare events such as extinction or long--range viral dispersal. On the other hand, cellular automata (perhaps with the exception of the whole--lung model by Moses \emph{et al.} \cite{moses_et_al_lung_covid_model}) should be considered microscopic models only of the small patch of cells which they comprise and --- depending on the structure and arrangement of the overall tissue --- not necessarily a description of dynamics on the macroscopic scale.
	
	We also showed that a substantial challenge in using agent--based models in particular is a dependence of the dynamics on the exact implementation of the model. As we discussed in Section~\ref{sec:papers_by_methodology}, \textbf{Wodarz \emph{et al.}} \cite{wodarz_et_al_complex_spatial_dynamics_oncolytic} and \textbf{Beauchemin \emph{et al.}} \cite{beauchemin_et_al_simple_cellular_automaton_model} have both shown that stochastic properties of cellular automaton models are influenced by the size of the grid. \added{In another study, Williams \emph{et al.} studied multicellular models in which the host tissue is discretised into cells and the virus population is represented by a density surface, and showed that for rates of viral diffusion akin to standard assumptions in the literature, the dynamics of the system was dependent on the resolution of spatial discretisation of the viral surface \cite{williams_et_al_spatial_discretisation}.} Berg \emph{et al.} \cite{berg_et_al_in_vitro_in_silico_multidimensional_oncolytic} and Moses \emph{et al.} \cite{moses_et_al_lung_covid_model} have also demonstrated that the dynamics of their models varied significantly depending on whether two or three spatial dimensions were considered. \added{Not all of these findings are necessarily limitations of the agent--based model framework}: the dependence of model dynamics on dimensionality, for example, is an important biological finding in its own right. This implementation--dependence clearly highlights the importance of biological fidelity in model construction, and moreover points to a need for careful study of various elements of the construction of agent--based models and how they might influence the dynamics. Key questions, such as the spatial arrangement of cells, the size and dimensionality of the spatial domain, or the \added{numerical simulation of complex, multi--scale systems}, have so far been largely unexplored in viral dynamical models, yet are crucial for ensuring robust implementation of agent--based models.
	
	\added{Another emerging trend in spatially--structured modelling is the growing importance of computational implementation. Recent model development, especially of large, complex systems, has increasingly shifted modelling language away from the mathematical and towards a software--oriented approach. In particular, we noted in Section~\ref{sec:agent_based} the increasing popularity of large viral dynamics ``simulators'', the development of which is usually facilitated by existing software packages for tissue dynamics such as Chaste \cite{pitt_francis_et_al_Chaste, mirams_et_al_chaste}, PhysiCell \cite{ghaffarizadeh_et_al_physicell}, Morpheus \cite{staruss_et_al_morpheus} and others. This direction in model development, on the one hand, is an important step towards modularity and submodel integration (illustrated especially in the work of Getz \emph{et al.} \cite{getz_rapid_community_driven_covid_model, islam_et_al_agent_based_lung_fibrosis}), but also poses challenges in model interpretability and reproducibility, including between software platforms. One aspect of this is because different software packages incorporate different dynamics which are ``baked in'' to the base tissue dynamics model, such as the Cellular Potts Model in CompuCell3D \cite{sego_et_al_covid_model, sego_et_al_multicellular_influenza, sego_et_al_cellularisation, aponte_serrano_RNA_virus_replication, gianlupi_et_al_multiscale_model_of_antiviral_timing} or the point force cell mechanics model of PhysiCell \cite{getz_rapid_community_driven_covid_model, islam_et_al_agent_based_lung_fibrosis, ghaffarizadeh_et_al_physicell}. This complicates comparison of models between software platforms, even if the assumed virus--host interactions are the same. In other fields of mathematical biology (systems biology in particular), efforts have increasingly been made to standardise and formalise modelling language through initiatives like Systems Biology Markup Language (SBML) \cite{hucka_et_al_SBML}, CellML \cite{cuellar_et_al_CellML} and BioNetGen \cite{harris_et_al_bionetgen}. While none of these approaches apply easily to spatially--structured systems (especially agent--based systems), as models grow increasingly complex and software--dependent, future development in standardisation approaches would aid in the communication and reproducibility of spatially--structured models of viral dynamics and of biological systems more generally.}

	\subsection{Application to experimental data}
	\label{sec:future_data}
	
	Perhaps the greatest obstacle to the wider usage of spatially--structured models of viral dynamics is the difficulty in calibrating such models to experimental data. Spatial models are inherently more complex and computationally expensive than their non--spatial counterparts and therefore both obtaining sufficient data \added{as well as} carrying out rigorous parameter estimation presents a significant challenge \cite{graw_and_perelson_spatial_hiv, graw_and_perelson_modeling_viral_spread, bauer_et_al_agent_based_models_virus, gallagher_spatial_spread}. \added{Perhaps as a consequence, in the vast majority of spatially--structured models, authors} rely on drawing at least some parameter estimates from elsewhere in the literature. However, the exact meaning of a particular parameter is likely to vary somewhat between models depending on what components and interactions are included. For example, parameter estimations between a model with an explicit eclipse phase and a model without one predict the same time interval between infection generations, but will predict different life spans for infected cells due to the latently infected period \cite{baccam_et_al_influenza_kinetics}. This parameter unreliability is especially notable when parameter estimates for spatial models are obtained using ODE models. We have already discussed examples of spatial and non--spatial models generating wildly different parameter values for the same overall dynamics (Strain \emph{et al.} \cite{strain_et_al_spatiotemporal_hiv}, \textbf{Mitchell \emph{et al.}} \cite{mitchell_et_al_high_replication_efficiency_H1N1}). As such, parameter estimates drawn from external sources cannot necessarily be treated as reliable for a given alternative application.
	
	One parameter that is almost always cited from other literature is the diffusion coefficient of extracellular virus. In many cases, this coefficient describes the rate of the only spatially--explicit phenomenon of the model (\textbf{Mitchell \emph{et al.}} \cite{mitchell_et_al_high_replication_efficiency_H1N1}, Beauchemin \emph{et al.} \cite{beauchemin_et_al_modeling_influenza_in_tissue}, Marzban \emph{et al.} \cite{marzban_et_al_hybrid_PDE_ABM_model}, for example), and therefore is by far the most difficult parameter to estimate. The viral diffusion coefficient is usually estimated using the Stokes--Einstein relation (Strain \emph{et al.} \cite{strain_et_al_spatiotemporal_hiv}, Beauchemin \emph{et al.} \cite{beauchemin_and_handel_influenza_review}, \textbf{Sego \emph{et al.}} \cite{sego_et_al_covid_model}, for example). For influenza viruses in plasma at body temperature, the diffusion coefficient has been calculated at approximately $3.18\times10^{-12}\text{m}^2\text{s}^{-1}$ \cite{beauchemin_et_al_modeling_influenza_in_tissue}. However, critically, this value is usually \emph{not} the one actually used in simulations. Instead, authors generally use a value some orders of magnitude smaller. There are varying justifications for this. Akpinar \emph{et al.} \cite{akpinar_et_al_spatiotemporal_viral_amplification} used a smaller diffusion coefficient to obtain qualitative agreement with experimentally--observed spatial features; Beauchemin \emph{et al.} \cite{beauchemin_et_al_modeling_influenza_in_tissue} concluded that a smaller diffusion coefficient provided a better match for the autocorrelation seen in experiments. \textbf{Sego \emph{et al.}} argued that, compared to diffusion in water, virions were likely to encounter significant additional drag in mucus and therefore suggested that the diffusion coefficient could be as much as six orders of magnitude smaller \cite{sego_et_al_covid_model}. Alternatively, You and Yin \cite{you_et_al_amplification_and_spread} and Yin and McCaskill \cite{yin_and_mccaskill_replication_of_viruses_in_a_growing_plaque} have suggested that diffusion is limited by adsorption to cells as virions spread in space. There is very little agreement among the literature as to what the true value of the diffusion coefficient should be under different conditions, yet this value is crucial. Holder \emph{et al.} \cite{holder_et_al_design_considerations_influenza} and Beauchemin \cite{beauchemin_well_mixed_assumption} have both demonstrated that the dynamics of spatially--structured viral infections hinge on the value of the viral diffusion coefficient. Williams \emph{et al.} showed that, in a multicellular model, the relative contributions of cell--to--cell and cell--free infection could only be distinguished for sufficiently large extracellular viral diffusion \cite{williams_et_al_inference}. Clearly, there is a need to quantify viral diffusion coefficients with increased accuracy. Fortunately, recent advances in imaging technology and analysis have hinted that more accurate quantification might soon be possible. \textbf{Imle \emph{et al.}} used image tracking to empirically quantify the mean squared displacement and thus diffusion coefficient of HIV virions in different media \cite{imle_et_al_hiv_3d_collagen_matrix}. The authors determined that the diffusion coefficient in suspension provided reasonable agreement with the Stokes--Einstein estimate, but spread was notably hampered for virions in a collagen matrix \cite{imle_et_al_hiv_3d_collagen_matrix}. \added{At least qualitatively, this confirms the assumption of slowed viral diffusion \emph{in vivo}, but moreover hints at the limited applicability of \emph{in vitro} viral spread in a within--host context.}
	
	\added{The mismatch between \emph{in vitro} observations and viral dynamics \emph{in vivo} is a broader challenge in the development of biologically relevant spatially--structured models. Data describing the spatial spread of viral plaques \emph{in vitro} has been readily available for decades \cite{koch_growing_viral_plaques}, and has been successfully used to fit a number of spatially--structured models (Holder \emph{et al.} \cite{holder_et_al_assessing_the_in_vitro_fitness}, Haseltine \emph{et al.} \cite{haseltine_et_al_image_guided_growth}, \textbf{Imle \emph{et al.}} \cite{imle_et_al_hiv_3d_collagen_matrix}, for example). However, as we mentioned above, the lack of realistic viral diffusion in the \emph{in vitro} context, as well as a simplified tissue structure and the lack of an immune response, means that such models may be of limited relevance to \emph{in vivo} dynamics. While spatially--structured data from infections \emph{in vivo} remains scarce, emerging genetic engineering and imaging technologies suggest possible new sources of information about how infections spread within living hosts. Several groups have engineered reporter viruses expressing fluorescent proteins, which can be detected by bioluminescence imaging in animal hosts, permitting the construction of precise maps of viral distribution \emph{in vivo} at various times over the course of infection \cite{tran_et_al_in_vivo_reporter_influenza, nogales_et_al_flourescent_and_bioluminescent_influenza_reporter, ye_et_al_covid_reporter_virus_in_mice}. Other groups have developed reporter mice which express fluorescent proteins as part of, for example, the innate immune response \cite{lienenklaus_et_al_ifn_beta_mouse, kumagai_et_al_alveolar_macrophages_ifn_alpha}. This likewise permits imaging of interferon distribution, for example, during viral infection. Other techniques, such as FDG-PET imaging, permit the visualisation of the extent of infection within living hosts (including human hosts) without the need for an engineered reporter virus, albeit at a lower resolution \cite{jonsson_et_al_molecular_imaging_ferrets_influenza, bray_et_al_molecular_imaging_influenza}. While these technologies provide useful descriptions of how the infection distribution evolves within the host, currently, data is usually available only at the macroscopic scale. In limited cases, some spatially--structured data can be gathered at the cell scale, but requires removing samples of infected tissue, and thus sacrificing the host \cite{manicassamy_et_al_gfp_reporter_influenza, fukuyama_et_al_color_flu}. As a consequence, it is not possible to collect multiple such measurements over the time course of a single infection. Nonetheless, such data may still be valuable in calibrating realistic spatially--structured models of viral infections \emph{in vivo}. One promising publication by Fukuyama and colleagues studied mice co-infected with four engineered influenza strains, each expressing differently--coloured fluorescent proteins, and examined samples of lung tissue taken at different stages in the infection \cite{fukuyama_et_al_color_flu}. Imaging showed not only temporal progression of the infection from the upper to the lower airways, but showed local clustering of infected cells with the same fluorescent tag, suggesting competition between local and distal infection mechanisms \cite{fukuyama_et_al_color_flu}. Emerging techniques such as these can therefore offer rich sources of spatiotemporal information about the spread of viral infection within the host, which will further aid in the development and calibration of biologically realistic models.}
			
	\added{A more practical obstacle in the calibration of spatially--structured models to data is in ensuring computational tractability of parameter fitting methods. Spatially--structured models are typically computationally expensive to simulate, which necessitates sophisticated, highly efficient parameter estimation methods --- often coupled with high--performance computing --- in order to conduct model calibration in feasible times \cite{bauer_et_al_agent_based_models_virus, alamoudi_et_al_fitmulticell}. Computational demands are even greater for stochastic models, for which it may be more difficult to determine the quality of fit for a given set of parameter values \cite{toni_et_al_PMC_ABC, kypraios_et_al_tutorial_intro_bayesian}. There are now a range of inference methods and software tools available for these parameter estimation tasks, including Bayesian and Approximate Bayesian Computation (ABC) methods which quantify parameter uncertainty (Stan \cite{stan_package}, pyABC \cite{klinger_et_al_pyABC}, ABCpy \cite{dutta_et_al_ABCpy}, for example). However, a barrier to wider uptake of this approach is a need for specific technical expertise in implementing and interpreting such methods, as well as a lack of adaptability in existing implementations \cite{alamoudi_et_al_fitmulticell}. However, just as there have been efforts to unify modelling languages under standards like SBML, several authors have also recently sought to introduce standards for parameter inference problems. PEtab, in particular, is a popular such format for parameter estimation on SBML models \cite{schmeister_et_al_PEtab}. Promisingly, efforts have also recently been made to introduce standard workflows for parameter inference on spatially--structured models in particular, such as Spatial Model Editor \cite{keegan_et_al_spatial_model_editor} and FitMultiCell \cite{alamoudi_et_al_fitmulticell}. Greater accessibility and interpretability of these advanced parameter estimation techniques will lead to better integration of experimental data, even with complex spatial models, as well as aid in reproducibility of results.}

	\section{Conclusion}
	\label{sec:conclusion}
	
	This review has demonstrated the broad suite of spatially--structured models of viral dynamics available, and discussed the important contributions they have made to our understanding of viral infections. Advancements in the field point to a important role for spatial structure in viral dynamics, and have offered insights into key mechanisms of virus--host interactions in a spatial context, across a wide range of virus and host types. Nonetheless, there are improvements still to be made. Key questions remain as to how the spatial spread of infections should be modelled in different contexts, and how those models should be implemented. This review highlights work still to be done in better coupling modelling work with experimental data, and in constructing models with application--appropriate implementation. Progress in this regard will result in robust, biologically realistic models, capable of offering important insights into viral illnesses and their treatment.

\section*{\added{Acknowledgements}}
	\added{TW's research is supported by an Australian Government Research Training Program (RTP) scholarship. JMM's research is support by the ARC (DP210101920). JMO's research is supported by the ARC (DP230100380, FT230100352). The funders had no role in study design, data collection and analysis, decision to publish, or preparation of the manuscript. The images in Figure \ref{fig:sources_of_spatial_structure} were created in BioRender. Osborne, J. (2025) \url{https://BioRender.com/t2kwkm3}.}

    \clearpage
    \bibliographystyle{unsrtnat}

{\footnotesize
        \begin{thebibliography}{245}
        	\providecommand{\natexlab}[1]{#1}
        	\providecommand{\url}[1]{\texttt{#1}}
        	\expandafter\ifx\csname urlstyle\endcsname\relax
        	\providecommand{\doi}[1]{doi: #1}\else
        	\providecommand{\doi}{doi: \begingroup \urlstyle{rm}\Url}\fi
        	
        	\bibitem[Kumberger et~al.(2018)Kumberger, Durso-Cain, Uprichard, Dahari, and
        	Graw]{kumberger_et_al_accounting_for_space_cell_to_cell}
        	Peter Kumberger, Karina Durso-Cain, Susan~L Uprichard, Harel Dahari, and
        	Frederik Graw.
        	\newblock Accounting for space---quantification of cell-to-cell transmission
        	kinetics using virus dynamics models.
        	\newblock \emph{Viruses}, 10\penalty0 (4):\penalty0 200, 2018.
        	\newblock \doi{10.3390/v10040200}.
        	
        	\bibitem[Imle et~al.(2019)Imle, Kumberger, Schnellb{\"a}cher, Fehr,
        	Carrillo-Bustamante, Ales, Schmidt, Ritter, Godinez, M{\"u}ller, Rohr,
        	Hamprecht, Schwarz, Graw, and Fackler]{imle_et_al_hiv_3d_collagen_matrix}
        	Andrea Imle, Peter Kumberger, Nikolas~D. Schnellb{\"a}cher, Jana Fehr, Paola
        	Carrillo-Bustamante, Janez Ales, Philip Schmidt, Christian Ritter, William~J.
        	Godinez, Barbara M{\"u}ller, Karl Rohr, Fred~A. Hamprecht, Ulrich~S. Schwarz,
        	Frederik Graw, and Oliver~T. Fackler.
        	\newblock Experimental and computational analyses reveal that environmental
        	restrictions shape {HIV}-1 spread in {3D} cultures.
        	\newblock \emph{Nature Communications}, 10\penalty0 (1):\penalty0 2144, 2019.
        	\newblock \doi{10.1038/s41467-019-09879-3}.
        	
        	\bibitem[Saeki and Sasaki(2018)]{saeki_and_sasaki_spatial_heterogeneity}
        	Koich Saeki and Akira Sasaki.
        	\newblock The role of spatial heterogeneity in the evolution of local and
        	global infections of viruses.
        	\newblock \emph{PLOS Computational Biology}, 14\penalty0 (1):\penalty0 1--20,
        	2018.
        	\newblock \doi{10.1371/journal.pcbi.1005952}.
        	
        	\bibitem[Kreger et~al.(2020)Kreger, Komarova, and
        	Wodarz]{kreger_et_al_synaptic_cell_to_cell}
        	Jesse Kreger, Natalia~L Komarova, and Dominik Wodarz.
        	\newblock Effect of synaptic cell-to-cell transmission and recombination on the
        	evolution of double mutants in {HIV}.
        	\newblock \emph{Journal of The Royal Society Interface}, 17\penalty0
        	(164):\penalty0 20190832, 2020.
        	\newblock \doi{10.1098/rsif.2019.0832}.
        	
        	\bibitem[Yakimovich et~al.(2016)Yakimovich, Yakimovich, Schmid, Mercer,
        	Sbalzarini, and Greber]{yakimovich_et_al_infectio}
        	Artur Yakimovich, Yauhen Yakimovich, Michael Schmid, Jason Mercer, Ivo~F
        	Sbalzarini, and Urs~F Greber.
        	\newblock Infectio: a generic framework for computational simulation of virus
        	transmission between cells.
        	\newblock \emph{mSphere}, 1\penalty0 (1), 2016.
        	\newblock \doi{10.1128/mSphere.00078-15}.
        	
        	\bibitem[Levin et~al.(2016)Levin, Forrest, Banerjee, Clay, Cannon, Moses, and
        	Koster]{levin_et_al_T_cell_search_influenza}
        	Drew Levin, Stephanie Forrest, Soumya Banerjee, Candice Clay, Judy Cannon,
        	Melanie Moses, and Frederick Koster.
        	\newblock A spatial model of the efficiency of {T} cell search in the
        	influenza-infected lung.
        	\newblock \emph{Journal of Theoretical Biology}, 398:\penalty0 52--63, 2016.
        	\newblock \doi{10.1016/j.jtbi.2016.02.022}.
        	
        	\bibitem[Kadolsky and
        	Yates(2015)]{kadolsky_and_yates_spatial_immune_surveillance}
        	Ulrich~D Kadolsky and Andrew~J Yates.
        	\newblock How is the effectiveness of immune surveillance impacted by the
        	spatial distribution of spreading infections?
        	\newblock \emph{Philosophical Transactions of the Royal Society B: Biological
        		Science}, 370\penalty0 (1675), 2015.
        	\newblock \doi{10.1098/rstb.2014.0289}.
        	
        	\bibitem[Sego et~al.(2020)Sego, Aponte-Serrano, Gianlupi, Heaps, Breithaupt,
        	Brusch, Crawshaw, Osborne, Quardokus, Plemper, and
        	Glazier]{sego_et_al_covid_model}
        	T.~J. Sego, Josua~O. Aponte-Serrano, Juliano~Ferrari Gianlupi, Samuel~R. Heaps,
        	Kira Breithaupt, Lutz Brusch, Jessica Crawshaw, James~M. Osborne, Ellen~M.
        	Quardokus, Richard~K. Plemper, and James~A. Glazier.
        	\newblock A modular framework for multiscale, multicellular, spatiotemporal
        	modeling of acute primary viral infection and immune response in epithelial
        	tissues and its application to drug therapy timing and effectiveness.
        	\newblock \emph{PLOS Computational Biology}, 16\penalty0 (12):\penalty0
        	e1008451, 2020.
        	\newblock \doi{10.1371/journal.pcbi.1008451}.
        	
        	\bibitem[Getz et~al.(2020)Getz, Wang, An, Becker, Cockrell, Collier, Craig,
        	Davis, Faeder, Versypt, Gianlupi, Glazier, Hamis, Heiland, Hillen, Hou,
        	Islam, Jenner, Kurtoglu, Liu, Macfarlane, Maygrundter, Morel, Narayanan,
        	Ozik, Pienaar, Rangamani, Shoemaker, Smith, and
        	Macklin]{getz_rapid_community_driven_covid_model}
        	Michael Getz, Yafei Wang, Gary An, Andrew Becker, Chase Cockrell, Nicholson
        	Collier, Morgan Craig, Courtney~L. Davis, James Faeder, Ashlee N.~Ford
        	Versypt, Juliano~F. Gianlupi, James~A. Glazier, Sara Hamis, Randy Heiland,
        	Thomas Hillen, Dennis Hou, Mohammad~Aminul Islam, Adrianne Jenner, Furkan
        	Kurtoglu, Bing Liu, Fiona Macfarlane, Pablo Maygrundter, Penelope~A Morel,
        	Aarthi Narayanan, Jonathan Ozik, Elsje Pienaar, Padmini Rangamani,
        	Jason~Edward Shoemaker, Amber~M. Smith, and Paul Macklin.
        	\newblock Iterative community-driven development of a {SARS-CoV-2} tissue
        	simulator [pre-print].
        	\newblock \emph{bioRxiv}, page 2020.04.02.019075, 2020.
        	\newblock \doi{10.1101/2020.04.02.019075}.
        	
        	\bibitem[Bauer et~al.(2009)Bauer, Beauchemin, and
        	Perelson]{bauer_et_al_agent_based_models_virus}
        	Amy~L. Bauer, Catherine A.~A. Beauchemin, and Alan~S. Perelson.
        	\newblock Agent-based modeling of host--pathogen systems: The successes and
        	challenges.
        	\newblock \emph{Information Sciences (New York)}, 179\penalty0 (10):\penalty0
        	1379--1389, 2009.
        	
        	\bibitem[Graw and Perelson(2013)]{graw_and_perelson_spatial_hiv}
        	Frederik Graw and Alan~S. Perelson.
        	\newblock Spatial aspects of {HIV} infection.
        	\newblock 2013.
        	
        	\bibitem[Strain et~al.(2002)Strain, Richman, Wong, and
        	Levine]{strain_et_al_spatiotemporal_hiv}
        	M.~C. Strain, D.~D. Richman, J.~K. Wong, and H.~Levine.
        	\newblock Spatiotemporal dynamics of {HIV} propagation.
        	\newblock \emph{Journal of Theoretical Biology}, 218\penalty0 (1):\penalty0
        	85--96, 2002.
        	\newblock \doi{https://doi.org/10.1006/jtbi.2002.3055}.
        	
        	\bibitem[Graw and Perelson(2016)]{graw_and_perelson_modeling_viral_spread}
        	Frederik Graw and Alan~S Perelson.
        	\newblock Modeling viral spread.
        	\newblock \emph{Annual Reviews in Virology}, 3\penalty0 (1):\penalty0 555--572,
        	2016.
        	\newblock \doi{10.1146/annurev-virology-110615-042249}.
        	
        	\bibitem[{Bocharov, G.} et~al.(2012){Bocharov, G.}, {Chereshnev, V.}, {Gainova,
        		I.}, {Bazhan, S.}, {Bachmetyev, B.}, {Argilaguet, J.}, {Martinez, J.}, and
        	{Meyerhans, A.}]{bocharov_review}
        	{Bocharov, G.}, {Chereshnev, V.}, {Gainova, I.}, {Bazhan, S.}, {Bachmetyev,
        		B.}, {Argilaguet, J.}, {Martinez, J.}, and {Meyerhans, A.}
        	\newblock Human immunodeficiency virus infection: from biological observations
        	to mechanistic mathematical modelling.
        	\newblock \emph{Mathematical Modelling of Natural Phenomena}, 7\penalty0
        	(5):\penalty0 78--104, 2012.
        	\newblock \doi{10.1051/mmnp/20127507}.
        	
        	\bibitem[Gallagher et~al.(2018)Gallagher, Brooke, Ke, and
        	Koelle]{gallagher_spatial_spread}
        	Molly~E Gallagher, Christopher~B Brooke, Ruian Ke, and Katia Koelle.
        	\newblock Causes and consequences of spatial within-host viral spread.
        	\newblock \emph{Viruses}, 10\penalty0 (11):\penalty0 627, 2018.
        	\newblock \doi{10.3390/v10110627}.
        	
        	\bibitem[Ghaemi et~al.(2023)Ghaemi, Nafiu, Tajkhorshid, Gruebele, and
        	Hu]{ghaemi_et_al_whole_cell_model_hbv}
        	Zhaleh Ghaemi, Oluwadara Nafiu, Emad Tajkhorshid, Martin Gruebele, and Jianming
        	Hu.
        	\newblock A computational spatial whole-cell model for hepatitis {B} viral
        	infection and drug interactions.
        	\newblock \emph{Scientific Reports}, 13\penalty0 (1):\penalty0 21392, 2023.
        	\newblock \doi{10.1038/s41598-023-45998-0}.
        	
        	\bibitem[Nakabayashi and Sasaki(2011)]{nakabayashi_et_al_hbv_in_cell}
        	Jun Nakabayashi and Akira Sasaki.
        	\newblock A mathematical model of the intracellular replication and within host
        	evolution of hepatitis type {B} virus: Understanding the long time course of
        	chronic hepatitis.
        	\newblock \emph{Journal of Theoretical Biology}, 269\penalty0 (1):\penalty0
        	318--329, 2011.
        	\newblock \doi{https://doi.org/10.1016/j.jtbi.2010.10.024}.
        	
        	\bibitem[Knodel et~al.(2019)Knodel, Targett-Adams, Grillo, Herrmann, and
        	Wittum]{knodel_et_al_surface_PDEs_HCV}
        	Markus~M Knodel, Paul Targett-Adams, Alfio Grillo, Eva Herrmann, and Gabriel
        	Wittum.
        	\newblock Advanced hepatitis {C} virus replication {PDE} models within a
        	realistic intracellular geometric environment.
        	\newblock \emph{International Journal of Environmental Research and Public
        		Health}, 16\penalty0 (3), 2019.
        	\newblock \doi{10.3390/ijerph16030513}.
        	
        	\bibitem[Celada and Seiden(1992)]{celada_and_seiden_comp_immune_model}
        	Franco Celada and Philip~E Seiden.
        	\newblock A computer model of cellular interactions in the immune system.
        	\newblock \emph{Immunology Today}, 13\penalty0 (2):\penalty0 56--62, 1992.
        	\newblock \doi{https://doi.org/10.1016/0167-5699(92)90135-T}.
        	
        	\bibitem[Grebennikov et~al.(2019)Grebennikov, Bouchnita, Volpert, Bessonov,
        	Meyerhans, and Bocharov]{grebennikov_et_al_spatial_lymphocytes}
        	Dmitry Grebennikov, Anass Bouchnita, Vitaly Volpert, Nikolay Bessonov, Andreas
        	Meyerhans, and Gennady Bocharov.
        	\newblock Spatial lymphocyte dynamics in lymph nodes predicts the cytotoxic {T}
        	cell frequency needed for {HIV} infection control.
        	\newblock \emph{Frontiers in Immunology}, 10:\penalty0 1213, 2019.
        	\newblock \doi{10.3389/fimmu.2019.01213}.
        	
        	\bibitem[Pinotti et~al.(2020)Pinotti, Ghanbarnejad, H{\"o}vel, and
        	Poletto]{pinotti_et_al_three_player_antigen}
        	Francesco Pinotti, Fakhteh Ghanbarnejad, Philipp H{\"o}vel, and Chiara Poletto.
        	\newblock Interplay between competitive and cooperative interactions in a
        	three-player pathogen system.
        	\newblock \emph{Royal Society Open Science}, 7\penalty0 (1):\penalty0 190305,
        	2020.
        	\newblock \doi{10.1098/rsos.190305}.
        	
        	\bibitem[Koch(1964)]{koch_growing_viral_plaques}
        	Arthur~L. Koch.
        	\newblock The growth of viral plaques during the enlargement phase.
        	\newblock \emph{Journal of Theoretical Biology}, 6\penalty0 (3):\penalty0
        	413--431, 1964.
        	\newblock \doi{https://doi.org/10.1016/0022-5193(64)90056-6}.
        	
        	\bibitem[Pandey(1989)]{pandey_1989_CA_model_HIV}
        	R.~B. Pandey.
        	\newblock Computer simulation of a cellular automata model for the immune
        	response in a retrovirus system.
        	\newblock \emph{Journal of Statistical Physics}, 54\penalty0 (3):\penalty0
        	997--1010, 1989.
        	\newblock \doi{10.1007/BF01019785}.
        	
        	\bibitem[Bernaschi and
        	Castiglione(2001)]{bernaschi_and_castiglione_design_and_implementation}
        	M~Bernaschi and F~Castiglione.
        	\newblock Design and implementation of an immune system simulator.
        	\newblock \emph{Computers in Biology and Medicine}, 31\penalty0 (5):\penalty0
        	303--331, 2001.
        	\newblock \doi{10.1016/s0010-4825(01)00011-7}.
        	
        	\bibitem[Zorzenon~dos Santos and
        	Coutinho(2001)]{dos_santos_and_countinho_dynamics_of_HIV_CA}
        	R~M Zorzenon~dos Santos and S~Coutinho.
        	\newblock Dynamics of {HIV} infection: a cellular automata approach.
        	\newblock \emph{Physical Review Letters}, 87\penalty0 (16):\penalty0 168102,
        	2001.
        	\newblock \doi{10.1103/PhysRevLett.87.168102}.
        	
        	\bibitem[Wu et~al.(2001)Wu, Byrne, Kirn, and
        	Wein]{wu_et_al_modelling_and_analysis_oncolytic}
        	J~T Wu, H~M Byrne, D~H Kirn, and L~M Wein.
        	\newblock Modeling and analysis of a virus that replicates selectively in tumor
        	cells.
        	\newblock \emph{Bulletin of Mathematical Biology}, 63\penalty0 (4):\penalty0
        	731--768, 2001.
        	\newblock \doi{10.1006/bulm.2001.0245}.
        	
        	\bibitem[Wein et~al.(2003)Wein, Wu, and
        	Kirn]{wein_et_al_validation_and_analysis_oncolytic}
        	Lawrence~M. Wein, Joseph~T. Wu, and David~H. Kirn.
        	\newblock Validation and analysis of a mathematical model of a
        	replication-competent oncolytic virus for cancer treatment: Implications for
        	virus design and delivery.
        	\newblock \emph{Cancer Research}, 63\penalty0 (6):\penalty0 1317--1324, 2003.
        	
        	\bibitem[Funk et~al.(2005)Funk, Jansen, Bonhoeffer, and
        	Killingback]{funk_et_al_spatial_models_virus_immune}
        	Georg~A. Funk, Vincent~A.A. Jansen, Sebastian Bonhoeffer, and Timothy
        	Killingback.
        	\newblock Spatial models of virus-immune dynamics.
        	\newblock \emph{Journal of Theoretical Biology}, 233\penalty0 (2):\penalty0
        	221--236, 2005.
        	\newblock \doi{10.1016/j.jtbi.2004.10.004}.
        	
        	\bibitem[Beauchemin et~al.(2005)Beauchemin, Samuel, and
        	Tuszynski]{beauchemin_et_al_simple_cellular_automaton_model}
        	Catherine Beauchemin, John Samuel, and Jack Tuszynski.
        	\newblock A simple cellular automaton model for influenza {A} viral infections.
        	\newblock \emph{Journal of Theoretical Biology}, 232\penalty0 (2):\penalty0
        	223--234, 2005.
        	\newblock \doi{10.1016/j.jtbi.2004.08.001}.
        	
        	\bibitem[Friedman et~al.(2006)Friedman, Tian, Fulci, Chiocca, and
        	Wang]{friedman_et_al_glioma_virotherapy}
        	Avner Friedman, Jianjun~Paul Tian, Giulia Fulci, E~Antonio Chiocca, and Jin
        	Wang.
        	\newblock Glioma virotherapy: effects of innate immune suppression and
        	increased viral replication capacity.
        	\newblock \emph{Cancer Research}, 66\penalty0 (4):\penalty0 2314--2319, 2006.
        	\newblock \doi{10.1158/0008-5472.CAN-05-2661}.
        	
        	\bibitem[Xiao et~al.(2006)Xiao, Shao, and
        	Chou]{xiao_et_al_cellular_automaton_HBV}
        	Xuan Xiao, Shi-Huang Shao, and Kuo-Chen Chou.
        	\newblock A probability cellular automaton model for hepatitis {B} viral
        	infections.
        	\newblock \emph{Biochemical and Biophysical Research Communications},
        	342\penalty0 (2):\penalty0 605--610, 2006.
        	\newblock \doi{10.1016/j.bbrc.2006.01.166}.
        	
        	\bibitem[Mok et~al.(2009)Mok, Stylianopoulos, Boucher, and
        	Jain]{mok_et_al_herpes_simplex_in_solid_tumour}
        	Wilson Mok, Triantafyllos Stylianopoulos, Yves Boucher, and Rakesh~K Jain.
        	\newblock Mathematical modeling of herpes simplex virus distribution in solid
        	tumors: implications for cancer gene therapy.
        	\newblock \emph{Clinical Cancer Research}, 15\penalty0 (7):\penalty0
        	2352--2360, 2009.
        	\newblock \doi{10.1158/1078-0432.CCR-08-2082}.
        	
        	\bibitem[Heilmann et~al.(2010)Heilmann, Sneppen, and
        	Krishna]{heilmann_et_al_2010_sustainability_of_virulence}
        	Silja Heilmann, Kim Sneppen, and Sandeep Krishna.
        	\newblock Sustainability of virulence in a phage-bacterial ecosystem.
        	\newblock \emph{Journal of Virology}, 84\penalty0 (6):\penalty0 3016--3022,
        	2010.
        	\newblock \doi{10.1128/jvi.02326-09}.
        	
        	\bibitem[Mitchell et~al.(2011)Mitchell, Levin, Forrest, Beauchemin, Tipper,
        	Knight, Donart, Layton, Pyles, Gao, Harrod, Perelson, and
        	Koster]{mitchell_et_al_high_replication_efficiency_H1N1}
        	Hugh Mitchell, Drew Levin, Stephanie Forrest, Catherine A~A Beauchemin,
        	Jennifer Tipper, Jennifer Knight, Nathaniel Donart, R~Colby Layton, John
        	Pyles, Peng Gao, Kevin~S Harrod, Alan~S Perelson, and Frederick Koster.
        	\newblock Higher level of replication efficiency of 2009 {(H1N1)} pandemic
        	influenza virus than those of seasonal and avian strains: kinetics from
        	epithelial cell culture and computational modeling.
        	\newblock \emph{Journal of Virology}, 85\penalty0 (2):\penalty0 1125--1135,
        	2011.
        	\newblock \doi{10.1128/JVI.01722-10}.
        	
        	\bibitem[Wodarz et~al.(2012)Wodarz, Hofacre, Lau, Sun, Fan, and
        	Komarova]{wodarz_et_al_complex_spatial_dynamics_oncolytic}
        	Dominik Wodarz, Andrew Hofacre, John~W. Lau, Zhiying Sun, Hung Fan, and
        	Natalia~L. Komarova.
        	\newblock Complex spatial dynamics of oncolytic viruses in vitro: Mathematical
        	and experimental approaches.
        	\newblock \emph{PLOS Computational Biology}, 8\penalty0 (6):\penalty0
        	e1002547--, 2012.
        	\newblock \doi{10.1371/journal.pcbi.1002547}.
        	
        	\bibitem[Heilmann et~al.(2012)Heilmann, Sneppen, and
        	Krishna]{heilmann_et_al_2012_coexistence_phage_and_bacteria}
        	Silja Heilmann, Kim Sneppen, and Sandeep Krishna.
        	\newblock Coexistence of phage and bacteria on the boundary of self-organized
        	refuges.
        	\newblock \emph{Proceedings of the National Academy of Sciences}, 109\penalty0
        	(31):\penalty0 12828--12833, 2012.
        	\newblock \doi{10.1073/pnas.1200771109}.
        	
        	\bibitem[Schiffer et~al.(2013{\natexlab{a}})Schiffer, Swan, Al~Sallaq, Magaret,
        	Johnston, Mark, Selke, Ocbamichael, Kuntz, Zhu, Robinson, Huang, Jerome,
        	Wald, and Corey]{schiffer_et_al_rapid_localised_spread}
        	Joshua~T Schiffer, David Swan, Ramzi Al~Sallaq, Amalia Magaret, Christine
        	Johnston, Karen~E Mark, Stacy Selke, Negusse Ocbamichael, Steve Kuntz, Jia
        	Zhu, Barry Robinson, Meei-Li Huang, Keith~R Jerome, Anna Wald, and Lawrence
        	Corey.
        	\newblock Rapid localized spread and immunologic containment define herpes
        	simplex virus-2 reactivation in the human genital tract.
        	\newblock \emph{eLife}, 2:\penalty0 e00288, 2013{\natexlab{a}}.
        	\newblock \doi{10.7554/eLife.00288}.
        	
        	\bibitem[Bocharov et~al.(2016)Bocharov, Meyerhans, Bessonov, Trofimchuk, and
        	Volpert]{bocharov_et_al_spatiotemporal_dynamics_virus}
        	Gennady Bocharov, Andreas Meyerhans, Nickolai Bessonov, Sergei Trofimchuk, and
        	Vitaly Volpert.
        	\newblock Spatiotemporal dynamics of virus infection spreading in tissues.
        	\newblock \emph{PLOS ONE}, 11\penalty0 (12):\penalty0 1--27, 2016.
        	\newblock \doi{10.1371/journal.pone.0168576}.
        	
        	\bibitem[Malinzi et~al.(2017)Malinzi, Eladdadi, and
        	Sibanda]{malinzi_2017_spatiotemporal_chemovirotherapy}
        	Joseph Malinzi, Amina Eladdadi, and Precious Sibanda.
        	\newblock Modelling the spatiotemporal dynamics of chemovirotherapy cancer
        	treatment.
        	\newblock \emph{Journal of Biological Dynamics}, 11\penalty0 (1):\penalty0
        	244--274, 2017.
        	\newblock \doi{10.1080/17513758.2017.1328079}.
        	
        	\bibitem[Friedman and
        	Lai(2018)]{friedman_and_lai_combination_oncolytic_checkpoint}
        	Avner Friedman and Xiulan Lai.
        	\newblock Combination therapy for cancer with oncolytic virus and checkpoint
        	inhibitor: A mathematical model.
        	\newblock \emph{PLOS ONE}, 13\penalty0 (2):\penalty0 1--21, 2018.
        	\newblock \doi{10.1371/journal.pone.0192449}.
        	
        	\bibitem[Eriksen et~al.(2018)Eriksen, Svenningsen, Sneppen, and
        	Mitarai]{eriksen_et_al_growing_microcolony}
        	Rasmus~Skytte Eriksen, Sine~L. Svenningsen, Kim Sneppen, and Namiko Mitarai.
        	\newblock A growing microcolony can survive and support persistent propagation
        	of virulent phages.
        	\newblock \emph{Proceedings of the National Academy of Sciences}, 115\penalty0
        	(2):\penalty0 337--342, 2018.
        	\newblock \doi{10.1073/pnas.1708954115}.
        	
        	\bibitem[Alzahrani et~al.(2019)Alzahrani, Eftimie, and
        	Trucu]{alzahrani_2019_multiscale_cancer_response}
        	Talal Alzahrani, Raluca Eftimie, and Dumitru Trucu.
        	\newblock Multiscale modelling of cancer response to oncolytic viral therapy.
        	\newblock \emph{Mathematical Biosciences}, 310:\penalty0 76--95, 2019.
        	\newblock \doi{https://doi.org/10.1016/j.mbs.2018.12.018}.
        	
        	\bibitem[Quirouette et~al.(2020)Quirouette, Younis, Reddy, and
        	Beauchemin]{quirouette_et_al_influenza_localisation_model}
        	Christian Quirouette, Nada~P Younis, Micaela~B Reddy, and Catherine A~A
        	Beauchemin.
        	\newblock A mathematical model describing the localization and spread of
        	influenza {A} virus infection within the human respiratory tract.
        	\newblock \emph{PLOS Computational Biology}, 16\penalty0 (4):\penalty0
        	e1007705, 2020.
        	\newblock \doi{10.1371/journal.pcbi.1007705}.
        	
        	\bibitem[Simmons et~al.(2020)Simmons, Bond, Koskella, Drescher, Bucci, and
        	Nadell]{simmons_et_al_biofilm_structure}
        	Emilia~L. Simmons, Matthew~C. Bond, Britt Koskella, Knut Drescher, Vanni Bucci,
        	and Carey~D. Nadell.
        	\newblock Biofilm structure promotes coexistence of phage-resistant and
        	phage-susceptible bacteria.
        	\newblock \emph{mSystems}, 5\penalty0 (3):\penalty0 10.1128/msystems.00877--19,
        	2020.
        	\newblock \doi{10.1128/msystems.00877-19}.
        	
        	\bibitem[Rodrigo et~al.(2014)Rodrigo, Zwart, and Elena]{rodrigo_plant_viruses}
        	Guillermo Rodrigo, Mark~P. Zwart, and Santiago~F. Elena.
        	\newblock Onset of virus systemic infection in plants is determined by speed of
        	cell-to-cell movement and number of primary infection foci.
        	\newblock \emph{Journal of The Royal Society Interface}, 11\penalty0
        	(98):\penalty0 20140555, 2014.
        	\newblock \doi{10.1098/rsif.2014.0555}.
        	
        	\bibitem[Baabdulla et~al.(2021)Baabdulla, Now, Park, Kim, Jung, Yoo, and
        	Hillen]{baabdulla_et_al_homogenisation_IAV}
        	Arwa~Abdulla Baabdulla, Hesung Now, Ju~An Park, Woo-Jong Kim, Sungjune Jung,
        	Joo-Yeon Yoo, and Thomas Hillen.
        	\newblock Homogenization of a reaction diffusion equation can explain influenza
        	{A} virus load data.
        	\newblock \emph{Journal of Theoretical Biology}, 527:\penalty0 110816, 2021.
        	\newblock \doi{https://doi.org/10.1016/j.jtbi.2021.110816}.
        	
        	\bibitem[Frank(2000)]{frank_within_host_spatial_dynamics}
        	Steven~A Frank.
        	\newblock Within-host spatial dynamics of viruses and defective interfering
        	particles.
        	\newblock \emph{Journal of Theoretical Biology}, 206\penalty0 (2):\penalty0
        	279--290, 2000.
        	\newblock \doi{https://doi.org/10.1006/jtbi.2000.2120}.
        	
        	\bibitem[Ait~Mahiout et~al.(2023)Ait~Mahiout, Bessonov, Kazmierczak, and
        	Volpert]{mahiout_2023_respiratory_viral_infections}
        	Latifa Ait~Mahiout, Nikolai Bessonov, Bogdan Kazmierczak, and Vitaly Volpert.
        	\newblock {Mathematical modeling of respiratory viral infection and
        		applications to SARS-CoV-2 progression}.
        	\newblock \emph{Mathematical Methods in the Applied Sciences}, 46\penalty0
        	(2):\penalty0 1740--1751, 2023.
        	\newblock \doi{https://doi.org/10.1002/mma.8606}.
        	
        	\bibitem[Amor and Fort(2010)]{amor_and_fort_virus_infection_speeds}
        	Daniel~R Amor and Joaquim Fort.
        	\newblock Virus infection speeds: theory versus experiment.
        	\newblock \emph{Physical Review E: Statistical, Nonlinear, and Soft Matter
        		Physics}, 82\penalty0 (6 Pt 1):\penalty0 061905, 2010.
        	\newblock \doi{10.1103/PhysRevE.82.061905}.
        	
        	\bibitem[{de Rioja} et~al.(2015){de Rioja}, Fort, and
        	Isern]{de_rioja_et_al_front_propagation_T7}
        	V.L. {de Rioja}, J.~Fort, and N.~Isern.
        	\newblock Front propagation speeds of t7 virus mutants.
        	\newblock \emph{Journal of Theoretical Biology}, 385:\penalty0 112--118, 2015.
        	\newblock \doi{https://doi.org/10.1016/j.jtbi.2015.08.005}.
        	
        	\bibitem[{Ait Mahiout} et~al.(2022){Ait Mahiout}, Mozokhina, Tokarev, and
        	Volpert]{mahiout_2022_virus_replication_and_competition}
        	L.~{Ait Mahiout}, A.~Mozokhina, A.~Tokarev, and V.~Volpert.
        	\newblock Virus replication and competition in a cell culture: {Application to
        		the SARS--CoV--2 variants}.
        	\newblock \emph{Applied Mathematics Letters}, 133:\penalty0 108217, 2022.
        	\newblock \doi{https://doi.org/10.1016/j.aml.2022.108217}.
        	
        	\bibitem[Tokarev et~al.(2022)Tokarev, Mozokhina, and
        	Volpert]{tokarev_et_al_fastest_autowave}
        	Alexey Tokarev, Anastasia Mozokhina, and Vitaly Volpert.
        	\newblock Competition of {SARS-CoV-2} variants in cell culture and tissue: Wins
        	the fastest viral autowave.
        	\newblock \emph{Vaccines (Basel)}, 10\penalty0 (7), 2022.
        	\newblock \doi{10.3390/vaccines10070995}.
        	
        	\bibitem[Holder et~al.(2011{\natexlab{a}})Holder, Simon, Liao, Abed, Bouhy,
        	Beauchemin, and Boivin]{holder_et_al_assessing_the_in_vitro_fitness}
        	Benjamin~P. Holder, Philippe Simon, Laura~E. Liao, Yacine Abed, Xavier Bouhy,
        	Catherine A.~A. Beauchemin, and Guy Boivin.
        	\newblock Assessing the in vitro fitness of an oseltamivir-resistant seasonal
        	{A/H1N1} influenza strain using a mathematical model.
        	\newblock \emph{PLOS ONE}, 6\penalty0 (3):\penalty0 1--11, 2011{\natexlab{a}}.
        	\newblock \doi{10.1371/journal.pone.0014767}.
        	
        	\bibitem[Alzahrani(2021)]{alzahrani_spatiotemporal_sars_cov_2}
        	Talal Alzahrani.
        	\newblock Spatio--temporal modeling of immune response to {SARS-CoV-2}
        	infection.
        	\newblock \emph{Mathematics}, 9\penalty0 (24), 2021.
        	\newblock \doi{10.3390/math9243274}.
        	
        	\bibitem[Vimalajeewa et~al.(2022)Vimalajeewa, Balasubramaniam, Berry, and
        	Barry]{vimalajeewa_virus_particle_propagation_respiratory_tract}
        	Dixon Vimalajeewa, Sasitharan Balasubramaniam, Donagh~P. Berry, and Gerald
        	Barry.
        	\newblock Virus particle propagation and infectivity along the respiratory
        	tract and a case study for {SARS-CoV-2}.
        	\newblock \emph{Scientific Reports}, 12\penalty0 (1):\penalty0 7666, 2022.
        	\newblock \doi{10.1038/s41598-022-11816-2}.
        	
        	\bibitem[Haseltine et~al.(2008)Haseltine, Lam, Yin, and
        	Rawlings]{haseltine_et_al_image_guided_growth}
        	Eric~L Haseltine, Vy~Lam, John Yin, and James~B Rawlings.
        	\newblock Image-guided modeling of virus growth and spread.
        	\newblock \emph{Bulletin of Mathematical Biology}, 70\penalty0 (6):\penalty0
        	1730--1748, 2008.
        	\newblock \doi{10.1007/s11538-008-9316-3}.
        	
        	\bibitem[Malinzi(2021)]{malinzi_oncolytic_virus_telegraph_equation}
        	Joseph Malinzi.
        	\newblock A mathematical model for oncolytic virus spread using the telegraph
        	equation.
        	\newblock \emph{Communications in Nonlinear Science and Numerical Simulation},
        	102:\penalty0 105944, 2021.
        	\newblock \doi{https://doi.org/10.1016/j.cnsns.2021.105944}.
        	
        	\bibitem[Malinzi et~al.(2015)Malinzi, Sibanda, and
        	Mambili-Mamboundou]{malinzi_et_al_2015_virotherapy_solid_tumour_invasion}
        	Joseph Malinzi, Precious Sibanda, and Hermane Mambili-Mamboundou.
        	\newblock Analysis of virotherapy in solid tumor invasion.
        	\newblock \emph{Mathematical Biosciences}, 263:\penalty0 102--110, 2015.
        	\newblock \doi{https://doi.org/10.1016/j.mbs.2015.01.015}.
        	
        	\bibitem[Dunia and Bonnecaze(2013)]{dunia_and_bonnecaze_spherical_organs}
        	Ricardo Dunia and Roger Bonnecaze.
        	\newblock Mathematical modeling of viral infection dynamics in spherical
        	organs.
        	\newblock \emph{Journal of Mathematical Biology}, 67\penalty0 (6):\penalty0
        	1425--1455, 2013.
        	\newblock \doi{10.1007/s00285-012-0593-y}.
        	
        	\bibitem[de~Rioja et~al.(2016)de~Rioja, Isern, and
        	Fort]{de_rioja_et_al_gliobastomas}
        	Victor~Lopez de~Rioja, Neus Isern, and Joaquim Fort.
        	\newblock A mathematical approach to virus therapy of glioblastomas.
        	\newblock \emph{Biology Direct}, 11\penalty0 (1):\penalty0 1, 2016.
        	\newblock \doi{10.1186/s13062-015-0100-7}.
        	
        	\bibitem[Amor and Fort(2014)]{amor_and_fort_cohabitation_reac_diff}
        	Daniel~R. Amor and Joaquim Fort.
        	\newblock Cohabitation reaction---diffusion model for virus focal infections.
        	\newblock \emph{Physica A: Statistical Mechanics and its Applications},
        	416:\penalty0 611--619, 2014.
        	\newblock \doi{https://doi.org/10.1016/j.physa.2014.08.023}.
        	
        	\bibitem[Fort and M\'{e}ndez(2002)]{fort_and_mendez_time_delayed_spread}
        	Joaquim Fort and Vicen{\c c} M\'{e}ndez.
        	\newblock Time-delayed spread of viruses in growing plaques.
        	\newblock \emph{Physical Review Letters}, 89\penalty0 (17):\penalty0 178101,
        	2002.
        	\newblock \doi{10.1103/PhysRevLett.89.178101}.
        	
        	\bibitem[Bessonov et~al.(2023)Bessonov, Bocharov, Mozokhina, and
        	Volpert]{bessonov_et_al_intracellular_spreading_viral_infection}
        	Nikolay Bessonov, Gennady Bocharov, Anastasiia Mozokhina, and Vitaly Volpert.
        	\newblock Viral infection spreading in cell culture with intracellular
        	regulation.
        	\newblock \emph{Mathematics}, 11\penalty0 (6), 2023.
        	\newblock \doi{10.3390/math11061526}.
        	
        	\bibitem[Graziano et~al.(2008)Graziano, Kettoola, Munshower, Stapleton, and
        	Towfic]{graziano_et_al_finite_element_HIV_model}
        	Frank~M Graziano, Samira~Y Kettoola, Judy~M Munshower, Jack~T Stapleton, and
        	George~J Towfic.
        	\newblock Effect of spatial distribution of {T-Cells} and {HIV} load on {HIV}
        	progression.
        	\newblock \emph{Bioinformatics}, 24\penalty0 (6):\penalty0 855--860, 2008.
        	\newblock \doi{10.1093/bioinformatics/btn008}.
        	
        	\bibitem[Reisch and Langemann(2022)]{reisch_and_langemann_automative_selection}
        	C.~Reisch and D.~Langemann.
        	\newblock Automative model selection and model certification for
        	reaction-diffusion equations.
        	\newblock \emph{IFAC-PapersOnLine}, 55\penalty0 (20):\penalty0 73--78, 2022.
        	\newblock \doi{https://doi.org/10.1016/j.ifacol.2022.09.074}.
        	
        	\bibitem[Liang et~al.(2023)Liang, Yang, Fan, and
        	Lo]{liang_et_al_patch_formation_virus_DIPs}
        	Qiantong Liang, Johnny Yang, Wai-Tong~Louis Fan, and Wing-Cheong Lo.
        	\newblock Patch formation driven by stochastic effects of interaction between
        	viruses and defective interfering particles.
        	\newblock \emph{PLOS Computational Biology}, 19\penalty0 (10):\penalty0 1--23,
        	2023.
        	\newblock \doi{10.1371/journal.pcbi.1011513}.
        	
        	\bibitem[Bailey et~al.(2013)Bailey, Kirk, Naik, Nace, Steele, Suksanpaisan, Li,
        	Federspiel, Peng, Kirk, and
        	Russell]{bailey_et_al_radial_expansion_intratumoural}
        	Kent Bailey, Amber Kirk, Shruthi Naik, Rebecca Nace, Michael~B. Steele, Lukkana
        	Suksanpaisan, Xing Li, Mark~J. Federspiel, Kah-Whye Peng, David Kirk, and
        	Stephen~J. Russell.
        	\newblock Mathematical model for radial expansion and conflation of
        	intratumoral infectious centers predicts curative oncolytic virotherapy
        	parameters.
        	\newblock \emph{PLOS ONE}, 8\penalty0 (9):\penalty0 1--11, 2013.
        	\newblock \doi{10.1371/journal.pone.0073759}.
        	
        	\bibitem[Mbopda et~al.(2021)Mbopda, Issa, Abdoulkary, Guiem, and
        	Fouda]{mbopda_et_al_pattern_formation}
        	B.~Tamko Mbopda, S.~Issa, S.~Abdoulkary, R.~Guiem, and H.~P.~Ekobena Fouda.
        	\newblock Pattern formations in nonlinear dynamics of hepatitis {B} virus.
        	\newblock \emph{The European Physical Journal Plus}, 136\penalty0 (5):\penalty0
        	586, 2021.
        	\newblock \doi{10.1140/epjp/s13360-021-01569-8}.
        	
        	\bibitem[Anekal et~al.(2009)Anekal, Zhu, Graham, and
        	Yin]{anekal_et_al_virus_spread_fluid_flow}
        	Samartha~G Anekal, Ying Zhu, Michael~D Graham, and John Yin.
        	\newblock Dynamics of virus spread in the presence of fluid flow.
        	\newblock \emph{Integrative Biology (Cambridge)}, 1\penalty0 (11-12):\penalty0
        	664--671, 2009.
        	\newblock \doi{10.1039/b908197f}.
        	
        	\bibitem[Timalsina et~al.(2017)Timalsina, Tian, and
        	Wang]{timalsina_et_al_math_comp_tumour_virotherapy}
        	Asim Timalsina, Jianjun~Paul Tian, and Jin Wang.
        	\newblock Mathematical and computational modeling for tumor virotherapy with
        	mediated immunity.
        	\newblock \emph{Bull Math Biol}, 79\penalty0 (8):\penalty0 1736--1758, 2017.
        	\newblock \doi{10.1007/s11538-017-0304-3}.
        	
        	\bibitem[Alzahrani et~al.(2020)Alzahrani, Eftimie, and
        	Trucu]{alzahrani_2020_moving_boundary_fusogenic}
        	Talal Alzahrani, Raluca Eftimie, and Dumitru Trucu.
        	\newblock Multiscale moving boundary modelling of cancer interactions with a
        	fusogenic oncolytic virus: The impact of syncytia dynamics.
        	\newblock \emph{Mathematical Biosciences}, 323:\penalty0 108296, 2020.
        	\newblock \doi{https://doi.org/10.1016/j.mbs.2019.108296}.
        	
        	\bibitem[Alsisi et~al.(2021)Alsisi, Eftimie, and Trucu]{alsisi_go_or_grow}
        	Abdulhamed Alsisi, Raluca Eftimie, and Dumitru Trucu.
        	\newblock Non--local multiscale approach for the impact of go or grow
        	hypothesis on tumour--viruses interactions.
        	\newblock \emph{Mathematical Biosciences and Engineering}, 18\penalty0
        	(5):\penalty0 5252--5284, 2021.
        	\newblock \doi{10.3934/mbe.2021267}.
        	
        	\bibitem[Alsisi et~al.(2022)Alsisi, Eftimie, and
        	Trucu]{alsisi_heterogeneous_ECM}
        	Abdulhamed Alsisi, Raluca Eftimie, and Dumitru Trucu.
        	\newblock Nonlocal multiscale modelling of tumour--oncolytic viruses
        	interactions within a heterogeneous fibrous/non--fibrous extracellular
        	matrix.
        	\newblock \emph{Mathematical Biosciences and Engineering}, 19\penalty0
        	(6):\penalty0 6157--6185, 2022.
        	\newblock \doi{10.3934/mbe.2022288}.
        	
        	\bibitem[Alsisi et~al.(2020)Alsisi, Eftimie, and
        	Trucu]{alsisi_nonlocal_multscale_approaches}
        	Abdulhamed Alsisi, Raluca Eftimie, and Dumitru Trucu.
        	\newblock Non--local multiscale approaches for tumour--oncolytic viruses
        	interactions.
        	\newblock \emph{Mathematics in Applied Sciences and Engineering}, 1\penalty0
        	(3):\penalty0 207--273, 2020.
        	
        	\bibitem[Kim et~al.(2014)Kim, Lee, Dmitrieva, Kim, Kaur, and
        	Friedman]{kim_et_al_choindroitinase_oncolytic}
        	Yangjin Kim, Hyun~Geun Lee, Nina Dmitrieva, Junseok Kim, Balveen Kaur, and
        	Avner Friedman.
        	\newblock Choindroitinase {ABC} {I}-mediated enhancement of oncolytic virus
        	spread and anti tumor efficacy: A mathematical model.
        	\newblock \emph{PLOS ONE}, 9\penalty0 (7):\penalty0 1--19, 2014.
        	\newblock \doi{10.1371/journal.pone.0102499}.
        	
        	\bibitem[Fukuyama et~al.(2015)Fukuyama, Katsura, Zhao, Ozawa, Ando, Shoemaker,
        	Ishikawa, Yamada, Neumann, Watanabe, Kitano, and
        	Kawaoka]{fukuyama_et_al_color_flu}
        	Satoshi Fukuyama, Hiroaki Katsura, Dongming Zhao, Makoto Ozawa, Tomomi Ando,
        	Jason~E. Shoemaker, Izumi Ishikawa, Shinya Yamada, Gabriele Neumann, Shinji
        	Watanabe, Hiroaki Kitano, and Yoshihiro Kawaoka.
        	\newblock Multi-spectral fluorescent reporter influenza viruses (color-flu) as
        	powerful tools for in vivo studies.
        	\newblock \emph{Nature Communications}, 6\penalty0 (1):\penalty0 6600, 2015.
        	\newblock \doi{10.1038/ncomms7600}.
        	
        	\bibitem[Manicassamy et~al.(2010)Manicassamy, Manicassamy, Belicha-Villanueva,
        	Pisanelli, Pulendran, and
        	Garc{\'\i}a-Sastre]{manicassamy_et_al_gfp_reporter_influenza}
        	Balaji Manicassamy, Santhakumar Manicassamy, Alan Belicha-Villanueva, Giuseppe
        	Pisanelli, Bali Pulendran, and Adolfo Garc{\'\i}a-Sastre.
        	\newblock Analysis of in vivo dynamics of influenza virus infection in mice
        	using a {GFP} reporter virus.
        	\newblock \emph{Proceedings of the National Academy of Sciences}, 107\penalty0
        	(25):\penalty0 11531--11536, 2010.
        	\newblock \doi{10.1073/pnas.0914994107}.
        	
        	\bibitem[Bray et~al.(2011)Bray, Lawler, Paragas, Jahrling, and
        	Mollura]{bray_et_al_molecular_imaging_influenza}
        	Mike Bray, James Lawler, Jason Paragas, Peter~B. Jahrling, and Daniel~J.
        	Mollura.
        	\newblock Molecular imaging of influenza and other emerging respiratory viral
        	infections.
        	\newblock \emph{The Journal of Infectious Diseases}, 203\penalty0
        	(10):\penalty0 1348--1359, 2011.
        	\newblock \doi{10.1093/infdis/jir038}.
        	
        	\bibitem[Baabdulla and
        	Hillen(2024)]{baabdulla_and_hillen_oscillations_spatial_oncolytic}
        	Arwa~Abdulla Baabdulla and Thomas Hillen.
        	\newblock Oscillations in a spatial oncolytic virus model.
        	\newblock \emph{Bulletin of Mathematical Biology}, 86\penalty0 (8):\penalty0
        	93, 2024.
        	\newblock \doi{10.1007/s11538-024-01322-z}.
        	
        	\bibitem[Li et~al.(2020)Li, Zhang, and Zhou]{li_et_al_spatially_antiviral_hcv}
        	Chentong Li, Yingying Zhang, and Yicang Zhou.
        	\newblock Spatially antiviral dynamics determines {HCV} in vivo replication and
        	evolution.
        	\newblock \emph{Journal of Theoretical Biology}, 503:\penalty0 110378, 2020.
        	\newblock \doi{10.1016/j.jtbi.2020.110378}.
        	
        	\bibitem[Orive et~al.(2005)Orive, Stearns, Kelly, Barfield, Smith, and
        	Holt]{orive_et_al_viral_infection_in_internally_structred_hosts}
        	Maria~E Orive, Miles~N Stearns, John~K Kelly, Michael Barfield, Marilyn~S
        	Smith, and Robert~D Holt.
        	\newblock Viral infection in internally structured hosts. {I. C}onditions for
        	persistent infection.
        	\newblock \emph{Journal of Theoretical Biology}, 232\penalty0 (4):\penalty0
        	453--466, 2005.
        	\newblock \doi{10.1016/j.jtbi.2004.08.023}.
        	
        	\bibitem[Ke et~al.(2020)Ke, Zitzmann, Ribeiro, and
        	Perelson]{ke_perelson_et_al_covid_upper_and_lower_rt}
        	Ruian Ke, Carolin Zitzmann, Ruy~M. Ribeiro, and Alan~S. Perelson.
        	\newblock Kinetics of {SARS-CoV-2} infection in the human upper and lower
        	respiratory tracts and their relationship with infectiousness [pre-print].
        	\newblock \emph{medRxiv}, page 2020.09.25.20201772, 2020.
        	\newblock \doi{10.1101/2020.09.25.20201772}.
        	
        	\bibitem[Goyal et~al.(2022)Goyal, Duke, Cardozo-Ojeda, and
        	Schiffer]{goyal_remdesivir_macaques}
        	Ashish Goyal, Elizabeth~R. Duke, E.~Fabian Cardozo-Ojeda, and Joshua~T.
        	Schiffer.
        	\newblock Mathematical modeling explains differential {SARS CoV-2} kinetics in
        	lung and nasal passages in remdesivir treated rhesus macaques.
        	\newblock \emph{iScience}, 25\penalty0 (6):\penalty0 104448, 2022.
        	\newblock \doi{10.1016/j.isci.2022.104448}.
        	
        	\bibitem[Voutouri et~al.(2021)Voutouri, Nikmaneshi, Hardin, Patel, Verma,
        	Khandekar, Dutta, Stylianopoulos, Munn, and
        	Jain]{voutouri_in_silico_dynamics_covid}
        	Chrysovalantis Voutouri, Mohammad~Reza Nikmaneshi, C.~Corey Hardin, Ankit~B.
        	Patel, Ashish Verma, Melin~J. Khandekar, Sayon Dutta, Triantafyllos
        	Stylianopoulos, Lance~L. Munn, and Rakesh~K. Jain.
        	\newblock In silico dynamics of {COVID-19} phenotypes for optimizing clinical
        	management.
        	\newblock \emph{Proceedings of the National Academy of Sciences}, 118\penalty0
        	(3), 2021.
        	\newblock \doi{10.1073/pnas.2021642118}.
        	
        	\bibitem[Byrne et~al.(2018)Byrne, Gantt, and
        	Coombs]{byrne_et_al_spatiotemporal_hsv2_and_hiv}
        	Catherine~M. Byrne, Soren Gantt, and Daniel Coombs.
        	\newblock Effects of spatiotemporal {HSV-2} lesion dynamics and antiviral
        	treatment on the risk of {HIV-1} acquisition.
        	\newblock \emph{PLOS Computational Biology}, 14\penalty0 (4):\penalty0 1--28,
        	2018.
        	\newblock \doi{10.1371/journal.pcbi.1006129}.
        	
        	\bibitem[Zhang and Liu(2005)]{zhang_et_al_massively_multi_agent_system}
        	Shiwu Zhang and Jiming Liu.
        	\newblock A massively multi-agent system for discovering {HIV}-immune
        	interaction dynamics.
        	\newblock In Toru Ishida, Les Gasser, and Hideyuki Nakashima, editors,
        	\emph{Massively Multi-Agent Systems I}, pages 161--173, 2005.
        	
        	\bibitem[Schiffer et~al.(2016{\natexlab{a}})Schiffer, Swan, Magaret, Corey,
        	Wald, Ossig, Ruebsamen-Schaeff, Stoelben, Timmler, Zimmermann, Melhem,
        	Van~Wart, Rubino, and Birkmann]{schiffer_et_al_hsv2_suppression_pritelivir}
        	Joshua~T Schiffer, David~A Swan, Amalia Magaret, Lawrence Corey, Anna Wald,
        	Joachim Ossig, Helga Ruebsamen-Schaeff, Susanne Stoelben, Burkhard Timmler,
        	Holger Zimmermann, Murad~R Melhem, Scott~A Van~Wart, Christopher~M Rubino,
        	and Alexander Birkmann.
        	\newblock Mathematical modeling of herpes simplex virus-2 suppression with
        	pritelivir predicts trial outcomes.
        	\newblock \emph{Science Translational Medicine}, 8\penalty0 (324):\penalty0
        	324ra15, 2016{\natexlab{a}}.
        	\newblock \doi{10.1126/scitranslmed.aad6654}.
        	
        	\bibitem[Schiffer et~al.(2016{\natexlab{b}})Schiffer, Swan, Magaret, Schacker,
        	Wald, and Corey]{schiffer_et_al_increased_hsv2_shedding}
        	Joshua~T. Schiffer, David~A. Swan, Amalia Magaret, Timothy~W. Schacker, Anna
        	Wald, and Lawrence Corey.
        	\newblock Mathematical modeling predicts that increased {HSV-2} shedding in
        	{HIV-1} infected persons is due to poor immunologic control in ganglia and
        	genital mucosa.
        	\newblock \emph{PLOS ONE}, 11\penalty0 (6):\penalty0 1--22, 2016{\natexlab{b}}.
        	\newblock \doi{10.1371/journal.pone.0155124}.
        	
        	\bibitem[Schiffer et~al.(2013{\natexlab{b}})Schiffer, Swan, Corey, and
        	Wald]{schiffer_et_al_rapid_viral_expansion}
        	Joshua~T Schiffer, David~A Swan, Lawrence Corey, and Anna Wald.
        	\newblock Rapid viral expansion and short drug half-life explain the incomplete
        	effectiveness of current herpes simplex virus 2-directed antiviral agents.
        	\newblock \emph{Antimicrobial Agents and Chemotherapy}, 57\penalty0
        	(12):\penalty0 5820--5829, 2013{\natexlab{b}}.
        	\newblock \doi{10.1128/AAC.01114-13}.
        	
        	\bibitem[Schiffer(2013)]{schiffer_hsv2_specific_cd8}
        	Joshua~T Schiffer.
        	\newblock Mucosal {HSV-2} specific {CD8+} {T}-cells represent containment of
        	prior viral shedding rather than a correlate of future protection.
        	\newblock \emph{Frontiers in Immunology}, 4:\penalty0 209, 2013.
        	\newblock \doi{10.3389/fimmu.2013.00209}.
        	
        	\bibitem[Dhankani et~al.(2014)Dhankani, Kutz, and
        	Schiffer]{dhankani_et_al_hsv2_genital_tract_shedding}
        	Varsha Dhankani, J.~Nathan Kutz, and Joshua~T. Schiffer.
        	\newblock Herpes simplex virus-2 genital tract shedding is not predictable over
        	months or years in infected persons.
        	\newblock \emph{PLOS Computational Biology}, 10\penalty0 (11):\penalty0 1--16,
        	2014.
        	\newblock \doi{10.1371/journal.pcbi.1003922}.
        	
        	\bibitem[Nakaoka et~al.(2016)Nakaoka, Iwami, and
        	Sato]{nakaoka_et_al_HIV_in_lymphoid_tissue_network}
        	Shinji Nakaoka, Shingo Iwami, and Kei Sato.
        	\newblock Dynamics of {HIV} infection in lymphoid tissue network.
        	\newblock \emph{Journal of Mathematical Biology}, 72\penalty0 (4):\penalty0
        	909--938, 2016.
        	\newblock \doi{10.1007/s00285-015-0940-x}.
        	
        	\bibitem[Cardozo et~al.(2014{\natexlab{a}})Cardozo, Zurakowski, and
        	Attoh-Okine]{cardozo_et_al_compartmental_analysis}
        	Erwing~Fabian Cardozo, Ryan Zurakowski, and Nii Attoh-Okine.
        	\newblock Analysis of {HIV-1} compartmental model parameters using {Bayesian
        		MCMC} estimation.
        	\newblock In \emph{2014 American Control Conference}, pages 2765--2770,
        	2014{\natexlab{a}}.
        	
        	\bibitem[Cardozo et~al.(2014{\natexlab{b}})Cardozo, Luo, Piovoso, and
        	Zurakowski]{cardozo_et_al_cryptic_viremia}
        	E~Fabian Cardozo, Rutao Luo, Michael~J Piovoso, and Ryan Zurakowski.
        	\newblock Spatial modeling of {HIV} cryptic viremia and {2-LTR} formation
        	during raltegravir intensification.
        	\newblock \emph{Journal of Theoretical Biology}, 345:\penalty0 61--69,
        	2014{\natexlab{b}}.
        	\newblock \doi{10.1016/j.jtbi.2013.12.020}.
        	
        	\bibitem[Jagarapu et~al.(2020)Jagarapu, Piovoso, and
        	Zurakowski]{jagarapu_et_al_integrated_spatial_dynamics_pharmacokinetic}
        	Aditya Jagarapu, Michael~J Piovoso, and Ryan Zurakowski.
        	\newblock An integrated spatial dynamics-pharmacokinetic model explaining poor
        	penetration of anti-retroviral drugs in lymph nodes.
        	\newblock \emph{Frontiers in Bioengineering and Biotechnology}, 8:\penalty0
        	667, 2020.
        	\newblock \doi{10.3389/fbioe.2020.00667}.
        	
        	\bibitem[Ollerton et~al.(2022)Ollerton, Folkvord, La~Mantia, Parry, Meditz,
        	McCarter, D'Aquila, and Connick]{ollerton_et_al_follicular}
        	Matthew~T. Ollerton, Joy~M. Folkvord, Andriana La~Mantia, David~A. Parry,
        	Amie~L. Meditz, Martin~D. McCarter, Richard~T. D'Aquila, and Elizabeth
        	Connick.
        	\newblock Follicular regulatory {T} cells eliminate {HIV-1-infected} follicular
        	helper {T} cells in an {IL-2} concentration dependent manner.
        	\newblock \emph{Frontiers in Immunology}, 13, 2022.
        	\newblock \doi{10.3389/fimmu.2022.878273}.
        	
        	\bibitem[Chung et~al.(2024)Chung, Connick, and
        	Wodarz]{chung_et_al_secondary_lymphoid}
        	Wen-Jian Chung, Elizabeth Connick, and Dominik Wodarz.
        	\newblock Human immunodeficiency virus dynamics in secondary lymphoid tissues
        	and the evolution of cytotoxic {T} lymphocyte escape mutants.
        	\newblock \emph{Virus Evolution}, 10\penalty0 (1):\penalty0 vead084, 2024.
        	\newblock \doi{10.1093/ve/vead084}.
        	
        	\bibitem[Akpinar et~al.(2016)Akpinar, Inankur, and
        	Yin]{akpinar_et_al_spatiotemporal_viral_amplification}
        	Fulya Akpinar, Bahar Inankur, and John Yin.
        	\newblock Spatial--temporal patterns of viral amplification and interference
        	initiated by a single infected cell.
        	\newblock \emph{Journal of Virology}, 90\penalty0 (16):\penalty0 7552--7566,
        	2016.
        	\newblock \doi{10.1128/jvi.00807-16}.
        	
        	\bibitem[Goyal and Murray(2016)]{goyal_and_murray_CCT_in_HBV}
        	Ashish Goyal and John~M. Murray.
        	\newblock Modelling the impact of cell-to-cell transmission in hepatitis {B}
        	virus.
        	\newblock \emph{PLOS ONE}, 11\penalty0 (8):\penalty0 1--22, 2016.
        	\newblock \doi{10.1371/journal.pone.0161978}.
        	
        	\bibitem[Bouchnita et~al.(2017)Bouchnita, Bocharov, Meyerhans, and
        	Volpert]{bouchnita_et_al_towards_multiscale_HIV}
        	Anass Bouchnita, Gennady Bocharov, Andreas Meyerhans, and Vitaly Volpert.
        	\newblock Towards a multiscale model of acute {HIV} infection.
        	\newblock \emph{Computation}, 5\penalty0 (1), 2017.
        	\newblock \doi{10.3390/computation5010006}.
        	
        	\bibitem[Smith(2018)]{smith_validated_models_of_immune_response}
        	Amber~M Smith.
        	\newblock Validated models of immune response to virus infection.
        	\newblock \emph{Current Opinion in Systems Biology}, 12:\penalty0 46--52, 2018.
        	\newblock \doi{10.1016/j.coisb.2018.10.005}.
        	
        	\bibitem[Yin and
        	Redovich(2018)]{yin_and_redovich_kinetic_modelling_virus_growth}
        	John Yin and Jacob Redovich.
        	\newblock Kinetic modeling of virus growth in cells.
        	\newblock \emph{Microbiology and Molecular Biology Reviews}, 82\penalty0 (2),
        	2018.
        	\newblock \doi{10.1128/MMBR.00066-17}.
        	
        	\bibitem[Holder et~al.(2011{\natexlab{b}})Holder, Liao, Simon, Boivin, and
        	Beauchemin]{holder_et_al_design_considerations_influenza}
        	Benjamin~P. Holder, Laura~E. Liao, Philippe Simon, Guy Boivin, and Catherine
        	A.~A. Beauchemin.
        	\newblock Design considerations in building in silico equivalents of common
        	experimental influenza virus assays.
        	\newblock \emph{Autoimmunity}, 44\penalty0 (4):\penalty0 282--293,
        	2011{\natexlab{b}}.
        	\newblock \doi{10.3109/08916934.2011.523267}.
        	
        	\bibitem[Rowlatt et~al.(2022)Rowlatt, Chaplain, Hughes, Gillespie, Dockrell,
        	Johannessen, and Bowness]{rowlatt_et_al_modelling_within_host_covid}
        	C.~F. Rowlatt, M.~A.~J. Chaplain, D.~J. Hughes, S.~H. Gillespie, D.~H.
        	Dockrell, I.~Johannessen, and R.~Bowness.
        	\newblock Modelling the within-host spread of {SARS-CoV-2} infection, and the
        	subsequent immune response, using a hybrid, multiscale, individual-based
        	model. {Part I: Macrophages} [pre-print].
        	\newblock \emph{bioRxiv}, page 2022.05.06.490883, 2022.
        	
        	\bibitem[Marzban et~al.(2021)Marzban, Han, Juh{\'a}sz, and
        	R{\"o}st]{marzban_et_al_hybrid_PDE_ABM_model}
        	Sadegh Marzban, Renji Han, N{\'o}ra Juh{\'a}sz, and Gergely R{\"o}st.
        	\newblock A hybrid {PDE--ABM} model for viral dynamics with application to
        	{SARS-CoV-2} and influenza.
        	\newblock \emph{Royal Society Open Science}, 8\penalty0 (11):\penalty0 210787,
        	2021.
        	\newblock \doi{10.1098/rsos.210787}.
        	
        	\bibitem[Bartha et~al.(2022)Bartha, Juh{\'a}sz, Marzban, Han, and
        	R{\"o}st]{bartha_et_al_in_silico_evalutation_paxlovid}
        	Ferenc~A Bartha, N{\'o}ra Juh{\'a}sz, Sadegh Marzban, Renji Han, and Gergely
        	R{\"o}st.
        	\newblock In silico evaluation of paxlovid's pharmacometrics for {SARS-CoV-2}:
        	A multiscale approach.
        	\newblock \emph{Viruses}, 14\penalty0 (5), 2022.
        	\newblock \doi{10.3390/v14051103}.
        	
        	\bibitem[Blahut et~al.(2021)Blahut, Quirouette, Feld, Iwami, and
        	Beauchemin]{blahut_et_al_hepatitis_c_two_modes_of_spread}
        	Kenneth Blahut, Christian Quirouette, Jordan~J. Feld, Shingo Iwami, and
        	Catherine A.~A. Beauchemin.
        	\newblock Quantifying the relative contribution of free virus and cell-to-cell
        	transmission routes to the propagation of hepatitis {C} virus infections in
        	vitro using an agent-based model [pre-print].
        	\newblock \emph{arXiv}, page 2102.05531, 2021.
        	
        	\bibitem[Castiglione et~al.(2007)Castiglione, Duca, Jarrah, Laubenbacher,
        	Hochberg, and Thorley-Lawson]{castiglione_et_al_EBV_immsim}
        	Filippo Castiglione, Karen Duca, Abdul Jarrah, Reinhard Laubenbacher, Donna
        	Hochberg, and David Thorley-Lawson.
        	\newblock Simulating {Epstein-Barr} virus infection with {C-ImmSim}.
        	\newblock \emph{Bioinformatics}, 23\penalty0 (11):\penalty0 1371--1377, 2007.
        	\newblock \doi{10.1093/bioinformatics/btm044}.
        	
        	\bibitem[Bezzi et~al.(1997)Bezzi, Celada, Ruffo, and
        	Seiden]{bezzi_et_al_transition_between_immune_and_disease_states}
        	Michele Bezzi, Franco Celada, Stefano Ruffo, and Philip~E. Seiden.
        	\newblock The transition between immune and disease states in a cellular
        	automaton model of clonal immune response.
        	\newblock \emph{Physica A: Statistical Mechanics and its Applications},
        	245\penalty0 (1):\penalty0 145--163, 1997.
        	\newblock \doi{https://doi.org/10.1016/S0378-4371(97)00290-2}.
        	
        	\bibitem[Kohler et~al.(2000)Kohler, Puzone, Seiden, and
        	Celada]{kohler_et_al_systematic_vaccine_complexity}
        	B~Kohler, R~Puzone, P~E Seiden, and F~Celada.
        	\newblock A systematic approach to vaccine complexity using an automaton model
        	of the cellular and humoral immune system. {I. Viral} characteristics and
        	polarized responses.
        	\newblock \emph{Vaccine}, 19\penalty0 (7-8):\penalty0 862--876, 2000.
        	\newblock \doi{10.1016/s0264-410x(00)00225-5}.
        	
        	\bibitem[Bernaschi and
        	Castiglione(2002)]{bernaschi_and_castiglione_escape_mutants}
        	M~Bernaschi and F~Castiglione.
        	\newblock Selection of escape mutants from immune recognition during {HIV}
        	infection.
        	\newblock \emph{Immunology \& Cell Biology}, 80\penalty0 (3):\penalty0
        	307--313, 2002.
        	\newblock \doi{10.1046/j.1440-1711.2002.01082.x}.
        	
        	\bibitem[Castiglione et~al.(2004)Castiglione, Poccia, D'Offizi, and
        	Bernaschi]{castiglione_et_al_mutation_fitness_viral_diversity}
        	Filippo Castiglione, Fabrizio Poccia, Gianpiero D'Offizi, and Massimo
        	Bernaschi.
        	\newblock Mutation, fitness, viral diversity, and predictive markers of disease
        	progression in a computational model of {HIV} type 1 infection.
        	\newblock \emph{AIDS Research and Human Retroviruses}, 20\penalty0
        	(12):\penalty0 1314--1323, 2004.
        	\newblock \doi{10.1089/aid.2004.20.1314}.
        	
        	\bibitem[Baldazzi et~al.(2006)Baldazzi, Castiglione, and
        	Bernaschi]{baldazzi_et_al_enhanced_agent_based_model}
        	V.~Baldazzi, F.~Castiglione, and M.~Bernaschi.
        	\newblock An enhanced agent based model of the immune system response.
        	\newblock \emph{Cellular Immunology}, 244\penalty0 (2):\penalty0 77--79, 2006.
        	\newblock \doi{https://doi.org/10.1016/j.cellimm.2006.12.006}.
        	
        	\bibitem[Fachada et~al.(2009)Fachada, Lopes, and
        	Rosa]{fachada_et_al_simulating_antigenic_drift_influenza_A}
        	Nuno Fachada, Vitor~V. Lopes, and Agostinho Rosa.
        	\newblock Simulating antigenic drift and shift in influenza {A}.
        	\newblock In \emph{Proceedings of the 2009 ACM Symposium on Applied Computing},
        	pages 2093--2100, 2009.
        	
        	\bibitem[Meier-Schellersheim and
        	Mack(1999)]{meierschellersheim_and_mack_SIMMUNE}
        	M.~Meier-Schellersheim and G.~Mack.
        	\newblock {SIMMUNE}, a tool for simulating and analyzing immune system behavior
        	[pre-print].
        	\newblock \emph{arXiv}, page cs/9903017, 1999.
        	
        	\bibitem[Yakimovich et~al.(2012)Yakimovich, Gumpert, Burckhardt, L{\"u}tschg,
        	Jurgeit, Sbalzarini, and Greber]{yakimovich_et_al_cell_free_human_adenovirus}
        	Artur Yakimovich, Heidi Gumpert, Christoph~J. Burckhardt, Verena~A.
        	L{\"u}tschg, Andreas Jurgeit, Ivo~F. Sbalzarini, and Urs~F. Greber.
        	\newblock Cell-free transmission of human adenovirus by passive mass transfer
        	in cell culture simulated in a computer model.
        	\newblock \emph{Journal of Virology}, 86\penalty0 (18):\penalty0 10123--10137,
        	2012.
        	\newblock \doi{10.1128/jvi.01102-12}.
        	
        	\bibitem[Shapiro et~al.(2008)Shapiro, Duca, Lee, Delgado-Eckert, Hawkins,
        	Jarrah, Laubenbacher, Polys, Hadinoto, and
        	Thorley-Lawson]{shapiro_et_al_virtual_EBV_simulation_mechanism}
        	M~Shapiro, K~A Duca, K~Lee, E~Delgado-Eckert, J~Hawkins, A~S Jarrah,
        	R~Laubenbacher, N~F Polys, V~Hadinoto, and D~A Thorley-Lawson.
        	\newblock A virtual look at {Epstein-Barr} virus infection: simulation
        	mechanism.
        	\newblock \emph{Journal of Theoretical Biology}, 252\penalty0 (4):\penalty0
        	633--648, 2008.
        	\newblock \doi{10.1016/j.jtbi.2008.01.032}.
        	
        	\bibitem[Duca et~al.(2007)Duca, Shapiro, Delgado-Eckert, Hadinoto, Jarrah,
        	Laubenbacher, Lee, Luzuriaga, Polys, and
        	Thorley-Lawson]{duca_et_al_virtual_EBV_biological_interpretations}
        	Karen~A Duca, Michael Shapiro, Edgar Delgado-Eckert, Vey Hadinoto, Abdul~S
        	Jarrah, Reinhard Laubenbacher, Kichol Lee, Katherine Luzuriaga, Nicholas~F
        	Polys, and David~A Thorley-Lawson.
        	\newblock A virtual look at {Epstein-Barr} virus infection: biological
        	interpretations.
        	\newblock \emph{PLOS Pathogens}, 3\penalty0 (10):\penalty0 1388--1400, 2007.
        	\newblock \doi{10.1371/journal.ppat.0030137}.
        	
        	\bibitem[Moses et~al.(2021)Moses, Hofmeyr, Cannon, Andrews, Gridley, Hinga,
        	Leyba, Pribisova, Surjadidjaja, Tasnim, and
        	Forrest]{moses_et_al_lung_covid_model}
        	Melanie~E. Moses, Steven Hofmeyr, Judy~L. Cannon, Akil Andrews, Rebekah
        	Gridley, Monica Hinga, Kirtus Leyba, Abigail Pribisova, Vanessa Surjadidjaja,
        	Humayra Tasnim, and Stephanie Forrest.
        	\newblock Spatially distributed infection increases viral load in a
        	computational model of {SARS-CoV-2} lung infection.
        	\newblock \emph{PLOS Computational Biology}, 17\penalty0 (12):\penalty0
        	e1009735--, 2021.
        	
        	\bibitem[Islam et~al.(2023)Islam, Getz, Macklin, and
        	Ford~Versypt]{islam_et_al_agent_based_lung_fibrosis}
        	Mohammad~Aminul Islam, Michael Getz, Paul Macklin, and Ashlee~N. Ford~Versypt.
        	\newblock An agent-based modeling approach for lung fibrosis in response to
        	{COVID-19}.
        	\newblock \emph{PLOS Computational Biology}, 19\penalty0 (12):\penalty0 1--27,
        	2023.
        	\newblock \doi{10.1371/journal.pcbi.1011741}.
        	
        	\bibitem[Becker et~al.(2020)Becker, An, and Cockrell]{becker_et_al_CIABM}
        	Andrew Becker, Gary An, and Chase Cockrell.
        	\newblock The cellular immunity agent based model ({CIABM}): Replicating the
        	cellular immune response to viral respiratory infection [pre-print].
        	\newblock \emph{bioRxiv}, page 663930, 2020.
        	\newblock \doi{10.1101/663930}.
        	
        	\bibitem[Cockrell and
        	An(2021)]{cockrell_and_an_comparative_computational_modelling}
        	Chase Cockrell and Gary An.
        	\newblock Comparative computational modeling of the bat and human immune
        	response to viral infection with the comparative biology immune agent based
        	model.
        	\newblock \emph{Viruses}, 13\penalty0 (8), 2021.
        	\newblock \doi{10.3390/v13081620}.
        	
        	\bibitem[Ghaffarizadeh et~al.(2018)Ghaffarizadeh, Heiland, Friedman,
        	Mumenthaler, and Macklin]{ghaffarizadeh_et_al_physicell}
        	Ahmadreza Ghaffarizadeh, Randy Heiland, Samuel~H. Friedman, Shannon~M.
        	Mumenthaler, and Paul Macklin.
        	\newblock {PhysiCell: An open source physics-based cell simulator for 3-D
        		multicellular systems}.
        	\newblock \emph{PLOS Computational Biology}, 14\penalty0 (2):\penalty0
        	e1005991, 2018.
        	
        	\bibitem[Sego et~al.(2021)Sego, Aponte-Serrano, Gianlupi, and
        	Glazier]{sego_et_al_cellularisation}
        	T.~J. Sego, Josua~O. Aponte-Serrano, Juliano~F. Gianlupi, and James~A. Glazier.
        	\newblock Generation of multicellular spatiotemporal models of population
        	dynamics from ordinary differential equations, with applications in viral
        	infection.
        	\newblock \emph{BMC Biology}, 19\penalty0 (1):\penalty0 196, 2021.
        	\newblock \doi{10.1186/s12915-021-01115-z}.
        	
        	\bibitem[Sego et~al.(2022)Sego, Mochan, Ermentrout, and
        	Glazier]{sego_et_al_multicellular_influenza}
        	T~J Sego, Ericka~D Mochan, G~Bard Ermentrout, and James~A Glazier.
        	\newblock A multiscale multicellular spatiotemporal model of local influenza
        	infection and immune response.
        	\newblock \emph{Journal of Theoretical Biology}, 532:\penalty0 110918, 2022.
        	\newblock \doi{10.1016/j.jtbi.2021.110918}.
        	
        	\bibitem[Aponte-Serrano et~al.(2021)Aponte-Serrano, Weaver, Sego, Glazier, and
        	Shoemaker]{aponte_serrano_RNA_virus_replication}
        	Josua~O. Aponte-Serrano, Jordan J.~A. Weaver, T.~J. Sego, James~A. Glazier, and
        	Jason~E. Shoemaker.
        	\newblock Multicellular spatial model of {RNA} virus replication and interferon
        	responses reveals factors controlling plaque growth dynamics.
        	\newblock \emph{PLOS Computational Biology}, 17\penalty0 (10):\penalty0 1--30,
        	2021.
        	\newblock \doi{10.1371/journal.pcbi.1008874}.
        	
        	\bibitem[Ferrari~Gianlupi et~al.(2022)Ferrari~Gianlupi, Mapder, Sego, Sluka,
        	Quinney, Craig, Stratford, and
        	Glazier]{gianlupi_et_al_multiscale_model_of_antiviral_timing}
        	Juliano Ferrari~Gianlupi, Tarunendu Mapder, T~J Sego, James~P Sluka, Sara~K
        	Quinney, Morgan Craig, Robert E~Jr Stratford, and James~A Glazier.
        	\newblock Multiscale model of antiviral timing, potency, and heterogeneity
        	effects on an epithelial tissue patch infected by {SARS-CoV-2}.
        	\newblock \emph{Viruses}, 14\penalty0 (3), 2022.
        	\newblock \doi{10.3390/v14030605}.
        	
        	\bibitem[Swat et~al.(2012)Swat, Thomas, Belmonte, Shirinifard, Hmeljak, and
        	Glazier]{swat_et_al_compucell3D}
        	Maciej~H. Swat, Gilberto~L. Thomas, Julio~M. Belmonte, Abbas Shirinifard,
        	Dimitrij Hmeljak, and James~A. Glazier.
        	\newblock Chapter 13 - {M}ulti-scale modeling of tissues using {CompuCell3D}.
        	\newblock 2012.
        	
        	\bibitem[Mirams et~al.(2013)Mirams, Arthurs, Bernabeu, Bordas, Cooper, Corrias,
        	Davit, Dunn, Fletcher, Harvey, Marsh, Osborne, Pathmanathan, Pitt-Francis,
        	Southern, Zemzemi, and Gavaghan]{mirams_et_al_chaste}
        	Gary~R. Mirams, Christopher~J. Arthurs, Miguel~O. Bernabeu, Rafel Bordas,
        	Jonathan Cooper, Alberto Corrias, Yohan Davit, Sara-Jane Dunn, Alexander~G.
        	Fletcher, Daniel~G. Harvey, Megan~E. Marsh, James~M. Osborne, Pras
        	Pathmanathan, Joe Pitt-Francis, James Southern, Nejib Zemzemi, and David~J.
        	Gavaghan.
        	\newblock Chaste: An open source {C++} library for computational physiology and
        	biology.
        	\newblock \emph{PLOS Computational Biology}, 9\penalty0 (3):\penalty0 1--8,
        	2013.
        	\newblock \doi{10.1371/journal.pcbi.1002970}.
        	
        	\bibitem[Pitt-Francis et~al.(2009)Pitt-Francis, Pathmanathan, Bernabeu, Bordas,
        	Cooper, Fletcher, Mirams, Murray, Osborne, Walter, Chapman, Garny, {van
        		Leeuwen}, Maini, Rodr{\'\i}guez, Waters, Whiteley, Byrne, and
        	Gavaghan]{pitt_francis_et_al_Chaste}
        	Joe Pitt-Francis, Pras Pathmanathan, Miguel~O. Bernabeu, Rafel Bordas, Jonathan
        	Cooper, Alexander~G. Fletcher, Gary~R. Mirams, Philip Murray, James~M.
        	Osborne, Alex Walter, S.~Jon Chapman, Alan Garny, Ingeborg~M.M. {van
        		Leeuwen}, Philip~K. Maini, Blanca Rodr{\'\i}guez, Sarah~L. Waters,
        	Jonathan~P. Whiteley, Helen~M. Byrne, and David~J. Gavaghan.
        	\newblock Chaste: A test-driven approach to software development for biological
        	modelling.
        	\newblock \emph{Computer Physics Communications}, 180\penalty0 (12):\penalty0
        	2452--2471, 2009.
        	\newblock \doi{https://doi.org/10.1016/j.cpc.2009.07.019}.
        	
        	\bibitem[Starru{\ss} et~al.(2014)Starru{\ss}, de~Back, Brusch, and
        	Deutsch]{staruss_et_al_morpheus}
        	J{\"o}rn Starru{\ss}, Walter de~Back, Lutz Brusch, and Andreas Deutsch.
        	\newblock {Morpheus: a user-friendly modeling environment for multiscale and
        		multicellular systems biology}.
        	\newblock \emph{Bioinformatics}, 30\penalty0 (9):\penalty0 1331--1332, 2014.
        	\newblock \doi{10.1093/bioinformatics/btt772}.
        	
        	\bibitem[Lavigne et~al.(2021)Lavigne, Russell, Sherry, and
        	Ke]{lavigne_et_al_interferon_signalling_ring_vaccination}
        	Michael~G. Lavigne, Hayley Russell, Barbara Sherry, and Ruian Ke.
        	\newblock {Autocrine and paracrine interferon signalling as `ring vaccination'
        		and `contact tracing' strategies to suppress virus infection in a host}.
        	\newblock \emph{Proceedings of the Royal Society of London. Series B:
        		Biological Sciences}, 288\penalty0 (1945):\penalty0 20203002, 2021.
        	\newblock \doi{10.1098/rspb.2020.3002}.
        	
        	\bibitem[Rodriguez-Brenes et~al.(2017)Rodriguez-Brenes, Hofacre, Fan, and
        	Wodarz]{rodriguez_brenes_et_al_virus_spread_from_low_moi}
        	Ignacio~A. Rodriguez-Brenes, Andrew Hofacre, Hung Fan, and Dominik Wodarz.
        	\newblock Complex dynamics of virus spread from low infection multiplicities:
        	Implications for the spread of oncolytic viruses.
        	\newblock \emph{PLOS Computational Biology}, 13\penalty0 (1):\penalty0
        	e1005241--, 2017.
        	
        	\bibitem[Pandey(1991)]{pandey_1991_CA_interacting_cellular_network}
        	R.B. Pandey.
        	\newblock Cellular automata approach to interacting cellular network models for
        	the dynamics of cell population in an early {HIV} infection.
        	\newblock \emph{Physica A: Statistical Mechanics and its Applications},
        	179\penalty0 (3):\penalty0 442--470, 1991.
        	\newblock \doi{https://doi.org/10.1016/0378-4371(91)90088-T}.
        	
        	\bibitem[Pandey and Stauffer(1990)]{pandey_and_stauffer_metastability}
        	R.~B. Pandey and D.~Stauffer.
        	\newblock Metastability with probabilistic cellular automata in an {HIV}
        	infection.
        	\newblock \emph{Journal of Statistical Physics}, 61\penalty0 (1):\penalty0
        	235--240, 1990.
        	\newblock \doi{10.1007/BF01013962}.
        	
        	\bibitem[Mielke and Pandey(1998)]{mielke_and_pandey_fuzzy_interaction}
        	Aaron Mielke and R.B Pandey.
        	\newblock A computer simulation study of cell population in a fuzzy interaction
        	model for mutating {HIV}.
        	\newblock \emph{Physica A: Statistical Mechanics and its Applications},
        	251\penalty0 (3):\penalty0 430--438, 1998.
        	\newblock \doi{https://doi.org/10.1016/S0378-4371(97)00576-1}.
        	
        	\bibitem[Mannion et~al.(2000{\natexlab{a}})Mannion, Ruskin, Pandey, and
        	Mannion]{mannion_et_al_effect_of_mutation}
        	Rachel Mannion, Heather Ruskin, R.B. Pandey, and Rachel Mannion.
        	\newblock Effect of mutation on helper {T}-cells and viral population: A
        	computer simulation model for {HIV}.
        	\newblock \emph{Theory in Biosciences}, 119\penalty0 (1):\penalty0 10--19,
        	2000{\natexlab{a}}.
        	\newblock \doi{https://doi.org/10.1078/1431-7613-00002}.
        	
        	\bibitem[Mannion et~al.(2000{\natexlab{b}})Mannion, Ruskin, and
        	Pandey]{mannion_et_al_monte_carlo_population_dynamics}
        	Rachel Mannion, Heather~J. Ruskin, and R.B. Pandey.
        	\newblock A {Monte Carlo} approach to population dynamics of cells in a {HIV}
        	immune response model.
        	\newblock \emph{Theory in Biosciences}, 119\penalty0 (2):\penalty0 145--155,
        	2000{\natexlab{b}}.
        	\newblock \doi{https://doi.org/10.1078/1431-7613-00006}.
        	
        	\bibitem[Kougias and
        	Schulte(1990)]{kougias_and_schulte_simulating_immune_response_to_HIV}
        	Ch.~F. Kougias and J.~Schulte.
        	\newblock Simulating the immune response to the {HIV-1} virus with cellular
        	automata.
        	\newblock \emph{Journal of Statistical Physics}, 60\penalty0 (1):\penalty0
        	263--273, 1990.
        	\newblock \doi{10.1007/BF01013677}.
        	
        	\bibitem[Wodarz and Levy(2011)]{wodarz_and_levy_hiv_two_modes_of_spread}
        	Dominik Wodarz and David~N. Levy.
        	\newblock Effect of different modes of viral spread on the dynamics of multiply
        	infected cells in human immunodeficiency virus infection.
        	\newblock \emph{Journal of the Royal Society Interface}, 8\penalty0
        	(55):\penalty0 289--300, 2011.
        	\newblock \doi{10.1098/rsif.2010.0266}.
        	
        	\bibitem[Reis et~al.(2010)Reis, Pacheco, Ennis, and
        	Dingli]{reis_et_al_in_silico_evolutionary_tumour_virotherapy}
        	Carlos~L Reis, Jorge~M Pacheco, Matthew~K Ennis, and David Dingli.
        	\newblock In silico evolutionary dynamics of tumour virotherapy.
        	\newblock \emph{Integrative Biology (Cambridge)}, 2\penalty0 (1):\penalty0
        	41--45, 2010.
        	\newblock \doi{10.1039/b917597k}.
        	
        	\bibitem[Bankhead~III et~al.(2011)Bankhead~III, Mancini, Sims, Baric, McWeeney,
        	and Sloot]{bankhead_et_al_simulation_framework_SARS}
        	Armand Bankhead~III, Emiliano Mancini, Amy~C Sims, Ralph~S Baric, Shannon
        	McWeeney, and Peter M~A Sloot.
        	\newblock A simulation framework to investigate in vitro viral infection
        	dynamics.
        	\newblock \emph{Procedia Computer Science}, 4:\penalty0 1798--1807, 2011.
        	\newblock \doi{10.1016/j.procs.2011.04.195}.
        	
        	\bibitem[Jafelice et~al.(2009)Jafelice, Bechara, Barros, Bassanezi, and
        	Gomide]{jafelice_et_al_CA_with_fuzzy_parameters}
        	R.~Motta Jafelice, B.F.Z. Bechara, L.C. Barros, R.C. Bassanezi, and F.~Gomide.
        	\newblock Cellular automata with fuzzy parameters in microscopic study of
        	positive {HIV} individuals.
        	\newblock \emph{Mathematical and Computer Modelling}, 50\penalty0 (1):\penalty0
        	32--44, 2009.
        	\newblock \doi{https://doi.org/10.1016/j.mcm.2009.01.008}.
        	
        	\bibitem[Immonen et~al.(2012)Immonen, Gibson, Leitner, Miller, Arts, Somersalo,
        	and Calvetti]{immonem_et_al_hybrid_stochastic_deterministic}
        	Taina Immonen, Richard Gibson, Thomas Leitner, Melanie~A Miller, Eric~J Arts,
        	Erkki Somersalo, and Daniela Calvetti.
        	\newblock A hybrid stochastic-deterministic computational model accurately
        	describes spatial dynamics and virus diffusion in {HIV-1} growth competition
        	assay.
        	\newblock \emph{Journal of Theoretical Biology}, 312:\penalty0 120--132, 2012.
        	\newblock \doi{10.1016/j.jtbi.2012.07.005}.
        	
        	\bibitem[Morselli et~al.(2023)Morselli, Delitala, and
        	Frascoli]{morselli_et_al_agent_based_and_continuum_models}
        	David Morselli, Marcello~Edoardo Delitala, and Federico Frascoli.
        	\newblock Agent-based and continuum models for spatial dynamics of infection by
        	oncolytic viruses.
        	\newblock \emph{Bulletin of Mathematical Biology}, 85\penalty0 (10):\penalty0
        	92, 2023.
        	\newblock \doi{10.1007/s11538-023-01192-x}.
        	
        	\bibitem[Jarrah et~al.(2004)Jarrah, Vastani, Duca, and
        	Laubenbacher]{jarrah_et_al_optimal_control_in_vitro_virus}
        	A.~Jarrah, H.~Vastani, K.~Duca, and R.~Laubenbacher.
        	\newblock An optimal control problem for in vitro virus competition.
        	\newblock In \emph{2004 43rd IEEE Conference on Decision and Control (CDC)},
        	pages 579--584, 2004.
        	
        	\bibitem[Precharattana et~al.(2010)Precharattana, Triampo, Modchang, Triampo,
        	and Lenbury]{precharattana_et_al_investigation_of_spatial_pattern_formation}
        	Monamorn Precharattana, Wannapong Triampo, Charin Modchang, Darapond Triampo,
        	and Yongwimon Lenbury.
        	\newblock Investigation of spatial pattern formation involving {CD4+} {T} cells
        	in {HIV/AIDS} dynamics by a stochastic cellular automata model.
        	\newblock \emph{International Journal of Mathematics and Computers in
        		Simulation}, 4, 2010.
        	
        	\bibitem[Macklin et~al.(2016)Macklin, Frieboes, Sparks, Ghaffarizadeh,
        	Friedman, Juarez, Jonckheere, and
        	Mumenthaler]{macklin_et_al_multicellular_review}
        	Paul Macklin, Hermann~B. Frieboes, Jessica~L. Sparks, Ahmadreza Ghaffarizadeh,
        	Samuel~H. Friedman, Edwin~F. Juarez, Edmond Jonckheere, and Shannon~M.
        	Mumenthaler.
        	\newblock Progress towards computational {3-D} multicellular systems biology.
        	\newblock 2016.
        	
        	\bibitem[Fletcher and Osborne(2022)]{fletcher_and_osborne_seven_challenges}
        	Alexander~G. Fletcher and James~M. Osborne.
        	\newblock Seven challenges in the multiscale modeling of multicellular tissues.
        	\newblock \emph{WIREs Mechanisms of Disease}, 14\penalty0 (1), 2022.
        	\newblock \doi{10.1002/wsbm.1527}.
        	
        	\bibitem[Okuda et~al.(2015)Okuda, Inoue, and
        	Adachi]{okuda_et_al_3D_vertex_model}
        	Satoru Okuda, Yasuhiro Inoue, and Taiji Adachi.
        	\newblock Three-dimensional vertex model for simulating multicellular
        	morphogenesis.
        	\newblock \emph{Biophysics and physicobiology}, 12:\penalty0 13 -- 20, 2015.
        	\newblock \doi{10.2142/biophysico.12.0\_13}.
        	
        	\bibitem[Paiva et~al.(2009)Paiva, Binny, Ferreira, and
        	Martins]{paiva_et_al_multiscale_oncolytic_virotherapy}
        	Leticia~R. Paiva, Christopher Binny, Silvio~C. Ferreira, and Marcelo~L.
        	Martins.
        	\newblock A multiscale mathematical model for oncolytic virotherapy.
        	\newblock \emph{Cancer Research}, 2009.
        	\newblock \doi{10.1158/0008-5472.CAN-08-2173}.
        	
        	\bibitem[Garrido~Zornoza et~al.(2024)Garrido~Zornoza, Mitarai, and
        	Haerter]{garrido_zornoza_et_al_stochastic_microbial_dispersal}
        	Miguel Garrido~Zornoza, Namiko Mitarai, and Jan~O. Haerter.
        	\newblock Stochastic microbial dispersal drives local extinction and global
        	diversity.
        	\newblock \emph{Royal Society Open Science}, 11\penalty0 (5):\penalty0 231301,
        	2024.
        	\newblock \doi{10.1098/rsos.231301}.
        	
        	\bibitem[Cangelosi et~al.(2017)Cangelosi, Means, and
        	Ho]{cangelosi_et_al_multiscale_HBV}
        	Quentin Cangelosi, Shawn~A. Means, and Harvey Ho.
        	\newblock A multi-scale spatial model of hepatitis-{B} viral dynamics.
        	\newblock \emph{PLOS ONE}, 12\penalty0 (12):\penalty0 1--28, 2017.
        	\newblock \doi{10.1371/journal.pone.0188209}.
        	
        	\bibitem[Shi et~al.(2008)Shi, Tridane, and
        	Kuang]{shi_et_al_viral_load_based_HIV}
        	Veronica Shi, Abdessamad Tridane, and Yang Kuang.
        	\newblock A viral load-based cellular automata approach to modeling {HIV}
        	dynamics and drug treatment.
        	\newblock \emph{Journal of Theoretical Biology}, 253\penalty0 (1):\penalty0
        	24--35, 2008.
        	\newblock \doi{10.1016/j.jtbi.2007.11.005}.
        	
        	\bibitem[Howat et~al.(2006)Howat, Barreca, O'Hare, Gog, and
        	Grenfell]{howat_et_al_modelling_dynamics_of_type_I_IFN}
        	Tom~J Howat, Cristina Barreca, Peter O'Hare, Julia~R Gog, and Bryan~T Grenfell.
        	\newblock Modelling dynamics of the type {I} interferon response to in vitro
        	viral infection.
        	\newblock \emph{Journal of the Royal Society Interface}, 3\penalty0
        	(10):\penalty0 699--709, 2006.
        	\newblock \doi{10.1098/rsif.2006.0136}.
        	
        	\bibitem[Durso-Cain et~al.(2021)Durso-Cain, Kumberger, Sch{\"a}lte, Fink,
        	Dahari, Hasenauer, Uprichard, and Graw]{durso_cain_hcv_dual_spread}
        	Karina Durso-Cain, Peter Kumberger, Yannik Sch{\"a}lte, Theresa Fink, Harel
        	Dahari, Jan Hasenauer, Susan~L Uprichard, and Frederik Graw.
        	\newblock {HCV} spread kinetics reveal varying contributions of transmission
        	modes to infection dynamics.
        	\newblock \emph{Viruses}, 13\penalty0 (7), 2021.
        	\newblock \doi{10.3390/v13071308}.
        	
        	\bibitem[Cai et~al.(2023)Cai, Zhao, and
        	Zhuge]{cai_et_al_spatial_dynamics_of_immune_repsonse}
        	Yanan Cai, Zhongrui Zhao, and Changjing Zhuge.
        	\newblock The spatial dynamics of immune response upon virus infection through
        	hybrid dynamical computational model.
        	\newblock \emph{Frontiers in Immunology}, 14, 2023.
        	\newblock \doi{10.3389/fimmu.2023.1257953}.
        	
        	\bibitem[Aristotelous et~al.(2022)Aristotelous, Chen, and
        	Forest]{aristotelous_et_al_covid_model}
        	Andreas~C Aristotelous, Alex Chen, and M~Gregory Forest.
        	\newblock A hybrid discrete--continuum model of immune responses to
        	{SARS-CoV-2} infection in the lung alveolar region, with a focus on
        	interferon induced innate response.
        	\newblock \emph{Journal of Theoretical Biology}, 555:\penalty0 111293, 2022.
        	\newblock \doi{10.1016/j.jtbi.2022.111293}.
        	
        	\bibitem[Chen et~al.(2022)Chen, Wessler, Daftari, Hinton, Boucher, Pickles,
        	Freeman, Lai, and Forest]{chen_sars_cov_2_in_lung}
        	Alexander Chen, Timothy Wessler, Katherine Daftari, Kameryn Hinton, Richard~C
        	Boucher, Raymond Pickles, Ronit Freeman, Samuel~K Lai, and M~Gregory Forest.
        	\newblock Modeling insights into {SARS-CoV-2} respiratory tract infections
        	prior to immune protection.
        	\newblock \emph{Biophysical Journal}, 121\penalty0 (9):\penalty0 1619--1631,
        	2022.
        	\newblock \doi{10.1016/j.bpj.2022.04.003}.
        	
        	\bibitem[Chen et~al.(2023)Chen, Wessler, and
        	Forest]{chen_et_al_antibody_protection}
        	Alex Chen, Timothy Wessler, and Gregory~M Forest.
        	\newblock Antibody protection from {SARS-CoV-2} respiratory tract exposure and
        	infection.
        	\newblock \emph{Journal of Theoretical Biology}, 557:\penalty0 111334, 2023.
        	\newblock \doi{10.1016/j.jtbi.2022.111334}.
        	
        	\bibitem[Zhang et~al.(2024)Zhang, Cao, Medlin, Pearson, Aristotelous, Chen,
        	Wessler, and
        	Forest]{zhang_et_al_computational_modelling_extereme_heterogeneity}
        	Leyi Zhang, Han Cao, Karen Medlin, Jason Pearson, Andreas~C. Aristotelous,
        	Alexander Chen, Timothy Wessler, and M.~Gregory Forest.
        	\newblock Computational modeling insights into extreme heterogeneity in
        	{COVID-19} nasal swab data.
        	\newblock \emph{Viruses}, 16\penalty0 (1), 2024.
        	\newblock \doi{10.3390/v16010069}.
        	
        	\bibitem[Pearson et~al.(2023)Pearson, Wessler, Chen, Boucher, Freeman, Lai,
        	Pickles, and Forest]{pearson_et_al_modelling_variability_in_sars_cov_2}
        	Jason Pearson, Timothy Wessler, Alex Chen, Richard~C. Boucher, Ronit Freeman,
        	Samuel~K. Lai, Raymond Pickles, and M.~Gregory Forest.
        	\newblock Modeling identifies variability in {SARS-CoV-2} uptake and eclipse
        	phase by infected cells as principal drivers of extreme variability in nasal
        	viral load in the 48 h post infection.
        	\newblock \emph{Journal of Theoretical Biology}, 565:\penalty0 111470, 2023.
        	\newblock \doi{https://doi.org/10.1016/j.jtbi.2023.111470}.
        	
        	\bibitem[Zhang et~al.(2023)Zhang, Cao, Medlin, Pearson, Aristotelous, Chen,
        	Wessler, and Forest]{zhang_et_al_global_sensitivity_analysis}
        	Leyi Zhang, Han Cao, Karen Medlin, Jason Pearson, Andreas Aristotelous,
        	Alexander Chen, Timothy Wessler, and M.~Gregory Forest.
        	\newblock Global sensitivity analysis of the onset of nasal passage infection
        	by {SARS-CoV-2} with respect to heterogeneity in host physiology and host
        	cell-virus kinetic interactions [pre-print].
        	\newblock \emph{bioRxiv}, page 2023.11.04.565660, 2023.
        	
        	\bibitem[Seich Al~Basatena et~al.(2013)Seich Al~Basatena, Chatzimichalis, Graw,
        	Frost, Regoes, and Asquith]{al_basatena_et_al_non_lytic_CD8}
        	Nafisa-Katrin Seich Al~Basatena, Konstantinos Chatzimichalis, Frederik Graw,
        	Simon D~W Frost, Roland~R Regoes, and Becca Asquith.
        	\newblock Can non-lytic {CD8+} {T} cells drive {HIV-1} escape?
        	\newblock \emph{PLoS Pathogens}, 9\penalty0 (11):\penalty0 e1003656, 2013.
        	\newblock \doi{10.1371/journal.ppat.1003656}.
        	
        	\bibitem[Jenner et~al.(2020)Jenner, Frascoli, Coster, and
        	Kim]{jenner_et_al_oncolytic_voronoi}
        	Adrianne~L Jenner, Federico Frascoli, Adelle C~F Coster, and Peter~S Kim.
        	\newblock Enhancing oncolytic virotherapy: Observations from a voronoi
        	cell-based model.
        	\newblock \emph{Journal of Theoretical Biology}, 485:\penalty0 110052, 2020.
        	\newblock \doi{10.1016/j.jtbi.2019.110052}.
        	
        	\bibitem[Itakura et~al.(2010)Itakura, Kurosaki, Itakura, Maekawa, Asahina,
        	Izumi, and
        	Enomoto]{itakura_et_al_reproducibility_chronic_virus_infection_model}
        	Jun Itakura, Masayuki Kurosaki, Yoshie Itakura, Sinya Maekawa, Yasuhiro
        	Asahina, Namiki Izumi, and Nobuyuki Enomoto.
        	\newblock Reproducibility and usability of chronic virus infection model using
        	agent-based simulation; comparing with a mathematical model.
        	\newblock \emph{Biosystems}, 99\penalty0 (1):\penalty0 70--78, 2010.
        	\newblock \doi{10.1016/j.biosystems.2009.09.001}.
        	
        	\bibitem[Graner and Glazier(1992)]{graner_and_glazier_CPM}
        	Francois Graner and James~A. Glazier.
        	\newblock Simulation of biological cell sorting using a two-dimensional
        	extended {Potts} model.
        	\newblock \emph{Physical Review Letters}, 69:\penalty0 2013--2016, 1992.
        	\newblock \doi{10.1103/PhysRevLett.69.2013}.
        	
        	\bibitem[Beauchemin et~al.(2006)Beauchemin, Forrest, and
        	Koster]{beauchemin_et_al_modeling_influenza_in_tissue}
        	Catherine Beauchemin, Stephanie Forrest, and Frederick~T. Koster.
        	\newblock Modeling influenza viral dynamics in tissue.
        	\newblock In Hugues Bersini and Jorge Carneiro, editors, \emph{Artificial
        		Immune Systems}, pages 23--36, 2006.
        	
        	\bibitem[Whitman et~al.(2020)Whitman, Dhanji, Hayot, Sealfon, and
        	Jayaprakash]{whitman_et_al_spatiotemporal_host_virus_influeza}
        	John Whitman, Aleya Dhanji, Fernand Hayot, Stuart~C. Sealfon, and Ciriyam
        	Jayaprakash.
        	\newblock Spatio-temporal dynamics of host-virus competition: A model study of
        	influenza {A}.
        	\newblock \emph{Journal of Theoretical Biology}, 484:\penalty0 110026, 2020.
        	\newblock \doi{10.1016/j.jtbi.2019.110026}.
        	
        	\bibitem[Jenner et~al.(2022)Jenner, Smalley, Goldman, Goins, Cobbs, Puchalski,
        	Chiocca, Lawler, Macklin, Goldman, and
        	Craig]{jenner_et_al_agent_based_glioblastoma}
        	Adrianne~L Jenner, Munisha Smalley, David Goldman, William~F Goins, Charles~S
        	Cobbs, Ralph~B Puchalski, E~Antonio Chiocca, Sean Lawler, Paul Macklin, Aaron
        	Goldman, and Morgan Craig.
        	\newblock Agent-based computational modeling of glioblastoma predicts that
        	stromal density is central to oncolytic virus efficacy.
        	\newblock \emph{iScience}, 25\penalty0 (6):\penalty0 104395, 2022.
        	\newblock \doi{10.1016/j.isci.2022.104395}.
        	
        	\bibitem[Williams et~al.(2023)Williams, McCaw, and
        	Osborne]{williams_et_al_spatial_discretisation}
        	Thomas Williams, James~M. McCaw, and James~M. Osborne.
        	\newblock Choice of spatial discretisation influences the progression of viral
        	infection within multicellular tissues.
        	\newblock \emph{Journal of Theoretical Biology}, 573:\penalty0 111592, 2023.
        	\newblock \doi{https://doi.org/10.1016/j.jtbi.2023.111592}.
        	
        	\bibitem[Perelson(2002)]{perelson_hiv_review}
        	Alan~S Perelson.
        	\newblock Modelling viral and immune system dynamics.
        	\newblock \emph{Nature Reviews Immunology}, 2\penalty0 (1):\penalty0 28--36,
        	2002.
        	\newblock \doi{10.1038/nri700}.
        	
        	\bibitem[Nowak and May(2000)]{nowak_and_may_virus_dynamics_book}
        	Martin Nowak and Robert~M. May.
        	\newblock \emph{Virus Dynamics: Mathematical Principles of Immunology and
        		Virology}.
        	\newblock Oxford University Press, 2000.
        	
        	\bibitem[Komarova et~al.(2013)Komarova, Anghelina, Voznesensky, Trinit{\'e},
        	Levy, and Wodarz]{komarova_et_al_relative_contribution_CC_in_hiv}
        	Natalia~L Komarova, Daniela Anghelina, Igor Voznesensky, Benjamin Trinit{\'e},
        	David~N Levy, and Dominik Wodarz.
        	\newblock Relative contribution of free-virus and synaptic transmission to the
        	spread of {HIV}-1 through target cell populations.
        	\newblock \emph{Biology Letters}, 9\penalty0 (1):\penalty0 20121049, 2013.
        	\newblock \doi{10.1098/rsbl.2012.1049}.
        	
        	\bibitem[Raza et~al.(2023)Raza, Arshed, Bakar, Shahzad, and
        	Inc]{raza_et_al_numerical_efficient_splitting_HIV}
        	Nauman Raza, Saima Arshed, Abu Bakar, Aamir Shahzad, and Mustafa Inc.
        	\newblock A numerical efficient splitting method for the solution of {HIV} time
        	periodic reaction--diffusion model having spatial heterogeneity.
        	\newblock \emph{Physica A: Statistical Mechanics and its Applications},
        	609:\penalty0 128385, 2023.
        	\newblock \doi{https://doi.org/10.1016/j.physa.2022.128385}.
        	
        	\bibitem[Marinho et~al.(2012)Marinho, Bacelar, and
        	Andrade]{marinho_et_al_PDE_HIV}
        	E.B.S. Marinho, F.S. Bacelar, and R.F.S. Andrade.
        	\newblock A model of partial differential equations for {HIV} propagation in
        	lymph nodes.
        	\newblock \emph{Physica A: Statistical Mechanics and its Applications},
        	391\penalty0 (1):\penalty0 132--141, 2012.
        	\newblock \doi{https://doi.org/10.1016/j.physa.2011.08.023}.
        	
        	\bibitem[Benyoussef et~al.(2003)Benyoussef, HafidAllah, ElKenz, Ez-Zahraouy,
        	and Loulidi]{benyoussef_et_al_dynamics_of_HIV_infection_2D}
        	A~Benyoussef, N.El HafidAllah, A~ElKenz, H~Ez-Zahraouy, and M~Loulidi.
        	\newblock Dynamics of {HIV} infection on {2D} cellular automata.
        	\newblock \emph{Physica A: Statistical Mechanics and its Applications},
        	322:\penalty0 506--520, 2003.
        	\newblock \doi{https://doi.org/10.1016/S0378-4371(02)01915-5}.
        	
        	\bibitem[Iwami et~al.(2015)Iwami, Takeuchi, Nakaoka, Mammano, Clavel, Inaba,
        	Kobayashi, Misawa, Aihara, Koyanagi, and Sato]{iwami_et_al_HIV_CC_estimate}
        	Shingo Iwami, Junko~S Takeuchi, Shinji Nakaoka, Fabrizio Mammano, Fran{\c c}ois
        	Clavel, Hisashi Inaba, Tomoko Kobayashi, Naoko Misawa, Kazuyuki Aihara,
        	Yoshio Koyanagi, and Kei Sato.
        	\newblock Cell-to-cell infection by {HIV} contributes over half of virus
        	infection.
        	\newblock \emph{eLife}, 4, 2015.
        	\newblock \doi{10.7554/eLife.08150}.
        	
        	\bibitem[Sloot et~al.(2002)Sloot, Chen, and
        	Boucher]{sloot_et_al_CA_drug_therapy_HIV}
        	Peter Sloot, Fan Chen, and Charles Boucher.
        	\newblock Cellular automata model of drug therapy for {HIV} infection.
        	\newblock In Stefania Bandini, Bastien Chopard, and Marco Tomassini, editors,
        	\emph{Cellular Automata}, pages 282--293, 2002.
        	
        	\bibitem[Bhatt et~al.(2022)Bhatt, Janzen, Daemen, and
        	Weissing]{bhatt_et_al_modelling_the_spatial_dynamics_oncolytic_virotherapy}
        	Darshak~Kartikey Bhatt, Thijs Janzen, Toos Daemen, and Franz~J. Weissing.
        	\newblock Modelling the spatial dynamics of oncolytic virotherapy in the
        	presence of virus-resistant tumour cells.
        	\newblock \emph{PLOS Computational Biology}, 18\penalty0 (12):\penalty0 1--21,
        	2022.
        	\newblock \doi{10.1371/journal.pcbi.1010076}.
        	
        	\bibitem[Grebennikov and
        	Bocharov(2019)]{grebennikov_and_bocharov_spatially_resolved_modelling}
        	Dmitry~S. Grebennikov and Gennady~A. Bocharov.
        	\newblock Spatially resolved modelling of immune responses following a
        	multiscale approach: from computational implementation to quantitative
        	predictions.
        	\newblock \emph{Russian Journal of Numerical Analysis and Mathematical
        		Modelling}, 34\penalty0 (5):\penalty0 253--260, 2019.
        	\newblock \doi{doi:10.1515/rnam-2019-0021}.
        	
        	\bibitem[Lawler et~al.(2017)Lawler, Speranza, Cho, and
        	Chiocca]{lawler_et_al_oncolytic_viruses_review}
        	Sean~E. Lawler, Maria-Carmela Speranza, Choi-Fong Cho, and E.~Antonio Chiocca.
        	\newblock Oncolytic viruses in cancer treatment: A review.
        	\newblock \emph{JAMA Oncology}, 3\penalty0 (6):\penalty0 841--849, 2017.
        	\newblock \doi{10.1001/jamaoncol.2016.2064}.
        	
        	\bibitem[Jebar et~al.(2015)Jebar, Errington-Mais, Vile, Selby, Melcher, and
        	Griffin]{jebar_et_al_clinical_oncolytics_hepatocellular}
        	Adel~H. Jebar, Fiona Errington-Mais, Richard~G. Vile, Peter~J. Selby, Alan~A.
        	Melcher, and Stephen Griffin.
        	\newblock Progress in clinical oncolytic virus-based therapy for hepatocellular
        	carcinoma.
        	\newblock \emph{Journal of General Virology}, 96\penalty0 (7):\penalty0
        	1533--1550, 2015.
        	\newblock \doi{https://doi.org/10.1099/vir.0.000098}.
        	
        	\bibitem[{Al-Johani} et~al.(2019){Al-Johani}, Simbawa, and
        	{Al-Tuwairqi}]{al_johani_et_al_spatiotemporal_dynamics_virotherapy}
        	Najwa {Al-Johani}, Eman Simbawa, and Salma {Al-Tuwairqi}.
        	\newblock Modeling the spatiotemporal dynamics of virotherapy and immune
        	response as a treatment for cancer.
        	\newblock \emph{Communications in Mathematical Biology and Neuroscience}, 2019.
        	
        	\bibitem[Simbawa et~al.(2020)Simbawa, Al-Johani, and
        	Al-Tuwairqi]{simbabwa_et_al_spatiotemporal_dynamics_oncolytic_radiovirotherapy}
        	Eman Simbawa, Najwa Al-Johani, and Salma Al-Tuwairqi.
        	\newblock Modeling the spatiotemporal dynamics of oncolytic viruses and
        	radiotherapy as a treatment for cancer.
        	\newblock \emph{Computational and Mathematical Methods in Medicine},
        	2020:\penalty0 3642654, 2020.
        	\newblock \doi{10.1155/2020/3642654}.
        	
        	\bibitem[Pooladvand et~al.(2021)Pooladvand, Yun, Yoon, Kim, and
        	Frascoli]{pooladvand_et_al_viral_infectivity_in_oncolytic_virotherapy_outcomes}
        	Pantea Pooladvand, Chae-Ok Yun, A.-Rum Yoon, Peter~S. Kim, and Federico
        	Frascoli.
        	\newblock The role of viral infectivity in oncolytic virotherapy outcomes: A
        	mathematical study.
        	\newblock \emph{Mathematical Biosciences}, 334:\penalty0 108520, 2021.
        	\newblock \doi{https://doi.org/10.1016/j.mbs.2020.108520}.
        	
        	\bibitem[Storey and
        	Jackson(2021)]{storey_and_jackson_ABM_combination_oncolytic}
        	Kathleen~M. Storey and Trachette~L. Jackson.
        	\newblock An agent-based model of combination oncolytic viral therapy and
        	anti-{PD}-1 immunotherapy reveals the importance of spatial location when
        	treating glioblastoma.
        	\newblock \emph{Cancers}, 13\penalty0 (21), 2021.
        	\newblock \doi{10.3390/cancers13215314}.
        	
        	\bibitem[Berg et~al.(2019)Berg, Offord, Kemler, Ennis, Chang, Paulik, Bajzer,
        	Neuhauser, and
        	Dingli]{berg_et_al_in_vitro_in_silico_multidimensional_oncolytic}
        	David~R. Berg, Chetan~P. Offord, Iris Kemler, Matthew~K. Ennis, Lawrence Chang,
        	George Paulik, Zeljko Bajzer, Claudia Neuhauser, and David Dingli.
        	\newblock In vitro and in silico multidimensional modeling of oncolytic tumor
        	virotherapy dynamics.
        	\newblock \emph{PLOS Computational Biology}, 15\penalty0 (3):\penalty0 1--18,
        	2019.
        	\newblock \doi{10.1371/journal.pcbi.1006773}.
        	
        	\bibitem[Makevnina(2018)]{makevnina_lung_atlas}
        	V~V Makevnina.
        	\newblock Solid-state modeling of human tracheobronchial tree for 23
        	generations of airways.
        	\newblock \emph{Journal of Physics: Conference Series}, 1124\penalty0
        	(3):\penalty0 031002, 2018.
        	\newblock \doi{10.1088/1742-6596/1124/3/031002}.
        	
        	\bibitem[Ayadi et~al.(2024)Ayadi, Frydman, and
        	Le]{ayadi_et_al_modelling_and_simulation_SARS_CoV_2}
        	Ali Ayadi, Claudia Frydman, and Quy~Thanh Le.
        	\newblock Modeling and simulation of the {SARS-CoV-2} lung infection and immune
        	response with {Cell-Devs}.
        	\newblock In \emph{Proceedings of the Winter Simulation Conference}, pages
        	1196--1207, 2024.
        	
        	\bibitem[Williams et~al.(2024{\natexlab{a}})Williams, McCaw, and
        	Osborne]{williams_et_al_lung}
        	Thomas Williams, James~M. McCaw, and James~M. Osborne.
        	\newblock Accounting for the geometry of the lung in respiratory viral
        	infections [pre-print].
        	\newblock \emph{arXiv}, page 2408.07618, 2024{\natexlab{a}}.
        	
        	\bibitem[Chakravarty et~al.(2023)Chakravarty, Panchagnula, and
        	Patankar]{chakravarty_et_al_virus_loaded_droplets}
        	Aranyak Chakravarty, Mahesh~V. Panchagnula, and Neelesh~A. Patankar.
        	\newblock Inhalation of virus-loaded droplets as a clinically plausible pathway
        	to deep lung infection.
        	\newblock \emph{Frontiers in Physiology}, 14, 2023.
        	\newblock \doi{10.3389/fphys.2023.1073165}.
        	
        	\bibitem[Mitarai et~al.(2016)Mitarai, Brown, and
        	Sneppen]{mitarai_et_al_population_dynamics_phage}
        	Namiko Mitarai, Stanley Brown, and Kim Sneppen.
        	\newblock Population dynamics of phage and bacteria in spatially structured
        	habitats using phage $\lambda$ and escherichia coli.
        	\newblock \emph{Journal of Bacteriology}, 198\penalty0 (12):\penalty0
        	1783--1793, 2016.
        	\newblock \doi{10.1128/jb.00965-15}.
        	
        	\bibitem[Valdez et~al.(2024)Valdez, Rodriguez-Roman, Sun, Levin, Weiss, and
        	Aranson]{valdez_et_al_heterogeneous_bacteria-phage}
        	Andres Valdez, Eduardo Rodriguez-Roman, Paul Sun, Bruce Levin, Howard Weiss,
        	and Igor Aranson.
        	\newblock Heterogeneous bacteria-phage competition generates complex colony
        	shapes [sneak peek].
        	\newblock \emph{iScience}, 2024.
        	
        	\bibitem[Mart{\'\i}nez-Calvo et~al.(2023)Mart{\'\i}nez-Calvo, Wingreen, and
        	Datta]{martinez-calvo_et_al_pattern_formation_bacteria_phage}
        	Alejandro Mart{\'\i}nez-Calvo, Ned~S Wingreen, and Sujit~S Datta.
        	\newblock Pattern formation by bacteria-phage interactions [pre-print].
        	\newblock \emph{bioRxiv}, page 2023.09.19.558479, 2023.
        	
        	\bibitem[Taylor et~al.(2017)Taylor, Penington, and
        	Weitz]{taylor_et_al_emergence_multiple_infections}
        	Bradford~P Taylor, Catherine~J Penington, and Joshua~S Weitz.
        	\newblock Emergence of increased frequency and severity of multiple infections
        	by viruses due to spatial clustering of hosts.
        	\newblock \emph{Phys Biol}, 13\penalty0 (6):\penalty0 066014, 2017.
        	\newblock \doi{10.1088/1478-3975/13/6/066014}.
        	
        	\bibitem[Hunter et~al.(2021)Hunter, Krishnan, Liu, M\"obius, and
        	Fusco]{hunter_et_al_virus_host_travelling_waves}
        	Michael Hunter, Nikhil Krishnan, Tongfei Liu, Wolfram M\"obius, and Diana
        	Fusco.
        	\newblock Virus-host interactions shape viral dispersal giving rise to distinct
        	classes of traveling waves in spatial expansions.
        	\newblock \emph{Physical Review X}, 11:\penalty0 021066, 2021.
        	\newblock \doi{10.1103/PhysRevX.11.021066}.
        	
        	\bibitem[Bull et~al.(2018)Bull, Christensen, Scott, Jack, Crandall, and
        	Krone]{bull_et_al_phage_bacterial_dynamics}
        	James~J Bull, Kelly~A Christensen, Carly Scott, Benjamin~R Jack, Cameron~J
        	Crandall, and Stephen~M Krone.
        	\newblock Phage-bacterial dynamics with spatial structure: Self organization
        	around phage sinks can promote increased cell densities.
        	\newblock \emph{Antibiotics (Basel)}, 7\penalty0 (1), 2018.
        	\newblock \doi{10.3390/antibiotics7010008}.
        	
        	\bibitem[Haerter and
        	Sneppen(2012)]{haerter_and_sneppen_spatial_structure_lamarckian}
        	Jan~O. Haerter and Kim Sneppen.
        	\newblock Spatial structure and {Lamarckian} adaptation explain extreme genetic
        	diversity at crispr locus.
        	\newblock \emph{mBio}, 3\penalty0 (4):\penalty0 10.1128/mbio.00126--12, 2012.
        	\newblock \doi{10.1128/mbio.00126-12}.
        	
        	\bibitem[Coberly et~al.(2009)Coberly, Wei, Sampson, Millstein, Wichman, and
        	Krone]{coberly_et_al_space_time_host_evolution}
        	L~Caitlin Coberly, Wei Wei, Koffi~Y Sampson, Jack Millstein, Holly~A Wichman,
        	and Stephen~M Krone.
        	\newblock Space, time, and host evolution facilitate coexistence of competing
        	bacteriophages: theory and experiment.
        	\newblock \emph{The American Naturalist}, 173\penalty0 (4):\penalty0 E121--38,
        	2009.
        	\newblock \doi{10.1086/597226}.
        	
        	\bibitem[Guedj et~al.(2013)Guedj, Dahari, Rong, Sansone, Nettles, Cotler,
        	Layden, Uprichard, and Perelson]{guedj_et_al_hcv_daclatasvir}
        	Jeremie Guedj, Harel Dahari, Libin Rong, Natasha~D Sansone, Richard~E Nettles,
        	Scott~J Cotler, Thomas~J Layden, Susan~L Uprichard, and Alan~S Perelson.
        	\newblock Modeling shows that the {NS5A} inhibitor daclatasvir has two modes of
        	action and yields a shorter estimate of the hepatitis {C} virus half-life.
        	\newblock \emph{Proceedings of the National Academy of Sciences}, 110\penalty0
        	(10):\penalty0 3991--3996, 2013.
        	\newblock \doi{10.1073/pnas.1203110110}.
        	
        	\bibitem[Dahari et~al.(2011)Dahari, Guedj, Perelson, and
        	Layden]{dahari_et_al_HCV_in_the_era_of_DAAs}
        	Harel Dahari, Jeremie Guedj, Alan~S Perelson, and Thomas~J Layden.
        	\newblock Hepatitis {C} viral kinetics in the era of direct acting antiviral
        	agents and {IL28B}.
        	\newblock \emph{Current Hepatology Reports}, 10\penalty0 (3):\penalty0
        	214--227, 2011.
        	\newblock \doi{10.1007/s11901-011-0101-7}.
        	
        	\bibitem[Canini and Perelson(2014)]{canini_and_perelson_viral_kinetic}
        	Laetitia Canini and Alan~S. Perelson.
        	\newblock Viral kinetic modeling: state of the art.
        	\newblock \emph{Journal of Pharmacokinetics and Pharmacodynamics}, 41\penalty0
        	(5):\penalty0 431--443, 2014.
        	\newblock \doi{10.1007/s10928-014-9363-3}.
        	
        	\bibitem[Sattentau(2008)]{sattenau_avoiding_the_void}
        	Quentin Sattentau.
        	\newblock Avoiding the void: cell-to-cell spread of human viruses.
        	\newblock \emph{Nature Reviews Microbiology}, 6\penalty0 (11):\penalty0
        	815--826, 2008.
        	\newblock \doi{10.1038/nrmicro1972}.
        	
        	\bibitem[Graw et~al.(2014)Graw, Balagopal, Kandathil, Ray, Thomas, Ribeiro, and
        	Perelson]{graw_et_al_inferring_HCV}
        	Frederik Graw, Ashwin Balagopal, Abraham~J. Kandathil, Stuart~C. Ray, David~L.
        	Thomas, Ruy~M. Ribeiro, and Alan~S. Perelson.
        	\newblock Inferring viral dynamics in chronically hcv infected patients from
        	the spatial distribution of infected hepatocytes.
        	\newblock \emph{PLOS Computational Biology}, 10\penalty0 (11):\penalty0 1--15,
        	2014.
        	\newblock \doi{10.1371/journal.pcbi.1003934}.
        	
        	\bibitem[Schiffer et~al.(2018)Schiffer, Swan, Prlic, and
        	Lund]{schiffer_et_al_HSV2_review}
        	Joshua~T. Schiffer, David~A. Swan, Martin Prlic, and Jennifer~M. Lund.
        	\newblock Herpes simplex virus-2 dynamics as a probe to measure the extremely
        	rapid and spatially localized tissue-resident {T}-cell response.
        	\newblock \emph{Immunological Reviews}, 285\penalty0 (1):\penalty0 113--133,
        	2018.
        	\newblock \doi{https://doi.org/10.1111/imr.12672}.
        	
        	\bibitem[Smith and Perelson(2011)]{smith_and_perelson_influenza_review}
        	Amber~M Smith and Alan~S Perelson.
        	\newblock Influenza {A} virus infection kinetics: quantitative data and models.
        	\newblock \emph{Wiley Interdisciplinary Reviews: Systems Biology and Medicine},
        	3\penalty0 (4):\penalty0 429--445, 2011.
        	\newblock \doi{10.1002/wsbm.129}.
        	
        	\bibitem[Beauchemin(2006)]{beauchemin_well_mixed_assumption}
        	Catherine Beauchemin.
        	\newblock Probing the effects of the well-mixed assumption on viral infection
        	dynamics.
        	\newblock \emph{Journal of Theoretical Biology}, 242\penalty0 (2):\penalty0
        	464--477, 2006.
        	\newblock \doi{10.1016/j.jtbi.2006.03.014}.
        	
        	\bibitem[Jansens et~al.(2020)Jansens, Tishchenko, and
        	Favoreel]{jansens_et_al_virus_long_distance_TNTs}
        	Robert J~J Jansens, Alexander Tishchenko, and Herman~W Favoreel.
        	\newblock Bridging the gap: Virus long-distance spread via tunneling nanotubes.
        	\newblock \emph{Journal of Virology}, 94\penalty0 (8), 2020.
        	\newblock \doi{10.1128/JVI.02120-19}.
        	
        	\bibitem[Kumar et~al.(2017)Kumar, Kim, Ranjan, Metcalfe, Cao, Mishina,
        	Gangappa, Guo, Boyden, Zaki, York, Garc{\'\i}a-Sastre, Shaw, and
        	Sambhara]{kumar_influenza_TNTs}
        	Amrita Kumar, Jin~Hyang Kim, Priya Ranjan, Maureen~G Metcalfe, Weiping Cao,
        	Margarita Mishina, Shivaprakash Gangappa, Zhu Guo, Edward~S Boyden, Sherif
        	Zaki, Ian York, Adolfo Garc{\'\i}a-Sastre, Michael Shaw, and Suryaprakash
        	Sambhara.
        	\newblock Influenza virus exploits tunneling nanotubes for cell-to-cell spread.
        	\newblock \emph{Scientific reports}, 7:\penalty0 40360--40360, 2017.
        	\newblock \doi{10.1038/srep40360}.
        	
        	\bibitem[Kongsomros et~al.(2021)Kongsomros, Manopwisedjaroen, Chaopreecha,
        	Wang, Borwornpinyo, and
        	Thitithanyanont]{kongsomros_et_al_trogocytosis_influenza}
        	Supasek Kongsomros, Suwimon Manopwisedjaroen, Jarinya Chaopreecha, Sheng-Fan
        	Wang, Suparerk Borwornpinyo, and Arunee Thitithanyanont.
        	\newblock Rapid and efficient cell-to-cell transmission of avian influenza
        	{H5N1} virus in {MDCK} cells is achieved by trogocytosis.
        	\newblock \emph{Pathogens}, 10\penalty0 (4), 2021.
        	\newblock \doi{10.3390/pathogens10040483}.
        	
        	\bibitem[Tiwari et~al.(2021)Tiwari, Koganti, Russell, Sharma, and
        	Shukla]{tiwari_et_al_TNTs_review}
        	Vaibhav Tiwari, Raghuram Koganti, Greer Russell, Ananya Sharma, and Deepak
        	Shukla.
        	\newblock Role of tunneling nanotubes in viral infection, neurodegenerative
        	disease, and cancer.
        	\newblock \emph{Frontiers in Immunology}, 12:\penalty0 2256, 2021.
        	\newblock \doi{10.3389/fimmu.2021.680891}.
        	
        	\bibitem[Graw et~al.(2015)Graw, Martin, Perelson, Uprichard, Dahari, and
        	Doms]{graw_et_al_hcv_cell_to_cell_stochastic}
        	Frederik Graw, Danyelle~N. Martin, Alan~S. Perelson, Susan~L. Uprichard, Harel
        	Dahari, and R.~W. Doms.
        	\newblock Quantification of hepatitis {C} virus cell-to-cell spread using a
        	stochastic modeling approach.
        	\newblock \emph{Journal of Virology}, 89\penalty0 (13):\penalty0 6551--6561,
        	2015.
        	\newblock \doi{10.1128/JVI.00016-15}.
        	
        	\bibitem[Williams et~al.(2024{\natexlab{b}})Williams, McCaw, and
        	Osborne]{williams_et_al_inference}
        	Thomas Williams, James~M. McCaw, and James~M. Osborne.
        	\newblock Spatial information allows inference of the prevalence of direct
        	cell--to--cell viral infection.
        	\newblock \emph{PLOS Computational Biology}, 20\penalty0 (7):\penalty0 1--35,
        	2024{\natexlab{b}}.
        	\newblock \doi{10.1371/journal.pcbi.1012264}.
        	
        	\bibitem[Zeng et~al.(2022)Zeng, Evans, King, Zheng, Oltz, Whelan, Saif,
        	Peeples, and Liu]{zeng_et_al_sars_cov_2_cell_to_cell}
        	Cong Zeng, John~P. Evans, Tiffany King, Yi-Min Zheng, Eugene~M. Oltz, Sean
        	P.~J. Whelan, Linda~J. Saif, Mark~E. Peeples, and Shan-Lu Liu.
        	\newblock {SARS-CoV-2} spreads through cell-to-cell transmission.
        	\newblock \emph{Proceedings of the National Academy of Sciences}, 119\penalty0
        	(1):\penalty0 e2111400119, 2022.
        	\newblock \doi{10.1073/pnas.2111400119}.
        	
        	\bibitem[Dixit and Perelson(2004)]{dixit_and_perelon_HIV_patients}
        	Narendra~M Dixit and Alan~S Perelson.
        	\newblock Complex patterns of viral load decay under antiretroviral therapy:
        	influence of pharmacokinetics and intracellular delay.
        	\newblock \emph{Journal of Theoretical Biology}, 226\penalty0 (1):\penalty0
        	95--109, 2004.
        	\newblock \doi{10.1016/j.jtbi.2003.09.002}.
        	
        	\bibitem[Segredo-Otero and
        	Sanju{\'a}n(2020)]{segredo_otero_et_al_spatial_structure_innnate_immunity}
        	Ernesto Segredo-Otero and Rafael Sanju{\'a}n.
        	\newblock The role of spatial structure in the evolution of viral innate
        	immunity evasion: A diffusion-reaction cellular automaton model.
        	\newblock \emph{PLOS Computational Biology}, 16\penalty0 (2):\penalty0 1--18,
        	2020.
        	\newblock \doi{10.1371/journal.pcbi.1007656}.
        	
        	\bibitem[Morselli et~al.(2024)Morselli, Delitalia, Jenner, and
        	Frascoli]{morselli_et_al_hybrid_discrete_immune}
        	David Morselli, Marcello~E. Delitalia, L.~Jenner, Adrienne, and Federico
        	Frascoli.
        	\newblock A hybrid discrete-continuum modelling approach for the interactions
        	of the immune system with oncolytic viral infections [pre-print].
        	\newblock \emph{arXiv}, page 2404.06459, 2024.
        	
        	\bibitem[Capit{\'a}n et~al.(2011)Capit{\'a}n, Cuesta, Manrubia, and
        	Aguirre]{capitan_et_al_severe_hindrance_viral_infection_propagation}
        	Jos{\'e}~A. Capit{\'a}n, Jos{\'e}~A. Cuesta, Susanna~C. Manrubia, and Jacobo
        	Aguirre.
        	\newblock Severe hindrance of viral infection propagation in spatially extended
        	hosts.
        	\newblock \emph{PLOS ONE}, 6\penalty0 (8):\penalty0 1--13, 2011.
        	\newblock \doi{10.1371/journal.pone.0023358}.
        	
        	\bibitem[Phan and Wodarz(2015)]{phan_and_wodarz_modelling_multiple_infection}
        	Dustin Phan and Dominik Wodarz.
        	\newblock Modeling multiple infection of cells by viruses: Challenges and
        	insights.
        	\newblock \emph{Mathematical Biosciences}, 264:\penalty0 21--28, 2015.
        	\newblock \doi{https://doi.org/10.1016/j.mbs.2015.03.001}.
        	
        	\bibitem[Yin(1994)]{yin_spatially_resolved_evolution}
        	John Yin.
        	\newblock Spatially resolved evolution of viruses.
        	\newblock \emph{Annals of the New York Academy of Sciences}, 745\penalty0
        	(1):\penalty0 399--408, 1994.
        	\newblock \doi{https://doi.org/10.1111/j.1749-6632.1994.tb44392.x}.
        	
        	\bibitem[Sardany\'{e}s et~al.(2008)Sardany\'{e}s, Elena, and
        	Sol\'{e}]{sardanyes_et_al_simple_quasispecies_survival_flattest}
        	Josep Sardany\'{e}s, Santiago~F Elena, and Ricard~V Sol\'{e}.
        	\newblock Simple quasispecies models for the survival-of-the-flattest effect:
        	The role of space.
        	\newblock \emph{Journal of Theoretical Biology}, 250\penalty0 (3):\penalty0
        	560--568, 2008.
        	\newblock \doi{10.1016/j.jtbi.2007.10.027}.
        	
        	\bibitem[Komarova(2007)]{komarova_viral_reproductive_strategies}
        	Natalia~L. Komarova.
        	\newblock Viral reproductive strategies: How can lytic viruses be
        	evolutionarily competitive?
        	\newblock \emph{Journal of Theoretical Biology}, 249\penalty0 (4):\penalty0
        	766--784, 2007.
        	\newblock \doi{https://doi.org/10.1016/j.jtbi.2007.09.013}.
        	
        	\bibitem[Cuesta et~al.(2011)Cuesta, Aguirre, Capit\'an, and
        	Manrubia]{cuesta_et_al_struggle_for_space}
        	Jos\'e~A. Cuesta, Jacobo Aguirre, Jos\'e~A. Capit\'an, and Susanna~C. Manrubia.
        	\newblock Struggle for space: Viral extinction through competition for cells.
        	\newblock \emph{Physical Review Letters}, 106:\penalty0 028104, 2011.
        	\newblock \doi{10.1103/PhysRevLett.106.028104}.
        	
        	\bibitem[Hucka et~al.(2003)Hucka, Finney, Sauro, Bolouri, Doyle, Kitano, Arkin,
        	Bornstein, Bray, Cornish-Bowden, Cuellar, Dronov, Gilles, Ginkel, Gor,
        	Goryanin, Hedley, Hodgman, Hofmeyr, Hunter, Juty, Kasberger, Kremling,
        	Kummer, Le~Nov{\`e}re, Loew, Lucio, Mendes, Minch, Mjolsness, Nakayama,
        	Nelson, Nielsen, Sakurada, Schaff, Shapiro, Shimizu, Spence, Stelling,
        	Takahashi, Tomita, Wagner, Wang, and {the rest of the SBML
        		Forum}]{hucka_et_al_SBML}
        	M.~Hucka, A.~Finney, H.~M. Sauro, H.~Bolouri, J.~C. Doyle, H.~Kitano, A.~P.
        	Arkin, B.~J. Bornstein, D.~Bray, A.~Cornish-Bowden, A.~A. Cuellar, S.~Dronov,
        	E.~D. Gilles, M.~Ginkel, V.~Gor, I.~I. Goryanin, W.~J. Hedley, T.~C. Hodgman,
        	J.-H. Hofmeyr, P.~J. Hunter, N.~S. Juty, J.~L. Kasberger, A.~Kremling,
        	U.~Kummer, N.~Le~Nov{\`e}re, L.~M. Loew, D.~Lucio, P.~Mendes, E.~Minch, E.~D.
        	Mjolsness, Y.~Nakayama, M.~R. Nelson, P.~F. Nielsen, T.~Sakurada, J.~C.
        	Schaff, B.~E. Shapiro, T.~S. Shimizu, H.~D. Spence, J.~Stelling,
        	K.~Takahashi, M.~Tomita, J.~Wagner, J.~Wang, and {the rest of the SBML
        		Forum}.
        	\newblock The systems biology markup language {(SBML)}: a medium for
        	representation and exchange of biochemical network models.
        	\newblock \emph{Bioinformatics}, 19\penalty0 (4):\penalty0 524--531, 2003.
        	\newblock \doi{10.1093/bioinformatics/btg015}.
        	
        	\bibitem[Cuellar et~al.(2003)Cuellar, Lloyd, Nielsen, Bullivant, Nickerson, and
        	Hunter]{cuellar_et_al_CellML}
        	Autumn~A. Cuellar, Catherine~M. Lloyd, Poul~F. Nielsen, David~P. Bullivant,
        	David~P. Nickerson, and Peter~J. Hunter.
        	\newblock An overview of cellml 1.1, a biological model description language.
        	\newblock \emph{SIMULATION}, 79\penalty0 (12):\penalty0 740--747, 2003.
        	\newblock \doi{10.1177/0037549703040939}.
        	
        	\bibitem[Harris et~al.(2016)Harris, Hogg, Tapia, Sekar, Gupta, Korsunsky,
        	Arora, Barua, Sheehan, and Faeder]{harris_et_al_bionetgen}
        	Leonard~A. Harris, Justin~S. Hogg, Jos{\'e}-Juan Tapia, John A.~P. Sekar,
        	Sanjana Gupta, Ilya Korsunsky, Arshi Arora, Dipak Barua, Robert~P. Sheehan,
        	and James~R. Faeder.
        	\newblock {BioNetGen} 2.2: advances in rule-based modeling.
        	\newblock \emph{Bioinformatics}, 32\penalty0 (21):\penalty0 3366--3368, 2016.
        	\newblock \doi{10.1093/bioinformatics/btw469}.
        	
        	\bibitem[Baccam et~al.(2006)Baccam, Beauchemin, Macken, Hayden, and
        	Perelson]{baccam_et_al_influenza_kinetics}
        	Prasith Baccam, Catherine Beauchemin, Catherine~A. Macken, Frederick~G. Hayden,
        	and Alan~S. Perelson.
        	\newblock Kinetics of influenza {A} virus infection in humans.
        	\newblock \emph{Journal of Virology}, 80\penalty0 (15):\penalty0 7590--7599,
        	2006.
        	\newblock \doi{10.1128/JVI.01623-05}.
        	
        	\bibitem[Beauchemin and Handel(2011)]{beauchemin_and_handel_influenza_review}
        	Catherine~AA Beauchemin and Andreas Handel.
        	\newblock A review of mathematical models of influenza {A} infections within a
        	host or cell culture: lessons learned and challenges ahead.
        	\newblock \emph{BMC Public Health}, 11\penalty0 (1):\penalty0 S7, 2011.
        	\newblock \doi{10.1186/1471-2458-11-S1-S7}.
        	
        	\bibitem[You and Yin(1999)]{you_et_al_amplification_and_spread}
        	L~You and J~Yin.
        	\newblock Amplification and spread of viruses in a growing plaque.
        	\newblock \emph{Journal of Theoretical Biology}, 200\penalty0 (4):\penalty0
        	365--373, 1999.
        	\newblock \doi{10.1006/jtbi.1999.1001}.
        	
        	\bibitem[Yin and
        	McCaskill(1992)]{yin_and_mccaskill_replication_of_viruses_in_a_growing_plaque}
        	J~Yin and J~S McCaskill.
        	\newblock Replication of viruses in a growing plaque: a reaction-diffusion
        	model.
        	\newblock \emph{Biophysical Journal}, 61\penalty0 (6):\penalty0 1540--1549,
        	1992.
        	\newblock \doi{10.1016/S0006-3495(92)81958-6}.
        	
        	\bibitem[Tran et~al.(2013)Tran, Moser, Poole, and
        	Mehle]{tran_et_al_in_vivo_reporter_influenza}
        	Vy~Tran, Lindsey~A. Moser, Daniel~S. Poole, and Andrew Mehle.
        	\newblock Highly sensitive real-time in vivo imaging of an influenza reporter
        	virus reveals dynamics of replication and spread.
        	\newblock \emph{Journal of Virology}, 87\penalty0 (24):\penalty0 13321--13329,
        	2013.
        	\newblock \doi{10.1128/JVI.02381-13}.
        	
        	\bibitem[Nogales et~al.(2019)Nogales, {\'A}vila-P{\'e}rez, Rangel-Moreno,
        	Chiem, DeDiego, Mart{\'\i}nez-Sobrido, and
        	Williams]{nogales_et_al_flourescent_and_bioluminescent_influenza_reporter}
        	Aitor Nogales, Gines {\'A}vila-P{\'e}rez, Javier Rangel-Moreno, Kevin Chiem,
        	Marta~L. DeDiego, Luis Mart{\'\i}nez-Sobrido, and Bryan R.~G. Williams.
        	\newblock A novel fluorescent and bioluminescent bireporter influenza a virus
        	to evaluate viral infections.
        	\newblock \emph{Journal of Virology}, 93\penalty0 (10):\penalty0 e00032--19,
        	2019.
        	\newblock \doi{10.1128/JVI.00032-19}.
        	
        	\bibitem[Ye et~al.(2021)Ye, Chiem, Park, Silvas, Morales~Vasquez, Sourimant,
        	Lin, Greninger, Plemper, Torrelles, Kobie, Walter, de~la Torre, and
        	Martinez-Sobrido]{ye_et_al_covid_reporter_virus_in_mice}
        	Chengjin Ye, Kevin Chiem, Jun-Gyu Park, Jesus~A. Silvas, Desarey
        	Morales~Vasquez, Julien Sourimant, Michelle~J. Lin, Alexander~L. Greninger,
        	Richard~K. Plemper, Jordi~B. Torrelles, James~J. Kobie, Mark~R. Walter,
        	Juan~Carlos de~la Torre, and Luis Martinez-Sobrido.
        	\newblock Analysis of {SARS-CoV-2} infection dynamic in vivo using
        	reporter-expressing viruses.
        	\newblock \emph{Proceedings of the National Academy of Sciences}, 118\penalty0
        	(41), 2021.
        	\newblock \doi{10.1073/pnas.2111593118}.
        	
        	\bibitem[Lienenklaus et~al.(2009)Lienenklaus, Cornitescu, Zietara,
        	{\L}yszkiewicz, Gekara, Jab{\l}{\'o}nska, Edenhofer, Rajewsky, Bruder,
        	Hafner, Staeheli, and Weiss]{lienenklaus_et_al_ifn_beta_mouse}
        	Stefan Lienenklaus, Marius Cornitescu, Natalia Zietara, Marcin {\L}yszkiewicz,
        	Nelson Gekara, Jadwiga Jab{\l}{\'o}nska, Frank Edenhofer, Klaus Rajewsky,
        	Dunja Bruder, Martin Hafner, Peter Staeheli, and Siegfried Weiss.
        	\newblock Novel reporter mouse reveals constitutive and inflammatory expression
        	of {IFN-beta} in vivo.
        	\newblock \emph{Journal of Immunology}, 183\penalty0 (5):\penalty0 3229--3236,
        	2009.
        	\newblock \doi{10.4049/jimmunol.0804277}.
        	
        	\bibitem[Kumagai et~al.(2007)Kumagai, Takeuchi, Kato, Kumar, Matsui, Morii,
        	Aozasa, Kawai, and Akira]{kumagai_et_al_alveolar_macrophages_ifn_alpha}
        	Yutaro Kumagai, Osamu Takeuchi, Hiroki Kato, Himanshu Kumar, Kosuke Matsui,
        	Eiichi Morii, Katsuyuki Aozasa, Taro Kawai, and Shizuo Akira.
        	\newblock Alveolar macrophages are the primary interferon-$\alpha$ producer in
        	pulmonary infection with {RNA} viruses.
        	\newblock \emph{Immunity}, 27\penalty0 (2):\penalty0 240--252, 2007.
        	\newblock \doi{https://doi.org/10.1016/j.immuni.2007.07.013}.
        	
        	\bibitem[Jonsson et~al.(2012)Jonsson, Camp, Wu, Zheng, Kraenzle, Biller,
        	Vanover, Chu, Ng, Proctor, Sherwood, Steffen, and
        	Mollura]{jonsson_et_al_molecular_imaging_ferrets_influenza}
        	Colleen~B. Jonsson, Jeremy~V. Camp, Albert Wu, Huaiyu Zheng, Jennifer~L.
        	Kraenzle, Ashley~E. Biller, Carol~D. Vanover, Yong-Kyu Chu, Chin~K. Ng, Mary
        	Proctor, Leslie Sherwood, Marlene~C. Steffen, and Daniel~J. Mollura.
        	\newblock Molecular imaging reveals a progressive pulmonary inflammation in
        	lower airways in ferrets infected with 2009 {H1N1} pandemic influenza virus.
        	\newblock \emph{PLOS ONE}, 7\penalty0 (7):\penalty0 1--12, 2012.
        	\newblock \doi{10.1371/journal.pone.0040094}.
        	
        	\bibitem[Alamoudi et~al.(2023)Alamoudi, Sch{\"a}lte, M{\"u}ller, Starru{\ss},
        	Bundgaard, Graw, Brusch, and Hasenauer]{alamoudi_et_al_fitmulticell}
        	Emad Alamoudi, Yannik Sch{\"a}lte, Robert M{\"u}ller, J{\"o}rn Starru{\ss},
        	Nils Bundgaard, Frederik Graw, Lutz Brusch, and Jan Hasenauer.
        	\newblock Fitmulticell: simulating and parameterizing computational models of
        	multi-scale and multi-cellular processes.
        	\newblock \emph{Bioinformatics}, 39\penalty0 (11):\penalty0 btad674, 2023.
        	\newblock \doi{10.1093/bioinformatics/btad674}.
        	
        	\bibitem[Toni et~al.(2009)Toni, Welch, Strelkowa, Ipsen, and
        	Stumpf]{toni_et_al_PMC_ABC}
        	Tina Toni, David Welch, Natalja Strelkowa, Andreas Ipsen, and Michael~P.H
        	Stumpf.
        	\newblock Approximate bayesian computation scheme for parameter inference and
        	model selection in dynamical systems.
        	\newblock \emph{Journal of the Royal Society Interface}, 6\penalty0
        	(31):\penalty0 187--202, 2009.
        	\newblock \doi{10.1098/rsif.2008.0172}.
        	
        	\bibitem[Kypraios et~al.(2017)Kypraios, Neal, and
        	Prangle]{kypraios_et_al_tutorial_intro_bayesian}
        	Theodore Kypraios, Peter Neal, and Dennis Prangle.
        	\newblock A tutorial introduction to {Bayesian} inference for stochastic
        	epidemic models using approximate bayesian computation.
        	\newblock \emph{Mathematical Biosciences}, 287:\penalty0 42--53, 2017.
        	\newblock \doi{10.1016/j.mbs.2016.07.001}.
        	
        	\bibitem[{Stan Development Team}(2023)]{stan_package}
        	{Stan Development Team}.
        	\newblock Stan modeling language users guide and reference manual, {Version
        		2.33 (R). Available from: https://mc-stan.org. Last accessed July 10, 2024},
        	2023.
        	
        	\bibitem[Klinger et~al.(2018)Klinger, Rickert, and
        	Hasenauer]{klinger_et_al_pyABC}
        	Emmanuel Klinger, Dennis Rickert, and Jan Hasenauer.
        	\newblock pyabc: distributed, likelihood-free inference.
        	\newblock \emph{Bioinformatics}, 34\penalty0 (20):\penalty0 3591--3593, 2018.
        	\newblock \doi{10.1093/bioinformatics/bty361}.
        	
        	\bibitem[Dutta et~al.(2017)Dutta, Schoengens, Onnela, and
        	Mira]{dutta_et_al_ABCpy}
        	Ritabrata Dutta, Marcel Schoengens, Jukka-Pekka Onnela, and Antonietta Mira.
        	\newblock {ABCpy}: A user-friendly, extensible, and parallel library for
        	approximate bayesian computation.
        	\newblock In \emph{Proceedings of the Platform for Advanced Scientific
        		Computing Conference}, 2017.
        	
        	\bibitem[Schmiester et~al.(2021)Schmiester, Sch{\"a}lte, Bergmann, Camba,
        	Dudkin, Egert, Fr{\"o}hlich, Fuhrmann, Hauber, Kemmer, Lakrisenko, Loos,
        	Merkt, M{\"u}ller, Pathirana, Raim{\'u}ndez, Refisch, Rosenblatt, Stapor,
        	St{\"a}dter, Wang, Wieland, Banga, Timmer, Villaverde, Sahle, Kreutz,
        	Hasenauer, and Weindl]{schmeister_et_al_PEtab}
        	Leonard Schmiester, Yannik Sch{\"a}lte, Frank~T. Bergmann, Tacio Camba, Erika
        	Dudkin, Janine Egert, Fabian Fr{\"o}hlich, Lara Fuhrmann, Adrian~L. Hauber,
        	Svenja Kemmer, Polina Lakrisenko, Carolin Loos, Simon Merkt, Wolfgang
        	M{\"u}ller, Dilan Pathirana, Elba Raim{\'u}ndez, Lukas Refisch, Marcus
        	Rosenblatt, Paul~L. Stapor, Philipp St{\"a}dter, Dantong Wang, Franz-Georg
        	Wieland, Julio~R. Banga, Jens Timmer, Alejandro~F. Villaverde, Sven Sahle,
        	Clemens Kreutz, Jan Hasenauer, and Daniel Weindl.
        	\newblock {PEtab} -- interoperable specification of parameter estimation
        	problems in systems biology.
        	\newblock \emph{PLOS Computational Biology}, 17\penalty0 (1):\penalty0 1--10,
        	2021.
        	\newblock \doi{10.1371/journal.pcbi.1008646}.
        	
        	\bibitem[Keegan et~al.(2025)Keegan, Caramizaru, Mack, Andriushchenko,
        	Schreiner, and Patel]{keegan_et_al_spatial_model_editor}
        	Liam Keegan, Horea Caramizaru, Harald Mack, Petr Andriushchenko, Henry
        	Schreiner, and Hrishikesh Patel.
        	\newblock spatial-model-editor/spatial-model-editor: 1.9.0 (1.9.0). {Zenodo.},
        	2025.
        	
        \end{thebibliography}
    }
    	
\end{document}